\documentclass[11pt,a4paper]{article}
\pdfoutput=1
\usepackage{jheppub}

\usepackage[english]{babel}
\usepackage{xspace}
\usepackage{epsfig}
\usepackage{amssymb,amsmath}
\usepackage{url,booktabs}
\usepackage{color}
\usepackage{soul}
\usepackage{slashed}
\usepackage{subfig}
\usepackage{xfrac}
\usepackage{wrapfig}
\usepackage{tikz}
\usepackage{booktabs}
\usepackage{hyperref}

\usepackage{array}
\newcolumntype{L}[1]{>{\raggedright\let\newline\\\arraybackslash\hspace{0pt}}m{#1}}
\newcolumntype{C}[1]{>{\centering\let\newline\\\arraybackslash\hspace{0pt}}m{#1}}
\newcolumntype{R}[1]{>{\raggedleft\let\newline\\\arraybackslash\hspace{0pt}}m{#1}}

\def\be{\begin{equation}}
\def\ee{\end{equation}}
\def\bea{\begin{eqnarray}}
\def\eea{\end{eqnarray}}
\def\beal{\begin{align}}
\def\eeal{\end{align}}

\newcommand{\Lag}{\ensuremath{\mathcal{L}}}
\newcommand{\Dcov}{\ensuremath{\textnormal{D}}}
\newcommand{\dInt}{\mathrm{d}}
\newcommand{\doubleprime}{{\prime\prime}}
\newcommand{\DM}{\textnormal{DM}}
\newcommand{\HL}{\textnormal{HL}}
\newcommand{\eff}{\textnormal{eff}}
\newcommand{\OMEGA}[1]{\ensuremath{h^2\Omega_{\textnormal{#1}}}}
\newcommand{\Zp}{\ensuremath{{Z^\prime}}}
\newcommand{\software}[1]{\texttt{#1}}
\newcommand{\experiment}[1]{\emph{#1}}
\newcommand{\mathfig}[1]{\vcenter{\hbox{#1}}}

\graphicspath{{Plots/}}
\relax

\title{Leptophilic dark matter from gauged lepton number: Phenomenology and gravitational wave signatures}

\author[a]{Eric Madge}
\author[a]{and Pedro Schwaller}

\affiliation[a]{PRISMA Cluster of Excellence \& Mainz Institute for Theoretical Physics, Johannes Gutenberg University, 55099 Mainz, Germany}
\emailAdd{eric.madge@uni-mainz.de}
\emailAdd{pedro.schwaller@uni-mainz.de}

\abstract{New gauge symmetries often appear in theories beyond the Standard Model. Here we study a model where lepton number is promoted to a gauge symmetry. Anomaly cancellation requires the introduction of additional leptons, the lightest of which is a natural leptophilic dark matter candidate. We perform a comprehensive study of both collider and dark matter phenomenology. Furthermore we find that the model exhibits a first order lepton number breaking phase transition in large regions of parameter space. The corresponding gravitational wave signal is computed, and its detectability at \experiment{LISA} and other future GW detectors assessed. Finally we comment on the complementarity of dark matter, collider and gravitational wave observables, and on the potential reach of future colliders.}

\arxivnumber{}

\begin{document}
\begin{flushright}
MITP/18-088
\end{flushright}
\maketitle
\newpage
\section{Introduction}
\label{sec:intro}
%
%
After more than 8 years of searches for new physics at the CERN \experiment{Large Hadron Collider (LHC)}, the Standard Model (SM) of particle physics remains undefeated. While new physics remains elusive, astrophysical observations have firmly established the existence of an additional form of matter, the so called dark matter (DM). The \experiment{LHC}, as well as direct searches for DM interactions with nuclei, place strong constraints on the interaction of the DM with quarks. On the other hand a sufficiently large interaction rate with some SM particles is required to explain the amount of DM observed in the universe, at least in the well motivated scenario where DM is a thermal relic. 

This suggests a DM candidate that dominantly couples to leptons, leptophilic dark matter~\cite{Fox:2008kb,Chang:2014tea,Bell:2014tta}. If lepton number is realised as a gauge symmetry, such a DM candidate arises naturally among the additional fermions that must be introduced to ensure anomaly cancellation~\cite{Schwaller:2013hqa}.\footnote{Various ways of anomaly free gauging of lepton number can be found in the literature~\cite{Foot:1989ts,FileviezPerez:2010gw,Dulaney:2010dj,FileviezPerez:2011pt,Duerr:2013dza,Chao:2010mp,Chang:2018vdd,Chang:2018nid,Perez:2014qfa,Ohmer:2015lxa}.} New physics that couples dominantly to leptons is also motivated by the persistence of the muon $g-2$ anomaly, in models of neutrino mass generation, and by the current hints for lepton flavour universality violation in B-physics observables. Some recent work in this direction can e.g.\ be found in~\cite{Agrawal:2014ufa,Bai:2014osa,Freitas:2014jla,Lee:2014tba,Cao:2014cda,Kile:2014jea,Aranda:2014zta,Jeong:2015bbi,Cavasonza:2016qem,DEramo:2017zqw,Rawat:2017fak,Duan:2017pkq,Chen:2018vkr}. 

Since we do not observe a massless gauge boson other than the photon, lepton number must be broken spontaneously at some higher scale. If the corresponding phase transition (PT) is first order, a stochastic gravitational wave (GW) signal will be generated in the early Universe, and might potentially be detectable at \experiment{LISA} or other future GW experiments. This first order PT could also provide the out-of-equilibrium condition necessary for successful baryogenesis, as was demonstrated recently in a model of non-abelian gauged lepton number~\cite{Fornal:2017owa}. 

Here, our goal is to identify regions of parameter space of the gauged lepton number model where the GW signal from the PT is sufficiently strong to be detectable by \experiment{LISA}, and where at the same time the model provides a DM candidate consistent with all current experimental constraints. The possibility to probe dark sector physics using GW signals was proposed in~\cite{Schwaller:2015tja} and explored further in~\cite{Jaeckel:2016jlh,Chala:2016ykx,Addazi:2016fbj,Baldes:2017rcu,Addazi:2017gpt,Tsumura:2017knk,Aoki:2017aws,Croon:2018erz,Baldes:2018emh}. Working with a renormalizable and perturbative model, we can predict the GW signal as well as the DM abundance, direct detection rates and collider constraints. Since we require that DM is a thermal relic, the dark sector is in thermal equilibrium with the SM during the PT, however the GW signal is largely independent of the nature of the electroweak PT. In a separate work, the scenario with a PT in a decoupled dark sector is considered~\cite{Breitbach:2018ddu}. 

Our paper is organised as follows: In section~\ref{sec:models}, we introduce the model and discuss constraints on the lepton number gauge coupling from RGE running. Dark matter properties and collider constraints are studied in sections~\ref{sec:dm} and~\ref{sec:pheno}, respectively. The lepton number breaking PT is investigated in section~\ref{sec:pt}, and the resulting GW signal and its detectability is computed in section~\ref{sec:gws}, before we present our conclusions.
%
%
\section{The model}
\label{sec:models}
%
%

The model considered here has been introduced in~\cite{Schwaller:2013hqa}. In this model, the SM gauge group is extended by an additional $U(1)_\ell$ lepton number gauge group under which all SM leptons including three generations of right-handed neutrinos carry unit charge, whereas the other SM fields are neutral. Lepton number is spontaneously broken by an SM singlet scalar field, giving mass to the lepton number gauge boson. Additional fermionic fields are added to cancel gauge anomalies. These additional fields are vector-like under the SM  gauge group.

\subsection{Gauge sector}
\label{sec:gauge}
%
The model is based on the gauge group $SU(3)_c \otimes SU(2)_W \otimes U(1)_Y \otimes U(1)_\ell$.
Omitting QCD, the gauge sector of the Lagrangian is given by
\be
	\Lag \supset - \frac{1}{4} W^{a}_{\mu\nu} W^{a\,\mu\nu} 
	             - \frac{1}{4} \hat{B}_{\mu\nu} \hat{B}^{\mu\nu}
	             - \frac{1}{4} \hat{Z}_{\ell\,\mu\nu} \hat{Z}_\ell^{\mu\nu}
	             + \frac{\epsilon}{2} \hat{B}_{\mu\nu} \hat{Z}_\ell^{\mu\nu}
	,
\ee
where $W^a$ and $\hat{B}$ are the gauge bosons of the SM weak and hypercharge gauge group respectively, and $\hat{Z}_\ell$ is the lepton number gauge boson.
The $\frac{\epsilon}{2} \hat{B}_{\mu\nu} \hat{Z}_\ell^{\mu\nu}$ term leads to kinetic mixing between the hypercharge and lepton number $U(1)$ gauge bosons.
The kinetic terms can be diagonalized by a $GL(2,\mathbb{R})$ transformation~\cite{Babu:1997st}
\be
  \begin{pmatrix}\hat{B}\\\hat{Z}_\ell\end{pmatrix} = 
	\begin{pmatrix}
		1 & \frac{\epsilon}{\sqrt{1-\epsilon^2}} \\
		0 & \frac{1}{\sqrt{1-\epsilon^2}}
	\end{pmatrix}
	\begin{pmatrix}B\\Z_\ell\end{pmatrix} 
	.
\ee 
The hats denote fields in the gauge basis and the unhatted fields are in the basis where the kinetic terms are diagonal and canonically normalized.

The model further features two scalar fields: the SM Higgs doublet transforming under $SU(2)_W \otimes U(1)_Y \otimes U(1)_\ell$ as $H \sim (2,1/2,0)$, and a complex scalar $\Phi \sim (1,0,L_\Phi)$ which is an SM singlet with lepton number $L_\Phi$. 
We let both fields acquire a vacuum expectation value (VEV) given by $\left<H\right> = (0,v_H/\sqrt{2})$ and $\left<\Phi\right> = v_\Phi/\sqrt{2}$, thus breaking the electroweak and lepton number gauge group $SU(2)_W \otimes U(1)_Y \otimes U(1)_\ell$ to $U(1)_{\textnormal{EM}}$ electromagnetism.
The gauge bosons then obtain masses from the kinetic terms of the scalar fields, with the covariant derivative given by
\be
\Dcov_\mu = \partial_\mu - i g_2 W^a_\mu T^a - i g_1 Y \hat{B}_\mu - i g_\ell L \hat{Z}_\ell.
\ee 
Here, $g_2$, $g_1$ and $g_\ell$ are the gauge couplings of the $SU(2)_W$, $U(1)_Y$ and $U(1)_\ell$ gauge groups respectively.
The $W$ mass is the same as in the SM, $m_W = \frac{1}{2} g_2 v_H$, whereas the mass matrix for the remaining gauge fields in the kinetic eigenbasis $(W^3, B, Z_\ell)$ is given by
\be
	M^2_{\textnormal{GB}} = 
	\begin{pmatrix}
		\frac{g_2^2 v_H^2}{4} & -\frac{g_1 g_2 v_H^2}{4}  & -\frac{\epsilon g_1 g_2 v_H^2}{4 \sqrt{1-\epsilon^2}} \\
		-\frac{g_1 g_2 v_H^2}{4}  & \frac{g_1^2 v_H^2}{4} & \frac{\epsilon g_1^2 v_H^2}{4 \sqrt{1-\epsilon^2}} \\
		-\frac{\epsilon g_1 g_2 v_H^2}{4 \sqrt{1-\epsilon^2}} & \frac{\epsilon g_1^2 v_H^2}{4 \sqrt{1-\epsilon^2}} & \frac{g_\ell^2 L_\Phi^2 v_\Phi^2}{1-\epsilon^2} + \frac{\epsilon g_1^2 v_H^2}{4 (1-\epsilon^2)} 
	\end{pmatrix}\ .
\ee
The upper-left $2\times 2$ submatrix is diagonalized by the SM weak mixing angle. 
If $\epsilon\neq 0$, the resulting $Z_\textnormal{SM}$ boson is mixing with the $Z_\ell$ boson.
The kinetic eigenstates are related to the physical mass eigenstates by
\be
  \begin{pmatrix} W^3 \\ B \\ Z_\ell \end{pmatrix}
  = \begin{pmatrix}
  	c_W c_\xi & s_W & -c_W s_\xi \\
  	- s_W c_\xi & c_W & s_W s_\xi \\
  	s_\xi & 0 & c_\xi
  \end{pmatrix}
  \begin{pmatrix} Z \\ A \\ Z^\prime \end{pmatrix}
  ,
\ee
where $c_W=\cos\theta_W = \frac{g_2}{\sqrt{g_1^2+g_2^2}}$, $s_W=\sin\theta_W = \frac{g_1}{\sqrt{g_1^2+g_2^2}}$, $c_\xi = \cos\xi$ and $s_\xi = \sin\xi$.
Defining $M_{Z_\textnormal{SM}}^2 = \frac{g_1^2+g_2^2}{4}v_H^2$, $M_{Z_\ell}^2 = g_\ell^2 L_\Phi^2 v_\Phi^2$, $M_B^2 = g_1^2 v_H^2 / 4$ and $\eta = \epsilon/\sqrt{1-\epsilon^2}$, the $Z-Z^\prime$ mixing angle and the neutral gauge boson masses are
\be
	\tan(2\xi) = \frac{2 M_{Z_\textnormal{SM}}^2 \sin\theta_W \epsilon \sqrt{1-\epsilon^2}}{M_{Z_\ell}^2 - M_{Z_\textnormal{SM}}^2 (1-\epsilon^2) + M_{Z_\textnormal{SM}}^2 \sin^2\theta_W \epsilon^2}
	\approx 2\epsilon \sin\theta_W  \frac{M_{Z_\textnormal{SM}}^2}{M_{Z_\ell}^2},
\ee
and
\begin{align}
	m_{Z^{(\prime)}}^2 &= \frac{1}{2}\left(M_{Z_\ell}^2 + M_{Z_{\rm SM}}^2 + \eta^2 M_B^2 \pm \sqrt{\left(M_{Z_\ell}^2 + M_{Z_{\rm SM}}^2 + \eta^2 M_B^2\right)^2 - 4 M_{Z_\ell}^2 M_{Z_{\rm SM}}^2 }\right) \\
	 &\approx M_{Z_{\rm SM}}^2 \,\left( M_{Z_\ell}^2\right) \,
,
\end{align}
where the approximate expressions are expansions up to linear order in $\epsilon$.
Note that the definition of the weak mixing angle $\theta_W$ and the electromagnetic coupling $e$ in terms of the SM gauge couplings $g_1$ and $g_2$ is not altered by the kinetic mixing.

\subsection{Scalar sector}
\label{sec:scalar}
%

The model has two scalar fields: the SM Higgs $H\sim(2,1/2,0)$ and the $U(1)_\ell$ breaking SM singlet scalar $\Phi\sim(1,0,L_\Phi)$, where we choose $L_\Phi=3$ as will be discussed in section~\ref{sec:fermions}. 
The corresponding potential is given by
\be
	\label{eq:potential}
	V(H,\Phi) = - \mu_H^2 H^\dagger H + \lambda_H \left(H^\dagger H\right)^2
	            - \mu_\Phi^2 \Phi^\dagger\Phi + \lambda_\Phi \left(\Phi^\dagger \Phi\right)^2
	            + \lambda_p\, H^\dagger H\, \Phi^\dagger \Phi
	.
\ee
Expanding the fields around their vacuum expectation values,
\be
	H = \begin{pmatrix} G^+ \\ \frac{1}{\sqrt{2}} \left( v_H + \hat{h} + \hat{G}^0\right) \end{pmatrix} \quad \textnormal{and} \quad
	\Phi = \frac{1}{\sqrt{2}} \left( v_\Phi + \hat{\phi} + \hat{\omega}^0\right)
	,
\ee
the would-be Nambu-Goldstone bosons $G^\pm$, $\hat{G}^0$ and $\hat{\omega}^0$ become the longitudinal degrees of freedom of the $W^\pm$, $Z$ and $Z^\prime$ gauge bosons.
The mass matrix for the remaining scalars is
\be
M^2_{H} = 
\begin{pmatrix}
	- \mu_H^2 + 3 \lambda_H v_H^2 + \frac{\lambda_p}{2} v_\Phi^2 & \lambda_p v_H v_\Phi \\
	\lambda_p v_H v_\Phi & - \mu_\Phi^2 + 3 \lambda_\Phi v_\Phi^2 + \frac{\lambda_p}{2} v_H^2   
\end{pmatrix}
.
\ee
The Higgs portal term $\lambda_{p}\,H^\dagger H\,\Phi^\dagger\Phi$ induces a mixing between the $\hat{h}$ and $\hat{\phi}$ fields. The mass eigenstates are defined by 
\be
	\label{eq:HiggsMixing}
	\begin{pmatrix} h \\ \phi \end{pmatrix}
	= \begin{pmatrix}
		\cos\theta_H & -\sin\theta_H \\
		\sin\theta_H & \cos\theta_H
	\end{pmatrix}
	\begin{pmatrix} \hat{h}\\ \hat{\phi} \end{pmatrix}
\ee
with the corresponding masses
\be
	m_{h,\phi}^2 = \left(\lambda_H v_H^2 + \lambda_\Phi v_\Phi^2\right) \pm \sqrt{\left(\lambda_H v_H^2 - \lambda_\Phi v_\Phi^2\right)^2 + \lambda_{p}^2 v_H^2 v_\Phi^2},
\ee
where we eliminated $\mu_H^2$ and $\mu_\Phi^2$ using the condition that the potential \eqref{eq:potential} has a minimum for $\hat h=v_H$ and $\hat\phi=v_\Phi$.
Here, $h$ is the SM-like Higgs with $m_h = 125$~GeV, and $\phi$ is the lepton number Higgs which will typically have a mass $m_\phi > m_h$ due to the VEV hierarchy imposed by \experiment{LEP} constraints (see section~\ref{sec:Zprime-constraints}).

\subsection{Fermion sector}
\label{sec:fermions}
%

With the SM fermion content only, lepton number is an anomalous symmetry.
The lepton-gravity $U(1)_\ell$ and pure lepton $[U(1)_\ell]^3$ anomalies are canceled by the presence of three generations of right-handed neutrinos $\nu_R\sim(1,0,1)$, whereas the cancellation of the remaining anomalies requires additional fermions. 
This can be realized in various ways (see e.g.~\cite{Foot:1989ts,FileviezPerez:2010gw,Dulaney:2010dj,FileviezPerez:2011pt,Duerr:2013dza,Chao:2010mp,Chang:2018vdd,Chang:2018nid,Perez:2014qfa,Ohmer:2015lxa}).
Here, we add two sets of chiral fermions that combine to transform vector-like under the SM gauge group, and thus do not spoil the cancellation of anomalies in the SM gauge sector:

\be
\begin{aligned}
	\label{eq:darkLeptons}
	\ell^\prime_L = \begin{pmatrix} N_L^\prime \\ E_L^\prime \end{pmatrix} \sim& \left(2,-\frac{1}{2},L^\prime\right), \hspace{2eM}&
	\ell^\doubleprime_R = \begin{pmatrix} N_R^\doubleprime \\ E_R^\doubleprime \end{pmatrix} \sim& \left(2,-\frac{1}{2},L^\doubleprime\right), \\
	\nu_R^\prime \sim& \left(1,0,L^\prime\right), &
	\nu_L^\doubleprime \sim& \left(1,0,L^\doubleprime\right), \\
	e_R^\prime \sim& \left(1,-1,L^\prime\right), &
	e_L^\doubleprime \sim& \left(1,-1,L^\doubleprime\right) .
\end{aligned}
\ee

The first set corresponds to a 4th generation of SM-like leptons but with lepton number $L^\prime$, whereas the second set has opposite chirality and lepton number $L^{\prime\prime}$.
Imposing the condition $L^\prime-L^\doubleprime = 3$, the remaining $[SU(2)_W]^2 \otimes U(1)_\ell$, $[U(1)_Y]^2 \otimes U(1)_\ell$ and $U(1)_Y \otimes [U(1)_\ell]^2$ anomalies cancel.

In order to write Yukawa terms for the additional fermions involving the lepton number breaking scalar $\Phi$, $L_\Phi=3$ must be chosen. The Yukawa sector is then given by
\be\begin{aligned}
	\label{eq:yukawa}
	\Lag \supset& 
	- c_\ell \bar{\ell}_R^\doubleprime \Phi  \ell_L^\prime - c_e \bar{e}_L^\doubleprime \Phi e_R^\prime -c_\nu \bar{\nu}_L^\doubleprime \Phi \nu_R^\prime
	\\
	&- y_e^\prime \bar{\ell}_L^\prime H e_R^\prime - y_e^\doubleprime \bar{\ell}_R^\doubleprime H e_L^\doubleprime
	- y_\nu^\prime \bar{\ell}_L^\prime \tilde{H} \nu_R^\prime - y_\nu^\doubleprime \bar{\ell}_R^\doubleprime \tilde{H} \nu_L^\doubleprime 
	+ \textnormal{h.c.} .
\end{aligned}\ee

Note that specific values of $L^\prime$ allow for additional terms in the Lagrangian. For example if $(L^\prime,L^\doubleprime) = (1,4)$ or $(-2,1)$, the dark fermions can mix with the SM ones, potentially leading to flavor changing neutral currents and threatening the stability of our DM candidate. 
Table~\ref{tab:excludedL} lists the lepton number charges that allow additional terms.
We will exclude these choices in the following.
For any other pair of real numbers with $L^\doubleprime = L^\prime +3$, the Yukawa interactions of the exotic leptons are fully described by \eqref{eq:yukawa}.

\begin{table}
	\renewcommand{\arraystretch}{1.5}
	\centering
	\begin{tabular}{c c l}
		\hline\hline
		\hspace*{1eM}$L'$\hspace*{1eM} & \hspace*{1eM}$L''$\hspace*{1eM} & $\Delta\mathcal{L}$ \\
		\hline
		-5 & -2 & $\bar \ell_R'' \Phi^\ast \ell_L\,,\ \bar e_L'' \Phi^\ast e_R\,,\ \bar \nu_L'' \Phi^\ast \nu_R$ \\
		-4 & -1 & $\bar \ell_R'' \tilde H \nu_R^c\,,\ \bar \nu_L'' H^\dagger \ell_L^c\,,\ \bar \nu_R' \Phi^\ast \nu_R^c$ \\
		-3 & 0 & $\bar \nu_L''  \nu_L''^c$ \\
		-2 & 1 & $\bar \ell_R''  \ell_L\,,\ \bar e_L''  e_R\,,\ \bar \nu_L''  \nu_R$ \\
		-3/2 & 3/2 & $\bar \ell_R'' \tilde H \nu_R'^c\,,\ \bar \nu_L'' H^\dagger \ell_L'^c\,,\ \bar \nu_L'' \Phi \nu_L''^c\,,\ \bar \nu_R' H^\dagger \ell_R''^c\,,\ \bar \nu_R' \Phi^\ast \nu_R'^c\,,\ \bar \ell_L' \tilde H \nu_L''^c$ \\
		-1 & 2 & $\bar \nu_R'  \nu_R^c$ \\
		0 & 3 & $\bar \nu_R'  \nu_R'^c$ \\
		1 & 4 & $\bar \ell_R'' \Phi \ell_L\,,\ \bar e_L'' \Phi e_R\,,\ \bar \nu_L'' \Phi \nu_R\,,\ \bar e_R' H^\dagger \ell_L\,,\ \bar \nu_R' \tilde{H}^\dagger \ell_L\,,\ \bar \ell_L' H e_R\,,\ \bar \ell_L' \tilde H \nu_R$ \\
		2 & 5 & $\bar \nu_R' \Phi \nu_R^c$ \\
		\hline\hline
	\end{tabular}
	\caption{Lepton number charge assignments that allow for additional Lagrangian terms.}
	\label{tab:excludedL}
\end{table}

After spontaneous symmetry breaking, mass terms for the additional fermions are generated,
\be
	\Lag \supset
	-\begin{pmatrix} \bar{N}_L^\prime & \bar{\nu}_L^\doubleprime \end{pmatrix} \mathcal{M}_{LR}^\nu \begin{pmatrix} N_R^\doubleprime \\ \nu_R^\prime \end{pmatrix} 
	- \begin{pmatrix} \bar{E}_L^\prime & \bar{e}_L^\doubleprime \end{pmatrix} \mathcal{M}_{LR}^e \begin{pmatrix} E_R^\doubleprime \\ e_R^\prime \end{pmatrix} 
	+ \textnormal{h.c.}
	,
\ee 
with the mass matrices given by
\be
	\mathcal{M}_{LR}^\nu = \frac{1}{\sqrt{2}} \begin{pmatrix}
		c_\ell^\ast v_\Phi & y_\nu^\prime v_H \\  
		y_\nu^{\doubleprime\ast} v_H & c_\nu v_\Phi
	\end{pmatrix},
	\hspace{1eM}
	\mathcal{M}_{LR}^e = \frac{1}{\sqrt{2}} \begin{pmatrix}
		c_\ell^\ast v_\Phi & y_e^\prime v_H \\  
		y_e^{\doubleprime\ast} v_H & c_e v_\Phi
	\end{pmatrix}
	.	
\ee
The matrices can be diagonalized via singular value decomposition, yielding the diagonal matrices $M_{D}^\nu = U_L^{\nu\dagger} M_{LR}^\nu U_R^\nu$ and $M_{D}^e = U_L^{e\dagger} M_{LR}^e U_R^e$, where $U_C^a$ are unitary matrices.

For simplicity, and to avoid $CP$ violating phases, let us assume that the Yukawa couplings $c_i$ and $y_i^{\prime(\prime)}$ are real.
The diagonalization matrices then become orthogonal.
The fermions combine to two charged ($e_4$ and $e_5$) and two neutral ($\nu_4$ and $\nu_5$) Dirac fields, which are given in terms of the original fields by  
\be\begin{aligned}
	\label{eq:fermionsMassEigenstates}
  \begin{pmatrix} \nu_4 \\ \nu_5\end{pmatrix} &=& 
  \begin{pmatrix}
  	\cos{\alpha_\nu} & \sin{\alpha_\nu} \\
  	-\sin{\alpha_\nu} & \cos{\alpha_\nu}
  \end{pmatrix}
  \begin{pmatrix} N_L^\prime \\ \nu_L^\doubleprime \end{pmatrix}
  &+
  \begin{pmatrix}
  	\cos{\beta_\nu} & \sin{\beta_\nu} \\
  	-\sin{\beta_\nu} & \cos{\beta_\nu}
  \end{pmatrix}
  \begin{pmatrix} N_R^\doubleprime \\ \nu_R^\prime \end{pmatrix}
  ,
  \\
  \begin{pmatrix} e_4 \\ e_5\end{pmatrix} &=&
  \begin{pmatrix}
  	\cos{\alpha_e} & \sin{\alpha_e} \\
  	-\sin{\alpha_e} & \cos{\alpha_e}
  \end{pmatrix}
  \begin{pmatrix} E_L^\prime \\ e_L^\doubleprime \end{pmatrix}
  &+
  \begin{pmatrix}
  	\cos{\beta_e} & \sin{\beta_e} \\
  	-\sin{\beta_e} & \cos{\beta_e}
  \end{pmatrix}
  \begin{pmatrix} E_R^\doubleprime \\ e_R^\prime \end{pmatrix}
  .
\end{aligned}\ee 

The right- and left-handed fields mix with different mixing angles unless we choose $y_\nu^\prime = y_\nu^\doubleprime$ and $y_e^\prime = y_e^\doubleprime$.
Thus, the resulting fermions in general are chiral with respect to both the SM and $U(1)_\ell$.

In the absence of the Yukawa terms~\eqref{eq:yukawa}, the Lagrangian exhibits a global $\left[U(1)\right]^6$ symmetry at the classical level, consisting of a $U(1)$ symmetry for each additional lepton field in~\eqref{eq:darkLeptons}. 
The $\Phi$ Yukawa terms break this to three $U(1)$ symmetries (one for the doublets, one for the charged singlets, and one for the neutral singlets), whereas the $H$ Yukawa terms (in the absence of the $\Phi$ Yukawas) break the symmetry to $U(1)_{L^\prime} \otimes U(1)_{L^\doubleprime}$.
Hence, small values of $c_i \ll 1$ and $y_i^{\prime(\prime)} \ll 1$ are technically natural, rendering vector-like masses $c_i v_\Phi \ll v_\Phi$.
Similarly, $y_{\nu\,i}^{\textnormal{SM}} \ll 1$ are technically natural.
As we will see later, $v_\Phi \gtrsim 2$~TeV, therefore we typically have $c_i v_\Phi \gg y_i^{\prime(\prime)} v_H$.
Consequently, the mixing angles $\alpha_{e/\nu}$ and $\beta_{e/\nu}$ are usually small, and the masses are approximately given by $m_{e_{4/5}} = c_{\ell/e} v_\phi /\sqrt{2}$ and $m_{\nu_{4/5}} = c_{\ell/\nu} v_\phi /\sqrt{2}$.

To simplify the discussion of the model we will restrict to the case of symmetric mass matrices, i.e.\ $y_{e/\nu} \equiv y_{e/\nu}^\prime = y_{e/\nu}^\doubleprime$, in the following, so that $\alpha_{e/\nu} = \beta_{e/\nu}$.
Let us further assume that $c_e = c_\ell$. The masses are then given by
\bea
	m_{\nu_{4/5}} &=& \frac{1}{2\sqrt{2}} \left( (c_\ell+c_\nu) v_\Phi \pm \sqrt{(c_\ell-c_\nu)^2 v_\Phi^2 + 4 y_\nu^2 v_H^2} \right) , \\
	m_{e_{4/5}} &=& \frac{1}{\sqrt{2}} \left( c_\ell v_\Phi \pm y_e v_H \right) .\label{eq:e45mass}
\eea
In particular, $m_{\nu_4} = (m_{e_4}+m_{e_5})/2$ if $y_\nu v_H \ll c_{l/\nu} v_\Phi$, and $e_4$ and $e_5$ are maximally mixed with $\alpha_e = \beta_e = \pi/4$.

Note that the SM neutrino masses in our model are pure Dirac masses generated from small Yukawa couplings to the SM Higgs doublet. Majorana mass terms are forbidden by the lepton number gauge symmetry, whereas mass terms arising from mixing with the exotic leptons would spoil the DM stability and are therefore avoided by suitable choices of the lepton number charges.

\subsection{RG running}
\label{sec:RGE}
%

Before exploring the phenomenology of the model, let us first consider the renormalization group running of the lepton number gauge coupling $g_\ell$.

The running of a gauge coupling $g$ is governed by the beta function
\be
	\beta = \frac{\partial\ g}{\partial \log\mu}\ .
\ee
For a $U(1)$ gauge group, the one-loop beta function is
\be
	\beta = \frac{g^3}{16 \pi^2} \left[ \frac{2}{3} \sum_f Q_f^2 + \frac{1}{3} \sum_s Q_s^2 \right]\ ,
\ee
where the sums run over all Weyl fermions and complex scalars charged under the gauge group with charge $Q_{f/s}$, respectively.

For the lepton number gauge group we get contributions from the SM leptons (with unit charge), the two additional generations of vector-like leptons (with charge $L'$ and $L''$), and the lepton-number-breaking scalar (with charge $L_\Phi$).
Thus,
\be
	\beta = \frac{g_\ell^3}{16 \pi^2} \left[  \frac{8}{3} \Big(N_f + L'^2 + L''^2\Big) + \frac{1}{3} L_\Phi^2 \right]
	= \frac{g_\ell^3}{16 \pi^2} \left[ 35 + 16 L' + \frac{16}{3} L'^2 \right] ,
\ee
where we used the lepton number charges $L'' = L' + 3$, $L_\Phi = 3$ and the number of SM flavors $N_f = 3$.

The dependence of the gauge coupling on the scale $\mu$ is consequently given by
\be
	g_\ell^2(\mu) =  \frac{g_0^2}{1- \frac{g_0^2}{8\pi^2} b \log\frac{\mu}{\mu_0}} \simeq g_0^2 \Big[1 + \frac{g_0^2}{8\pi^2} b \log\frac{\mu}{\mu_0}\Big]\ ,
	\label{eq:running}
\ee
where $g_0 = g_\ell(\mu_0)$ and $b = \left[ 35 + 16 L' + \frac{16}{3} L'^2 \right]$.
$U(1)_\ell$ has a Landau pole when the second term in the bracket in equation~\eqref{eq:running} is of order unity, i.e.\ at the scale $\mu = \Lambda$ with
\be
	\Lambda = \mu_0 \exp\left(\frac{8 \pi^2}{b g_0^2}\right)\ .
\ee

\begin{figure}
	\centering
	\includegraphics[width=.65\textwidth]{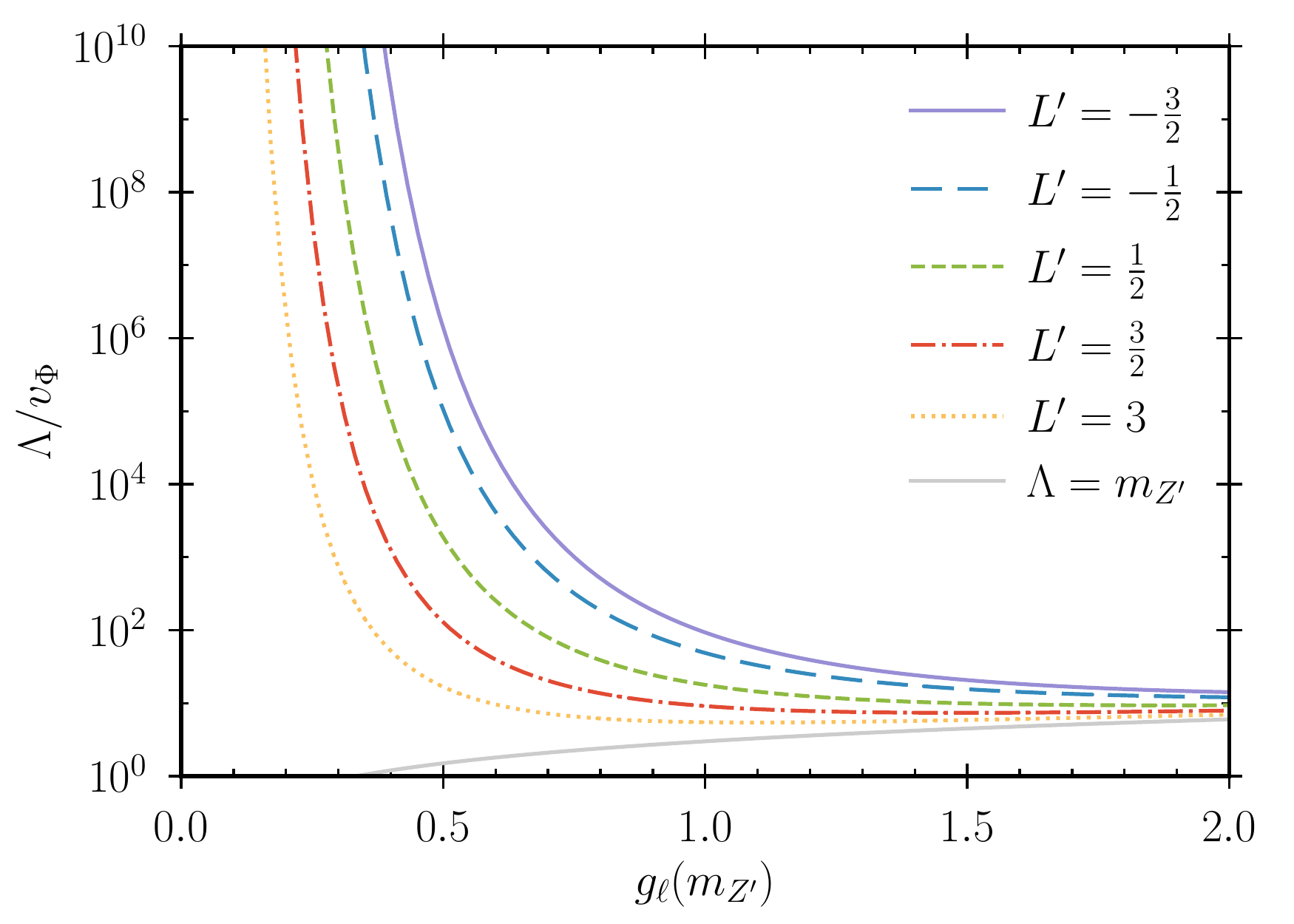}
	\caption{
		Landau pole $\Lambda$ normalized to the scalar VEV $v_\Phi$ as a function of $g_0 = g_\ell(m_\Zp)$ for different values of the charge $L^\prime$. 
		The gray, solid line indicates the value of the \Zp\ mass corresponding to $g_0$.
	}
	\label{fig:LandauPole}
\end{figure}

We now choose $\mu_0 = m_\Zp = 3 g_0 v_\Phi$.
Figure~\ref{fig:LandauPole} shows the Landau pole $\Lambda$ normalized to the scalar VEV $v_\Phi$ as a function of $g_0 = g_\ell(m_\Zp)$ for different values of the charge $L^\prime$.
Certainly, we want the Landau pole to occur significantly above the $\Zp$ mass and above $v_\Phi$, otherwise the validity of our perturbative results would be questionable. 
This requires choosing $g_\ell(m_\Zp) \lesssim 0.5$ for most values of $L^\prime$. 
The slowest running is obtained for $L^\prime = -3/2$, which we however excluded since it allows for Majorana mass terms.

To prevent the gauge coupling to run into a Landau pole at low scales, we choose $L^\prime = -1/2$ in the remainder of this paper.\footnote{Note that this particular choice is mostly for aesthetic reasons. We could have chosen any real number not listed in table~\ref{tab:excludedL}. Further requiring $\Lambda \lesssim 100$~TeV for $g_\ell=1$ and $v_\Phi = 2$~TeV restricts the viable choices to $L^\prime\in[-5/2,-1/2]$.}
In this case $g_\ell(m_\Zp) \approx 1$ is acceptable, with the Landau pole located almost two orders of magnitude above $v_\Phi$. 
For $v_\Phi=2$~TeV this implies an upper bound of $m_\Zp \lesssim 6$~TeV for the mass of the $Z^\prime$ boson. 
For most of the considerations that follow, the exact value of $L^\prime$ merely matters anyway.
An exception are the dark matter observables considered in section~\ref{sec:dm}, where we therefore also consider different values for $L^\prime$.

\section{Leptophilic dark matter}
\label{sec:dm}
%

Provided that we avoid the specific choices of $L^\prime$ and $L^\doubleprime$ discussed in section~\ref{sec:fermions}, the model features a global $U(1)_{L^\prime+L^\doubleprime}$ symmetry under which all SM fermions are neutral whereas the additional leptons have unit charge, and which is free of anomalies.
This symmetry persists when the EW and $U(1)_\ell$ gauge symmetries are broken and ensures the stability of the lightest dark lepton.
If neutral, it is a candidate for dark matter. 

For the remainder we identify $\nu_5$ as the DM candidate (which can always be achieved by defining the mixing angles in \eqref{eq:fermionsMassEigenstates} accordingly) and relabel it as $\nu_\DM \equiv \nu_5$.
Since direct detection experiments exclude DM candidates with unsuppressed couplings to the SM $Z$ boson, $\nu_\DM$ should be composed predominantly of the SM singlets $\nu_L^\doubleprime$ and $\nu_R^\prime$.
Consequently $\alpha_\nu$ and $\beta_\nu$ should be small.
In particular, this requires $c_\nu < c_\ell$. Finally, at least one of $y_\nu^{\prime}$ and $y_\nu^{\prime\prime} $ should be non-vanishing, otherwise an additional global $U(1)$ symmetry remains unbroken and the next-to-lightest dark sector state (either $\nu_4$ or $e_{4/5}$) would be stable as well. 

The model is implemented in \software{FeynRules}~\cite{Alloul:2013bka} and subsequently \software{mircOMEGAs}~\cite{Belanger:2018ccd} was used to calculate the relic density as well as direct and indirect detection constraints.
Unless specified otherwise, we use the parameters listed in table~\ref{tab:defaultParameters}.
We here relabelled $\theta_\DM \equiv \alpha_\nu = \beta_\nu$.

\begin{table}
	\centering
	\renewcommand{\arraystretch}{1.5}
	\begin{tabular}{cccc}
		\hline\hline
		\makebox[3cm]{scalar sector} & \makebox[3cm]{gauge sector} & \multicolumn{2}{c}{fermion sector} \\\hline
		$v_\Phi = 2$\ TeV     & $m_{Z^\prime} = 1.5$\ TeV   & $m_\DM= 640$\ GeV      & $m_{e_4} = 2.0$\ TeV \\
		$m_\phi = 2.5$\ TeV   & $\epsilon = 0$              & $\sin(\theta_\DM) = 0$ & $m_{e_5} = 1.5$\ TeV \\
		$\sin(\theta_H) = 0 $ & $ L^\prime = - \frac{1}{2}$ & &  \\\hline\hline
	\end{tabular}
	\caption{Default values for the model parameters (assuming $y_{\nu/e}^\prime=y_{\nu/e}^\doubleprime$ and $c_e=c_l$) used throughout this paper, unless specified otherwise. For negligible $\sin(\theta_\DM)$, the mostly-doublet, heavy neutrino mass is given by $m_{\nu_4} \simeq (m_{e_4}+m_{e_5})/2 = 1.75$~TeV. } 
	\label{tab:defaultParameters}
\end{table}

\subsection{Relic abundance}
\label{sec:abundance}

\begin{figure}
	\centering
	\begin{minipage}[c]{.32\textwidth}
		\centering
		\subfloat[][]{\includegraphics[scale=.8]{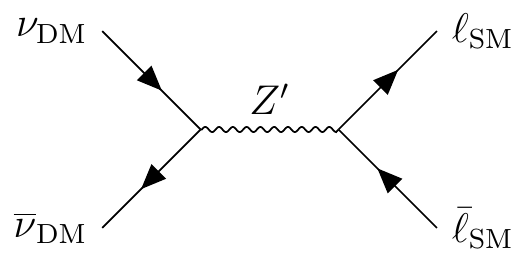}\label{fig:RD_DM-ann}}
	\end{minipage}
	\begin{minipage}[c]{.32\textwidth}
		\centering
		\subfloat[][]{\includegraphics[scale=.8]{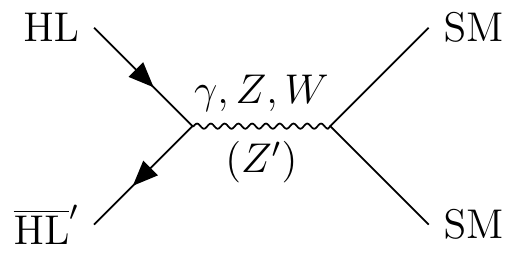}\label{fig:RD_HL-ann}}
	\end{minipage}
	\begin{minipage}[c]{.32\textwidth}
		\centering
		\subfloat[][]{\includegraphics[scale=.8]{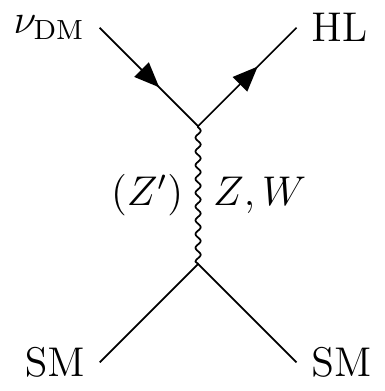}\label{fig:RD_co-ann}}
	\end{minipage}
	\caption{Processes contributing to the depletion of the DM relic abundance.}
\end{figure}

Assuming that $\nu_\DM$ is a thermal relic, its abundance is predominantly set by its annihilation cross section to two leptons through a $s$-channel \Zp\ (figure~\ref{fig:RD_DM-ann}). 
Other possible annihilation channels are annihilation to gauge or scalar bosons through an intermediate $h$ or $\phi$, or to fermions via a $Z$ boson.
The former is suppressed by the Higgs-scalar mixing, whereas the latter can arise from $Z-\Zp$ mixing or by an admixture of the SM component in the DM.
The doublet-singlet mixing further allows for $t$-channel annihilation to two bosons, and a small mass splitting between the DM and the other exotic leptons can lead to coannihilation.
See~\cite{Schwaller:2013hqa} for more details.

\begin{figure}
	\centering
	\subfloat[][$g_\ell = 0.1$, $\lambda_\Phi=0.5$]{\includegraphics[width=.49\textwidth]{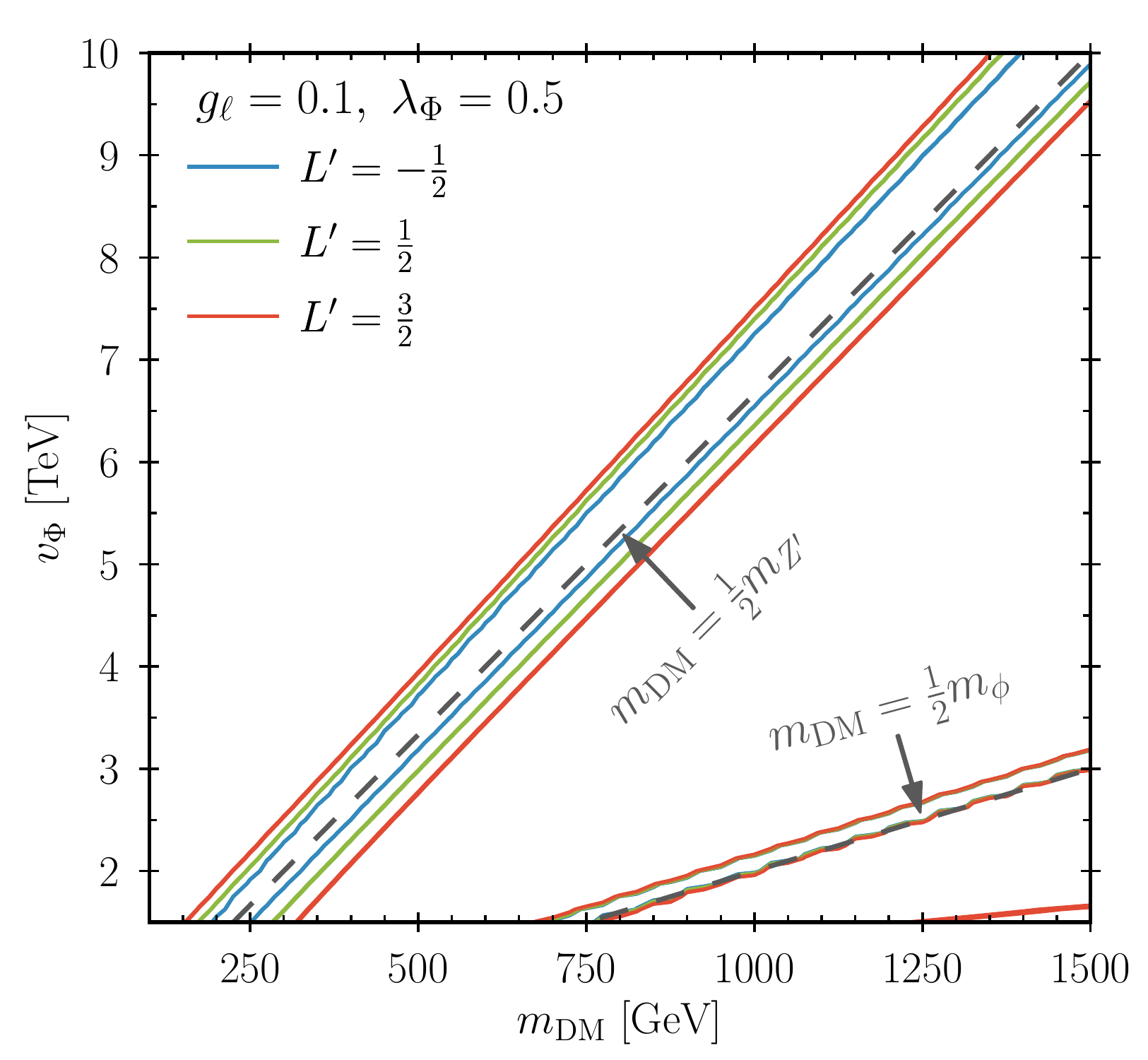}\label{fig:RDV}}
	\hfill
	\subfloat[][$v_\Phi = 2$ TeV]{\includegraphics[width=.49\textwidth]{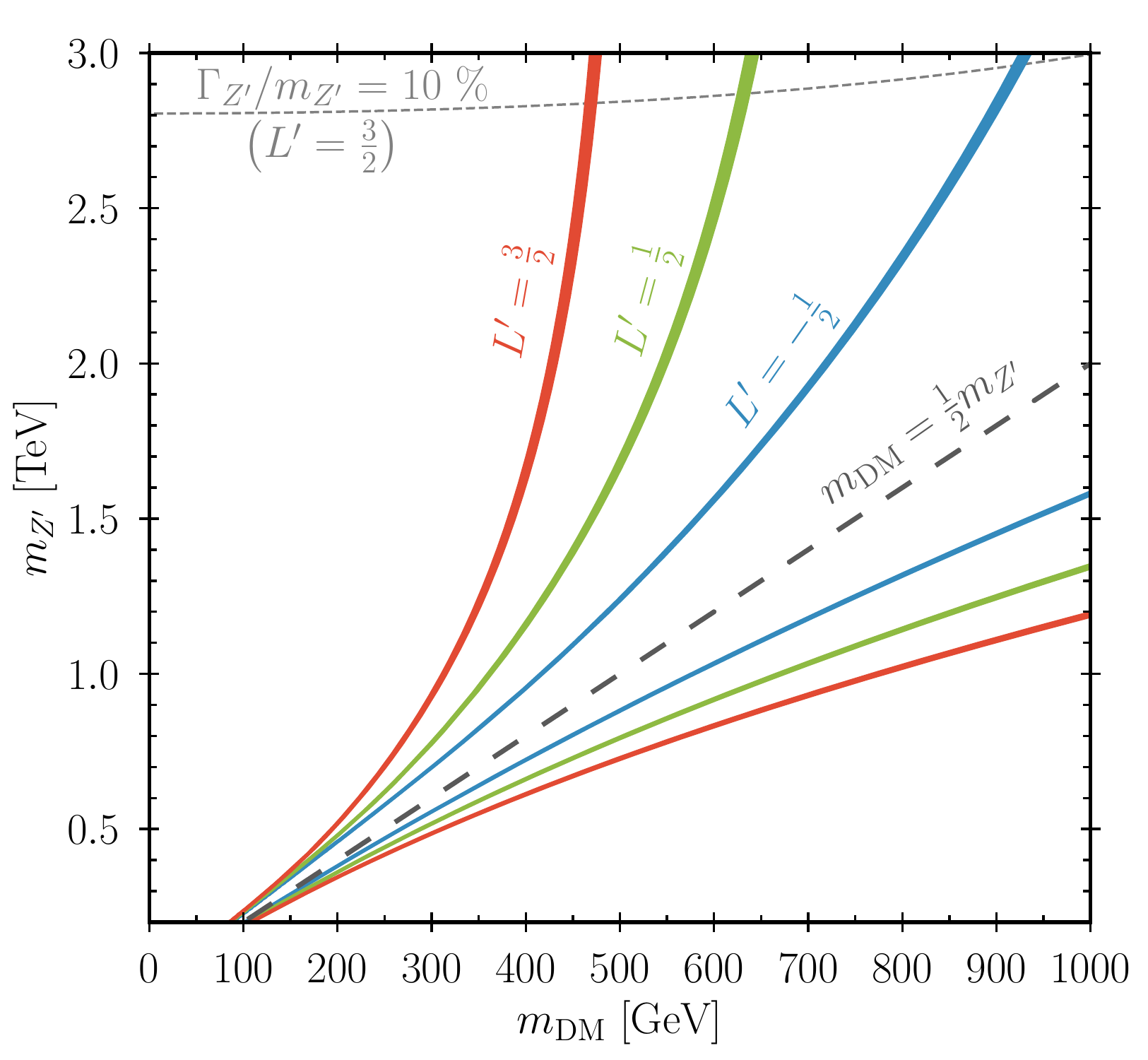}\label{fig:RDL}}
	\caption{Parameter regions reproducing the DM relic density $\OMEGA{DM} = 0.1198 \pm 0.0015$ measured by \experiment{Planck}~\cite{Ade:2015xua} within two standard deviations for different values of $L^\prime$, fixing the gauge coupling (left) or scalar VEV (right). The gray, dashed lines in the right plot indicate the parameters for which the $Z^\prime$ width exceeds 10\,\% of its mass.}\label{fig:DMRelicDensity}
\end{figure}

The parameter regions in the $m_\DM - v_\Phi$ and $m_\DM - m_{\Zp}$ planes that reproduce the DM relic abundance of $\OMEGA{DM} = 0.1198 \pm 0.0015$ measured by the \experiment{Planck} satellite~\cite{Ade:2015xua} are shown in figure~\ref{fig:DMRelicDensity} for different values of $L^\prime$.
We assume a lepton number gauge coupling of $g_\ell = 0.1$ and a scalar self-coupling of $\lambda_\Phi = 0.5$ as well as Yukawa couplings $c_\ell = 1.5$ and $y_e = 0$ when scanning over the VEV (left panel), and a scalar VEV of $v_\Phi = 2$~TeV when varying the \Zp\ mass (right panel).
The remaining parameters are set to the values specified in table~\ref{tab:defaultParameters}.

The colored regions yield a DM abundance that lies within two standard deviations around the \experiment{Planck} measurement.
For each value of $m_{\Zp}$ we typically obtain two viable values for the DM mass, one below and one above the $m_\DM = m_\Zp/2$ resonance. 
In figure~\ref{fig:RDV}, we can in addition also see the scalar resonance at $m_\DM = m_\phi/2$.
To guide the eye, the resonances are indicated by dashed, dark  gray lines.

Since the $\Zp$ predominantly decays into $\nu_\DM$ and the other heavy leptons, its width increases with $L^\prime$. 
For $L^\prime=-1/2$, the \Zp\ is rather narrow and the DM mass is restricted to values close to half of the \Zp\ mass.
For larger charges, the resonance becomes broader and the DM mass can be lower.
The light gray, dashed line in figure~\ref{fig:RDL} indicates the regions in which the width of the \Zp\ exceeds 10\,\% of its mass for $L'=3/2$.
 
\begin{figure}
	\centering
	\includegraphics[width=.5\textwidth]{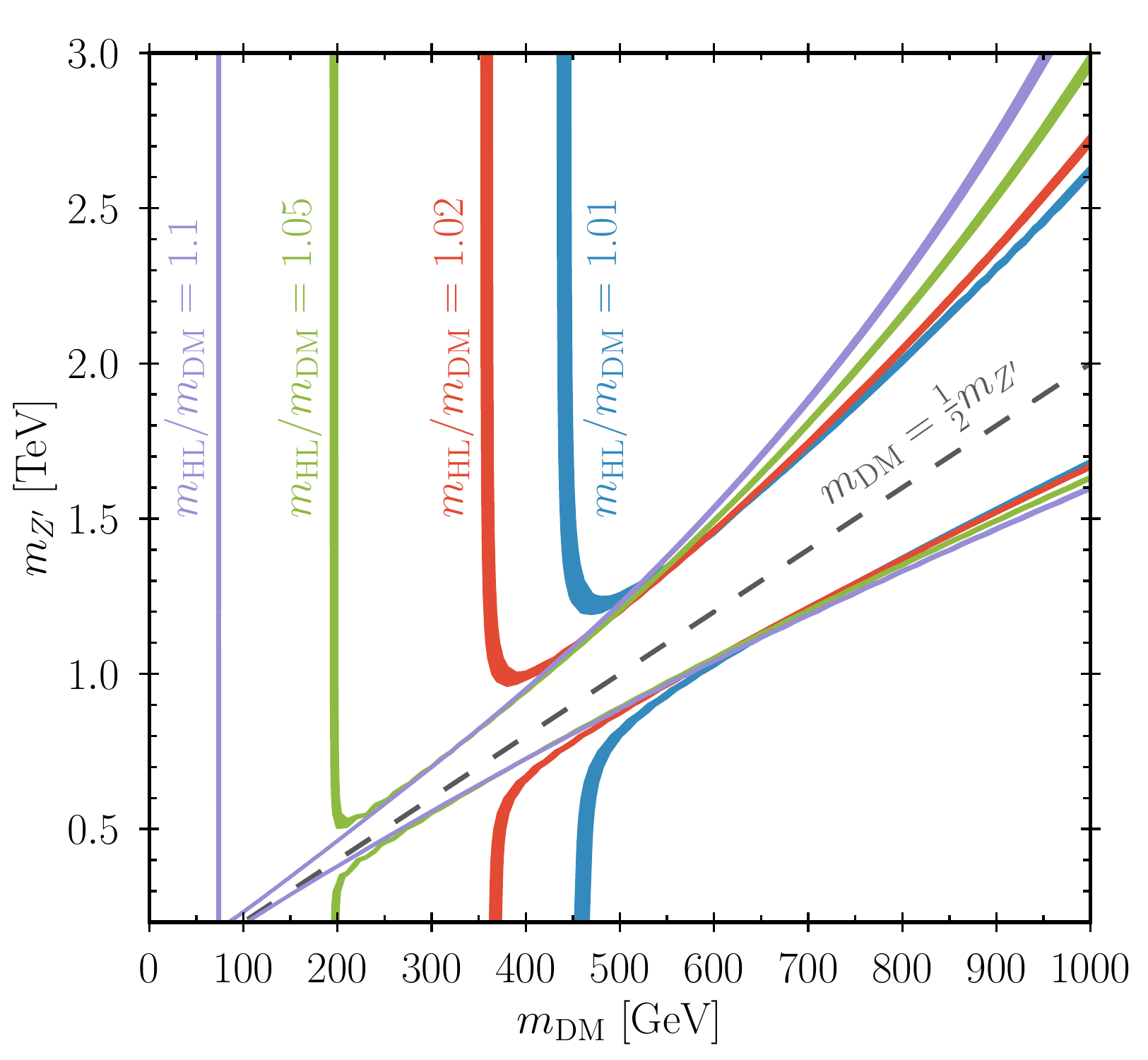}
	\caption{Same as figure~\ref{fig:RDL} for $L^\prime = -1/2$, but including co-annihilation with the heavy leptons for $\Delta_m = 1 - 10\ \%$, assuming that $e_4$, $e_5$ and $\nu_4$ have equal mass $m_\mathrm{HL}$. The colored regions reproduce the measured relic abundance, the colors correspond to different values of $m_\mathrm{HL}/m_\DM = 1 + \Delta_m$.}
 	\label{fig:RDCA}
\end{figure}

The dependence of the DM relic density on the masses of the non-DM heavy leptons $e_4$, $e_5$, and $\nu_4$ is shown in figure~\ref{fig:RDCA}, assuming that they all have the same mass $m_\mathrm{HL}$. 
The colored regions again yield the measured DM abundance, now assuming $L^\prime=-1/2$. The colors correspond to a relative mass splitting $\Delta_m \equiv (m_\HL - m_\DM)/m_\DM$ of $1\,\%$~(blue), $2\,\%$~(red), $5\,\%$~(green), and $10\,\%$~(purple) between the DM and the heavy leptons.

For high DM masses, the heavy lepton masses affect the relic density only by changing the \Zp\ width.
However, for lower DM masses the abundance is no longer set by annihilation of the DM to SM leptons as depicted in figure~\ref{fig:RD_DM-ann}, but via co-annihilation.
In this case, the heavy lepton (HL) abundance is depleted by annihilation of $e_4$, $e_5$ and $\nu_4$ to SM particles through electroweak processes as shown in figure~\ref{fig:RD_HL-ann}.
This depletion is transferred to the DM abundance by the electroweak processes depicted in figure~\ref{fig:RD_co-ann} in which a DM particle scatters off SM particles and changes into a heavy lepton.
These processes require $\sin\theta_\DM \neq 0$, but we need this assumption anyways to ensure that there is only a single DM component.
As the diagram~\ref{fig:RD_HL-ann} is dominanted by electroweak processes, the relic density in this regime is independent of the \Zp\ mass.
Figure~\ref{fig:RDCA} assumes $\sin\theta_\DM = 0$, however modifying the mixing within the range allowed by direct detection (see section~\ref{sec:detection}) does not alter the result. 

Varying the remaining parameters of the model only has a minor effect on these results. 
The scalar mass $m_\phi$ and the $h-\phi$ mixing angle $\theta_H$ only have an effect in the region of the $\phi$ resonance $m_\DM = m_\phi/2$ (or the $h$ resonance). 

\subsection{Direct and indirect detection}
\label{sec:detection}

Direct detection experiments strongly constrain DM couplings to the SM $Z$ boson via scattering off nuclei.
For small values of the kinetic mixing parameter $\epsilon$, the coupling is given by
\be
  \mathfig{\includegraphics{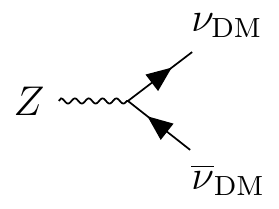}}\quad
	s_\DM^2 \frac{i e}{2 c_W s_W} \gamma_\mu 
	+ s_\xi \frac{i g_\ell}{2} \gamma_\mu \left[ (L^\prime + L^\doubleprime) + (L^\prime - L^\doubleprime) (c_\DM^2-s_\DM^2) \gamma_5\right],
\ee
where $s_W = \sin\theta_W$, $s_\xi=\sin\xi$, $s_\DM = \sin\theta_\DM$ and  $c_\DM = \cos\theta_\DM$.
The first term originates from the heavy doublet neutrino $N=N_L^\prime+N_R^\doubleprime$ (which has vector-like couplings to the SM $Z$) mixing into the DM, whereas the second part comes from the chiral $\DM-\Zp$ coupling. The axial part of the latter is also modified by the DM mixing via the $\overline{\nu}_4 \nu_4 Z^\prime$ vertex, the vector part remains untouched by $\theta_\DM$ since it here enters as $(c_\DM^2+s_\DM^2)$.

\begin{figure}
	\centering
	\subfloat[][kinetic mixing parameter $\epsilon$ ($\theta_\DM = 0$)]{\includegraphics[width=.49\textwidth]{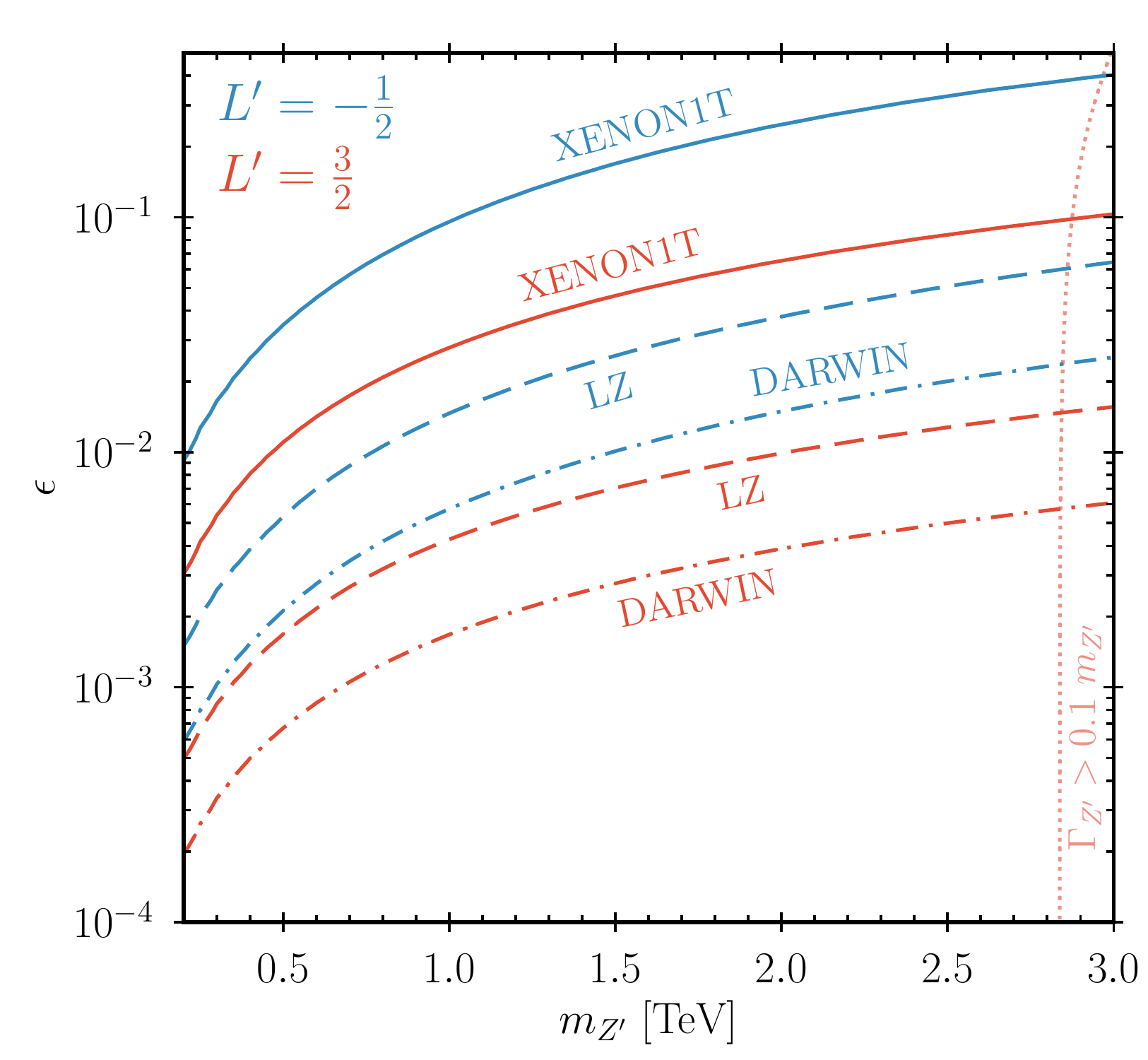}\label{fig:DDeps}}
	\hfill
	\subfloat[][DM mixing angle $\sin\theta_\DM$ ($\epsilon = 0$)]{\includegraphics[width=.49\textwidth]{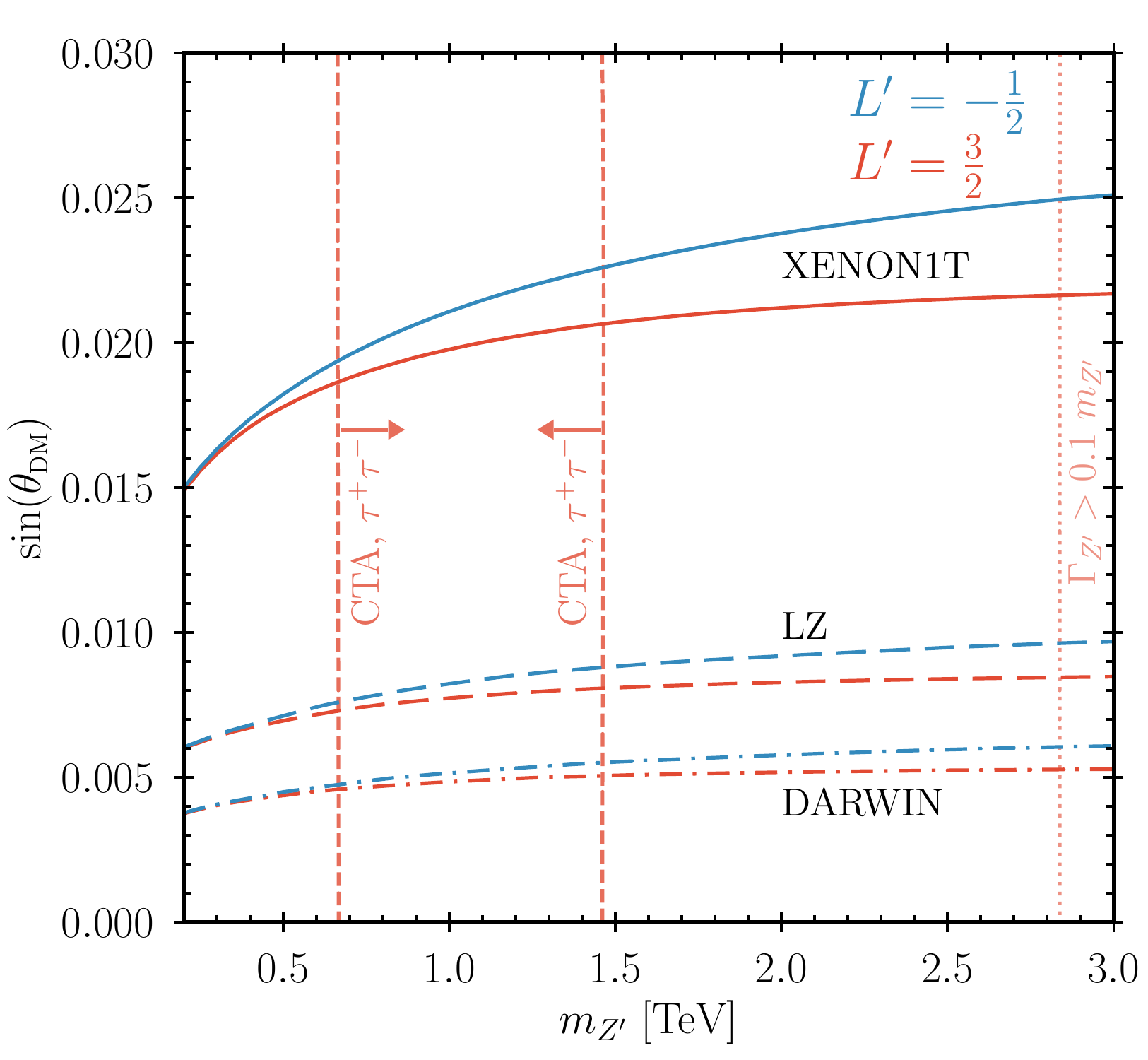}\label{fig:DDsinDM}}
	\caption{Direct detection limits on the $Z-\Zp$ and $\nu_\DM - \nu_4$ mixing for $L^\prime=-1/2$ (blue) and $L^\prime=3/2$ (red). Current constraints from the \experiment{XENON1T} experiment~\cite{Aprile:2018dbl} are shown as solid lines, the long-dashed lines indicate the projected sensitivity of \experiment{LZ}~\cite{Akerib:2018lyp}, and the dash-dotted lines correspond to \experiment{DARWIN}~\cite{Aalbers:2016jon}. The DM mass is fixed by the requirement to reproduce the \experiment{Planck} relic density, all other parameters are set according to table~\ref{tab:defaultParameters}. The short-dashed lines in~\ref{fig:DDsinDM} indicate the prospective reach of the \experiment{CTA}~\cite{Carr:2015hta} indirect detection experiment. The light, dotted lines show the region where the \Zp\ width grows above 10\,\% of the mass.}
	\label{fig:DDconstraints}
\end{figure}

Figure \ref{fig:DDconstraints} shows the constraints on the $Z-\Zp$ and \DM\ mixing as a function of the \Zp\ mass, obtained from direct detection limits on spin-independent DM-nucleus scattering. 
The solid lines correspond to current constraints from the \experiment{XENON1T} experiment based on one tonne times year of data acquisition~\cite{Aprile:2018dbl}, the  long dashed lines indicate the prospective sensitivity of \experiment{LZ} \cite{Akerib:2018lyp}, and the dash-dotted lines show the projected reach of \experiment{DARWIN}~\cite{Aalbers:2016jon}.
At each parameter point the DM mass is fixed to a value reproducing $\OMEGA{\DM} = 0.1198$ (chosing the value below the \Zp\ resonance), the charges are taken to be $L^\prime=-1/2$ (blue) or $L^\prime=3/2$ (red), and the VEV is set to $v_\Phi = 2$~TeV.

Current direct detection experiments can probe kinetic mixing parameters in the percent range and DM mixing angles of $\sin\theta_\DM \sim 0.015 - 0.025$, depending on the \Zp\ mass.
With \experiment{LZ}, DM mixing angles of $\sin\theta_\DM \sim 0.006 - 0.01$ and kinetic mixing in the sub-percent range can be reached.
\experiment{DARWIN} can prospectively exclude $\sin\theta_\DM \sim 0.004 - 0.006$, and sub-per-mill kinetic mixing for $m_\Zp \lesssim 1$~TeV.
The constraints for $L^\prime=3/2$ are stronger than for $L^\prime=-1/2$ since the latter case leads to higher DM masses, whereas the former case gives DM masses below 500~GeV.

\begin{figure}
	\centering
	\subfloat[][$m_  \phi = 2.5$~TeV]{\includegraphics[width=.49\textwidth]{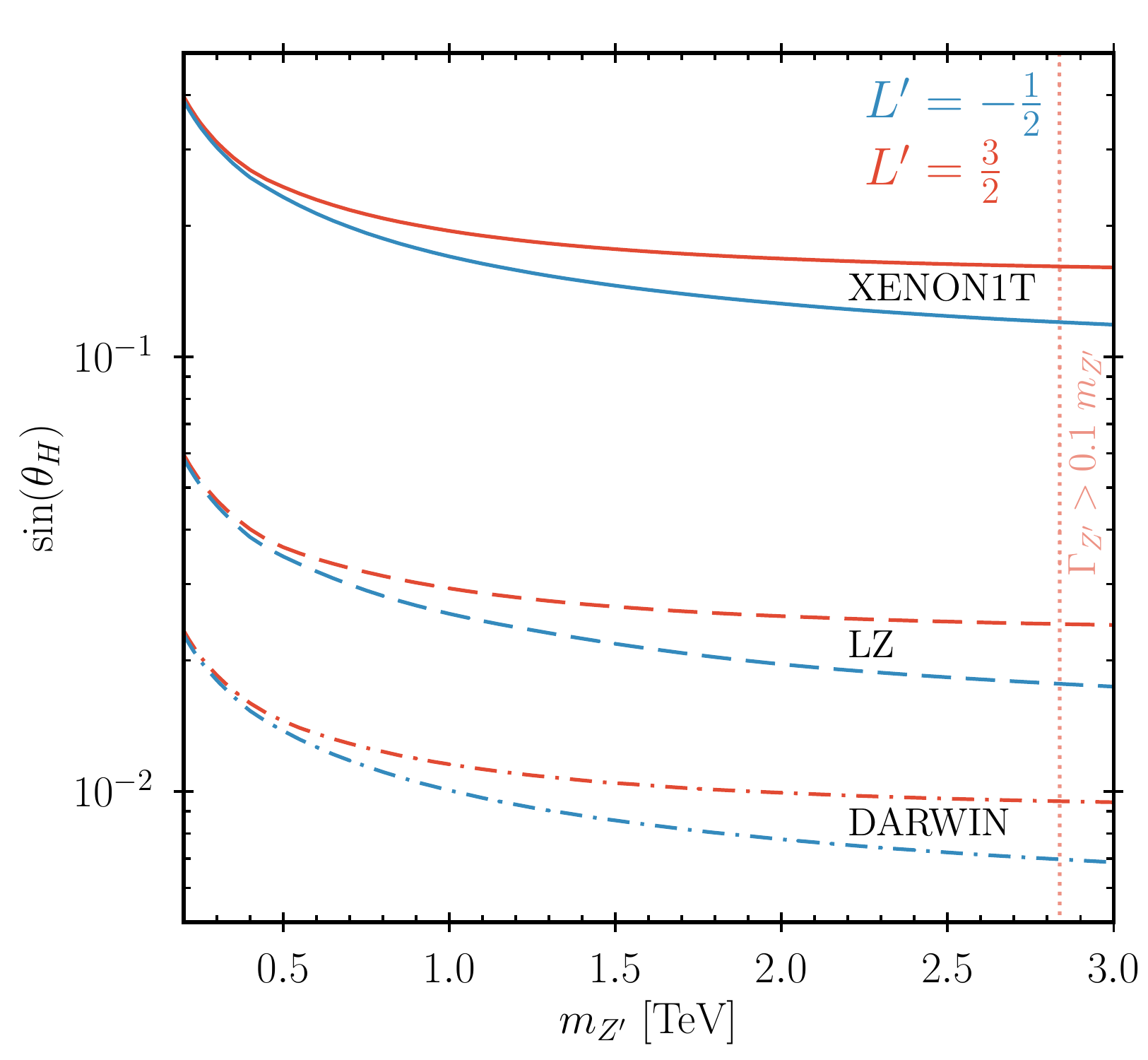}}
	\hfill
	\subfloat[][$L^\prime = -1/2$]{\includegraphics[width=.49\textwidth]{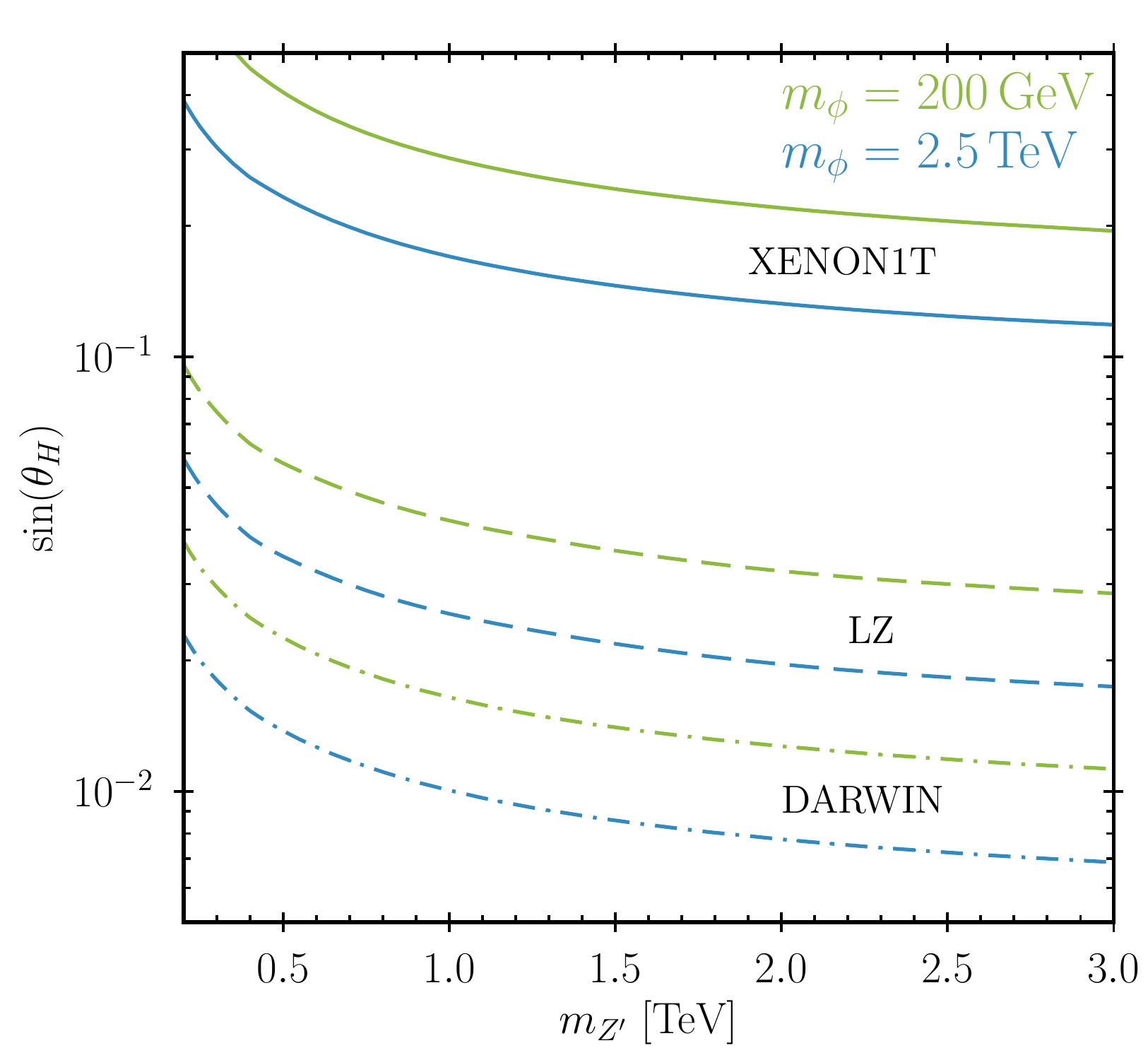}}
	\caption{Direct detection limits on the Higgs mixing angle $\theta_H$ for lepton number charges (left plot, $m_\phi = 2.5$~TeV) of  $L^\prime=-1/2$ (blue) and $L^\prime=3/2$ (red), and scalar masses (right plot, $L^\prime = -1/2$) of $m_\phi=200$~GeV (green) and $m_\phi=2.5$~TeV (blue). The current constraints from the~\experiment{XENON1T} experiment \cite{Aprile:2018dbl} are shown as solid lines, the long-dashed and dash-dotted lines indicate the projected sensitivities of \experiment{LZ}~\cite{Akerib:2018lyp} and \experiment{DARWIN}~\cite{Aalbers:2016jon}, respectively. As before, the DM mass is set to the value reproducing the measured relic abundance.}
	\label{fig:DDsinH}
\end{figure}

Beside $Z$-mediated DM-quark interactions, nuclear scattering can also proceed via Higgs (or $\phi$) exchange, either through Higgs-scalar mixing or by direct DM-Higgs couplings.
However, since the Higgs only weakly couples to nuclei, the direct detection constraints on the Higgs mixing angle are much weaker than on $\theta_\DM$ or $\xi$. 
For $m_\phi = 2.5$~TeV, \experiment{XENON1T} can currently exclude $\sin\theta_H\gtrsim0.1-0.4$. \experiment{LZ} and \experiment{DARWIN} can prospectively probe $\sin\theta_H \sim 0.02-0.06$ and $\sin\theta_H \sim 0.007 - 0.02$, respectively.
Here, direct detection experiments are more sensitive for $L^\prime=-1/2$ since the Yukawa couplings are proportional to the DM mass, i.e. we benefit from the higher DM masses in the $L^\prime=-1/2$ case. 
For lower $\phi$ masses on the other hand, the scattering cross section is reduced due to $\phi-h$ interference effects, leading to weaker direct detection limits.
The corresponding constraints are shown in figure~\ref{fig:DDsinH}.

A further, indirect way of probing DM is via its annihilation to SM particles.
Since the observation of charged particles suffers from large uncertainties associated with their propagation through the Galactic halo, we here only consider indirect detection constraints from $\gamma$-ray searches.
As photons travel unperturbed by Galactic magnetic fields, they can be traced back to their production site, allowing for constraints on DM annihilation by observing photons from regions with a high DM density.

In our model the DM typically annihilates through the lepton number gauge boson $Z^\prime$ into SM leptons with equal branching ratios.
Photons are thus predominantly produced as secondary products from annihilation to charged leptons.
Direct production of monochromatic photons is possible via annihilation through the scalar bosons $h$ and $\phi$, however, this is suppressed unless resonant.

We tested the annihilation of thermally produced DM (i.e.\ satisfying the relic density constraint) in our model against current limits from observations of dwarf spheroidal galaxies by \experiment{MAGIC} and \experiment{Fermi-LAT}~\cite{Ahnen:2016qkx}, and of the inner Galactic halo by \experiment{H.E.S.S.}~\cite{Abdallah:2016ygi}, as well as against $\gamma$ line searches from \experiment{Fermi-LAT}~\cite{Ackermann:2015lka} and \experiment{H.E.S.S.}~\cite{Rinchiuso:2017kfn}.
The strongest limits come from secondary produced photons from annihilation into $\tau$ leptons.
However, the current sensitivity reaches the level of the annihilation cross section required for thermal production (which in addition is reduced by the branching ratio of 1/6 into tauons) only for DM below 100~GeV, which, even for light $Z^\prime$ masses, is below the dark matter masses predicted by our model (cf.\ figure~\ref{fig:DMRelicDensity}).
This also holds for the projected sensitivity of \experiment{Fermi-LAT}, assuming a 15-year data set of 60 dwarf spheroidal galaxies~\cite{Charles:2016pgz}.
On the other hand, a next-generation $\gamma$-ray observatory such as the \experiment{Cherenkov Telescope Array (CTA)}~\cite{Carr:2015hta} will be able to exclude \Zp\ masses between roughly 670~GeV and 1.46~TeV if $L^\prime=3/2$.
The corresponding limit is indicated by the vertical, dashed lines in figure~\ref{fig:DDsinDM}.

In principle, our model can furthermore be probed through its neutrino sector.
Modifications of the neutrino interactions with the SM leptons arise from $Z^\prime$ exchange or kinetic mixing, and DM-neutrino interactions can be mediated by a $Z$ or $Z'$ boson. 
Neutrino couplings to the lepton number breaking scalar and SM Higgs boson are suppressed by the neutrino Yukawa couplings.
For the range of $Z^\prime$ and DM masses considered here, no constraints are obtained from current data~\cite{Campo:2017nwh,Esteban:2018ppq}.

\section{Collider phenomenology}
\label{sec:pheno}

While new physics that couples directly to quarks and gluons is nowadays severly constrained by direct searches at the \experiment{LHC}, the situation is different for the leptophilic new physics model we are considering here.
In this model constraints predominantly arise from a combination of \experiment{LEP} limits as well as direct and indirect \experiment{LHC} measurements, such as e.g.\ Higgs data. 
In the following we present an overview of the most important constraints on the lepton number gauge boson, the extended Higgs sector and the new leptons introduced in our model, and comment on the prospects for detection at the high-luminosity \experiment{LHC} and future colliders. 

\subsection[$Z'$ constraints]{$\boldsymbol{Z'}$ constraints}
\label{sec:Zprime-constraints}

As in the absence of kinetic mixing the lepton number gauge boson does not couple to quarks, the strongest constraints on the $Z^\prime$ boson come from \experiment{LEP~II}.
These exclude $Z^\prime$ masses below the maximal \experiment{LEP} center-of-mass energy of 209~GeV (except for tiny gauge couplings $g_\ell <  10^{-2}$~\cite{Carena:2004xs,Patrignani:2016xqp}) and severely restrict the lepton number breaking VEV $v_\Phi$ through 4-lepton contact interactions.

A heavy $Z'$ induces effective contact interactions between electrons and charged (SM) leptons.
Since the $Z'$ couplings to SM fermions are vector-like, the corresponding contact interactions are given by (neglecting kinetic mixing)
\be
	\Lag_\mathrm{eff} \supset 
		- \frac{g_\ell^2}{2 m_{Z'}^2}\, \bar{e} \gamma_\mu e\, \bar{e} \gamma^\mu e 
		- \frac{g_\ell^2}{m_{Z'}^2}\, \bar{e} \gamma_\mu e\, \bar{\mu} \gamma^\mu \mu 
		- \frac{g_\ell^2}{m_{Z'}^2}\, \bar{e} \gamma_\mu e\, \bar{\tau} \gamma^\mu \tau\ ,
\ee
which interfere destructively with the SM $Z$ boson for center-of-mass energies above the $Z$-pole.
\experiment{LEP} puts a 95\,\% lower bound of $\Lambda > 20$~TeV~\cite{Schael:2013ita} on the scale suppressing this contact interaction, related to the model parameters by $\Lambda^2 = 4\pi\,\frac{m_{Z'}^2}{g_\ell^2}$.
Using $m_{Z^\prime} = L_\Phi g_\ell v_\Phi$ with $L_\Phi = 3$, this gives a lower bound on the scalar VEV of
\be
	\label{eq:VEVconstraint}
	v_\Phi \gtrsim 1880\,\textnormal{GeV} .
\ee
To be conservative we set the VEV to $v_\Phi = 2$~TeV in the following. 

Future $e^+ e^-$ colliders have the potential to substantially tighten these bounds. 
For instance, the \experiment{ILC} with a center-of-mass energy of $\sqrt{s} = 1$~TeV can basically exclude $Z'$ masses below 1~TeV (unless the gauge coupling is below $g_\ell \lesssim 7.6\times 10^{-8}$), and constrain the VEV to be above $v_\Phi \gtrsim 15$~TeV at 90\,\% CL using muon contact interactions~\cite{Freitas:2014jla}.

At the \experiment{LHC}, the $Z'$ is rather hard to produce.
In particular, in the absence of kinetic mixing it is predominantly produced by pair-producing SM leptons that radiate off a $Z'$.
The $Z'$ can then be detected from its decays to charged SM leptons. 

\begin{figure}
	\begin{minipage}[t]{.48\textwidth}
		\centering
		\includegraphics[width=\textwidth]{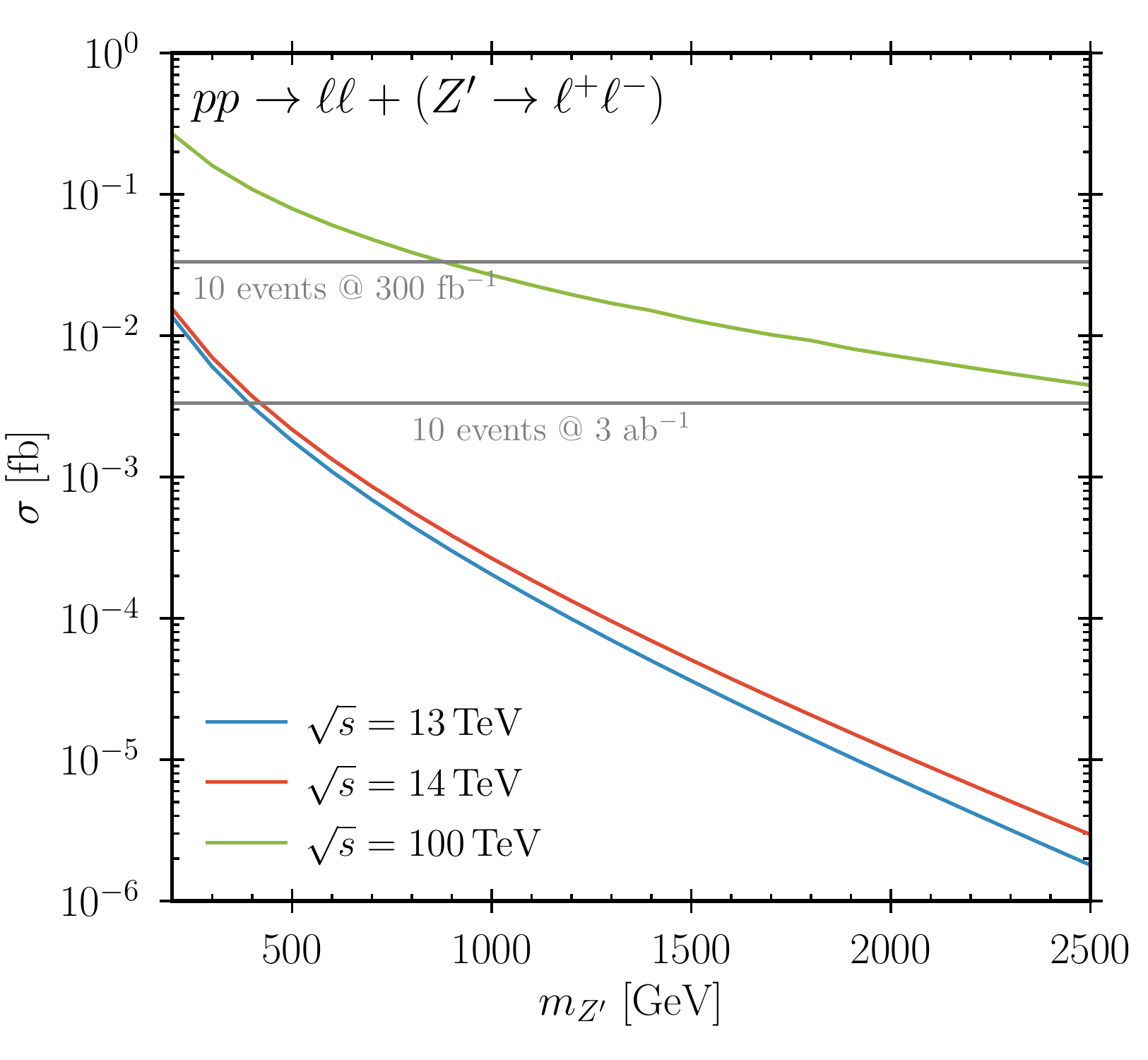}
		\caption{$Z'$ production cross section in the absence of kinetic mixing for \experiment{LHC-13} (blue), \experiment{LHC-14} (red), and a 100~TeV collider (green).}
		\label{fig:LHC-Zp-eps0}
	\end{minipage}
	\hfill
	\begin{minipage}[t]{.48\textwidth}
		\centering
		\includegraphics[width=\textwidth]{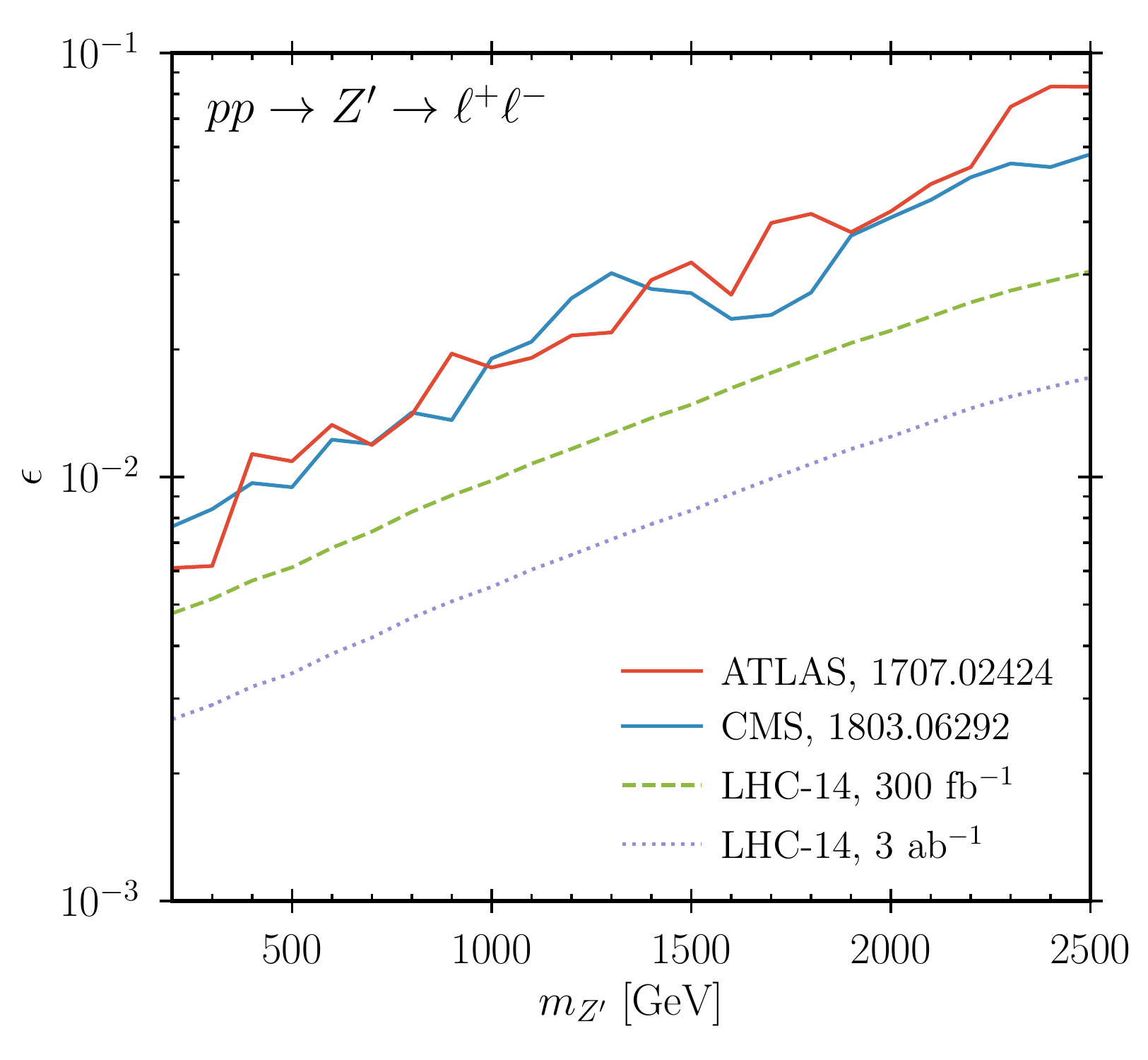}
		\caption{Current \experiment{LHC} limits in the $Z'$ mass vs.\ kinetic mixing parameter plane from \experiment{ATLAS} (solid red) and \experiment{CMS} (solid blue), as well as projections for the 14~TeV~\experiment{LHC} with $300$~fb${}^{-1}$ (dashed green) and $3$~ab${}^{-1}$ (dotted purple).}
		\label{fig:LHC-Zp-eps}
	\end{minipage}
\end{figure}

To obtain a rough estimate of the detection prospects, we calculate the parton level cross section for $p p \longrightarrow \ell \ell Z'$ with \software{CalcHEP 3.6}~\cite{Belyaev:2012qa}, where $\ell$ can be any SM lepton, including neutrinos. 
For the decay we use the narrow width approximation, assuming that the $Z'$ decays into SM leptons only.
In this case, the corresponding branching ratio to charged leptons is $\mathrm{Br}\left(Z'\rightarrow \ell^+ \ell^-\right) = 50\,\%$.
The cross section as a function of the $Z'$ mass is shown in figure~\ref{fig:LHC-Zp-eps0} for the \experiment{LHC} with center-of-mass energies of 13~TeV (blue) and 14~TeV (red), as well as for a 100~TeV collider (green).
The gray lines indicate the cross sections that would produce 10 $Z'$s assuming integrated luminosities of $300$~fb${}^{-1}$ and $3$~ab${}^{-1}$.

With the current \experiment{LHC} data, no constraints can be put on the $Z'$ mass if kinetic mixing is absent. 
With 300~fb${}^{-1}$, the \experiment{LHC} would not produce a sufficient amount of $Z'$s. At the high luminosity \experiment{LHC}, $Z'$ masses below $\sim 400$~GeV can be reached.
The prospects for a 100~TeV collider are more promising. 
With 3~ab${}^{-1}$, more than 10 events are produced for masses up to 2.5~TeV, extending the reach into the multi-TeV region. 

In the presence of kinetic mixing the situation is different.
The lepton number gauge boson then couples to quarks with couplings proportional to the kinetic mixing parameter $\epsilon$ and the quark hypercharge.
It can thus be produced directly in proton-proton collisions and be searched for as a dilepton resonance, giving constraints in the $m_{Z'}$ vs.\ $\epsilon$ plane.

Figure~\ref{fig:LHC-Zp-eps} shows constraints from current searches for dilepton resonances using 36~fb${}^{-1}$ of data collected at a center-of-mass energy of $\sqrt{s} = 13$~TeV by \experiment{ATLAS}~\cite{Aaboud:2017buh} (solid red line) and \experiment{CMS}~\cite{Sirunyan:2018exx} (solid blue line).
Projections for the \experiment{LHC} at a center-of-mass energy of $\sqrt{s}=14$~TeV with an integrated luminosity of 300~fb${}^{-1}$ (dashed green) and 3~ab${}^{-1}$ (dotted purple) are also shown.
Currently, kinetic mixing can be probed in the percent range, the \experiment{HL-LHC} can prospectively reach the sub-percent range for light $Z'$.
Again, the cross sections have been calculated with \software{CalcHEP}~\cite{Belyaev:2012qa}, assuming a narrow $Z'$ width with a branching ratio of $1/3$ to light leptons ($e$ or $\mu$).
The projections have been obtained assuming that the limits on cross section ratios provided by \experiment{CMS}~\cite{Sirunyan:2018exx} do not change when increasing the center-of-mass energy from 13~TeV to 14~TeV, and that the exclusion reach scales with the square root of the luminosity.

The kinetic mixing can also be probed via its effects on SM precision measurements at electron-positron colliders~\cite{Hook:2010tw}. However, as these effects are suppressed for high \Zp\ masses, the \experiment{LHC} provides the strongest constraints in the mass range considered here.

\subsection{Higgs constraints}
\label{sec:Higgs-constraints}

The scalar sector of our model is subject to constraints from measurements of the properties of the 125~GeV Higgs boson at \experiment{ATLAS} and \experiment{CMS}, as well as from null-results of searches for scalar bosons at different masses. 

In our model, the SM Higgs properties can be modified by three effects: the $h-\phi$ mixing which modifies all SM Higgs couplings, modifications of Higgs couplings to electroweak gauge bosons (in particular to two photons) by loops of heavy charged leptons, and decays to BSM states (if kinematically accessible).
However, given the lower bound on the lepton-number-breaking VEV~\eqref{eq:VEVconstraint}, the new states are typically too heavy for the SM Higgs to decay into, so that the last effect is absent in most of the parameter space.

The mixing between the lepton-number-breaking scalar and the SM Higgs boson given by equation~\eqref{eq:HiggsMixing} reduces the Higgs couplings to SM fields by  $\cos\theta_H$. 
\experiment{ATLAS} and \experiment{CMS} provide limits on modifications of Higgs couplings compared to the SM values in terms of signal strengths, defined by
\be
	\mu_X = \frac{\sigma\left(p p \longrightarrow h\right) \times \mathrm{Br}\left(h\longrightarrow X\right)}{\sigma^\mathrm{SM}\left(p p \longrightarrow h\right) \times \mathrm{Br}^\mathrm{SM}\left(h\longrightarrow X\right)}\ .
\ee
Neglecting additional Higgs decay channels and further modifications of loop-induced Higgs couplings discussed below, the production cross sections are modified by a factor $\cos^2\theta_H$, whereas the branching ratios remain unchanged as the cosine factors in the partial and total widths cancel. Thus, the signal strengths are $\mu = \cos^2\theta_H$. 
An estimate of the limit on the mixing angle can be obtained from the global signal strength.
The current \experiment{CMS} measurement is $\mu = 1.17 \pm 0.10$~\cite{CMS:2018lkl}.
This gives a 95\,\% exclusion of
\be
	\label{eq:GlobalSignalStrengthLimit}
	|\sin\theta_H| < 0.16\,.
\ee

Loops of the dark electrons $e_4$ and $e_5$ can contribute sizeably to the $h \longrightarrow \gamma \gamma$ and $h \longrightarrow Z \gamma$ rate.
In the SM, these rates are given by~\cite{Djouadi:2005gi}
\begin{align}
	\label{eq:Higgs-diphoton}
	\Gamma(h\longrightarrow\gamma\gamma) =&\ \frac{\alpha^2 m_h^3}{256 \pi^3 v_H^2} \left| \sum_f N_c^f Q_f^2 A_{1/2}(\tau_f) + A_1(\tau_W)  \right|^2\,,\\
	\label{eq:Higgs-Zgamma}
	\Gamma(h\longrightarrow Z\gamma) =&\ \frac{\alpha m_W^2 m_h^3}{128 \pi^4 v_H^4} \left(1-\frac{m_Z^2}{m_h^2}\right)^3 \left| \sum_f N_c^f \frac{Q_f \hat v_f}{c_W} A_{1/2}(\tau_f,\lambda_f) + A_1(\tau_W,\lambda_W)  \right|^2\,,
\end{align}
where $\alpha$ is the electromagnetic coupling constant, $\tau_i = 4 m_i^2/m_h^2$, and $\lambda_i = 4 m_i^2/m_Z^2$.
The expressions for the form factors $A_s$ for a spin $s$ particle running in the loop can be found in~\cite{Djouadi:2005gi}.
The sums run over all charged fermions that couple to the Higgs. $N_c^f$ is the color-representation of the fermion, $Q_f$ is its electric charge, and $\hat v_f$ is the fermion's (reduced) vector-coupling to the $Z$ boson.
In the SM, the dominant contribution from fermions comes from the top quark with $N_c^t=3$, $Q_t = 2/3$, and $\hat v_t = 1- \frac{8}{3} s_W^2$.

Equations~\eqref{eq:Higgs-diphoton}~and~\eqref{eq:Higgs-Zgamma} assume that the fermions couple to the Higgs with a vertex factor proportional to their masses. The top quark in the SM for instance has a vertex factor $- i \frac{y_t}{\sqrt{2}} = - i \frac{m_t}{v_H}$.
This is not true for the heavy charged leptons.
The corresponding vertex factors are
\begin{equation}
	\mathfig{\includegraphics{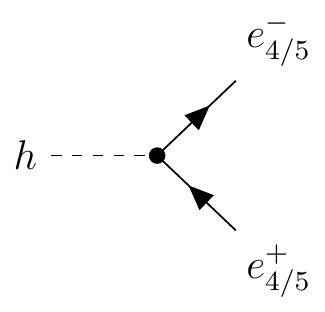}}\qquad - \frac{i}{\sqrt{2}} \mathcal{Y}_{h e_{4/5}},\qquad \mathcal{Y}_{h e_{4/5}} = \pm c_H y_e - s_H c_\ell\,,
\end{equation}
for the SM-like Higgs.\footnote{The corresponding interactions of $\phi$ are obtained by replacing $c_H \rightarrow s_H$ and $s_H \rightarrow - c_H$.}
The correct result can then be obtained by rescaling the heavy lepton contributions by a factor $\frac{\mathcal{Y}_{s f} v_H}{\sqrt{2} m_f}$, where $s = h, \phi$ and $f=e_4, e_5$.
Due to the scalar mixing, the SM contributions further get a factor of $c_H$ or $s_H$ for $h$ and $\phi$, respectively.
We thus obtain
\begin{align}
	\Gamma(h\longrightarrow\gamma\gamma) =&\ \frac{\alpha^2 m_h^3}{256 \pi^3 v_H^2} \bigg| 
		\frac{4}{3} c_H A_{1/2}(\tau_t) 
		+ c_H  A_1(\tau_W)
		\notag\\&\hspace{3cm}
		+ \frac{\mathcal{Y}_{h e_4} v_H}{\sqrt{2} m_{e_4}} A_{1/2}(\tau_{e_4}) 
		+ \frac{\mathcal{Y}_{h e_5} v_H}{\sqrt{2} m_{e_5}} A_{1/2}(\tau_{e_5})  
	\bigg|^2\,,\label{eq:h2aa}\\
	\Gamma(h\longrightarrow Z\gamma) =&\ \frac{\alpha m_W^2 m_h^3}{128 \pi^4 v_H^4} \left(1-\frac{m_Z^2}{m_h^2}\right)^3 \bigg| 
		\frac{6+16 s_W^2}{3 c_W} c_H A_{1/2}(\tau_t,\lambda_t) + c_H A_1(\tau_W,\lambda_W)  
		\notag\\&\hspace{4cm}
		+ \frac{1-4 s_W^2}{c_W} \frac{\mathcal{Y}_{h e_4} v_H}{\sqrt{2} m_{e_4}} A_{1/2}(\tau_{e_4},\lambda_{e_4}) 
		\notag\\&\hspace{4cm}
		+ \frac{1-4 s_W^2}{c_W} \frac{\mathcal{Y}_{h e_5} v_H}{\sqrt{2} m_{e_5}} A_{1/2}(\tau_{e_5},\lambda_{e_5})  
	\bigg|^2\,.\label{eq:h2Za}
\end{align}
The corresponding width for the $\phi$ scalar can be obtained by replacing $m_h \rightarrow m_\phi$, $c_H \rightarrow s_H$, $\mathcal{Y}_{h e_i} \rightarrow \mathcal{Y}_{\phi e_i}$, and $\tau_i = 4 m_i^2/m_\phi^2$.
The leading QCD corrections can be included by multiplying the top contribution by $\left(1-\frac{\alpha_s}{\pi}\right)$~\cite{Djouadi:2005gi}.

\begin{figure}
	\begin{minipage}[t]{.48\textwidth}
		\centering
		\includegraphics[width=\textwidth]{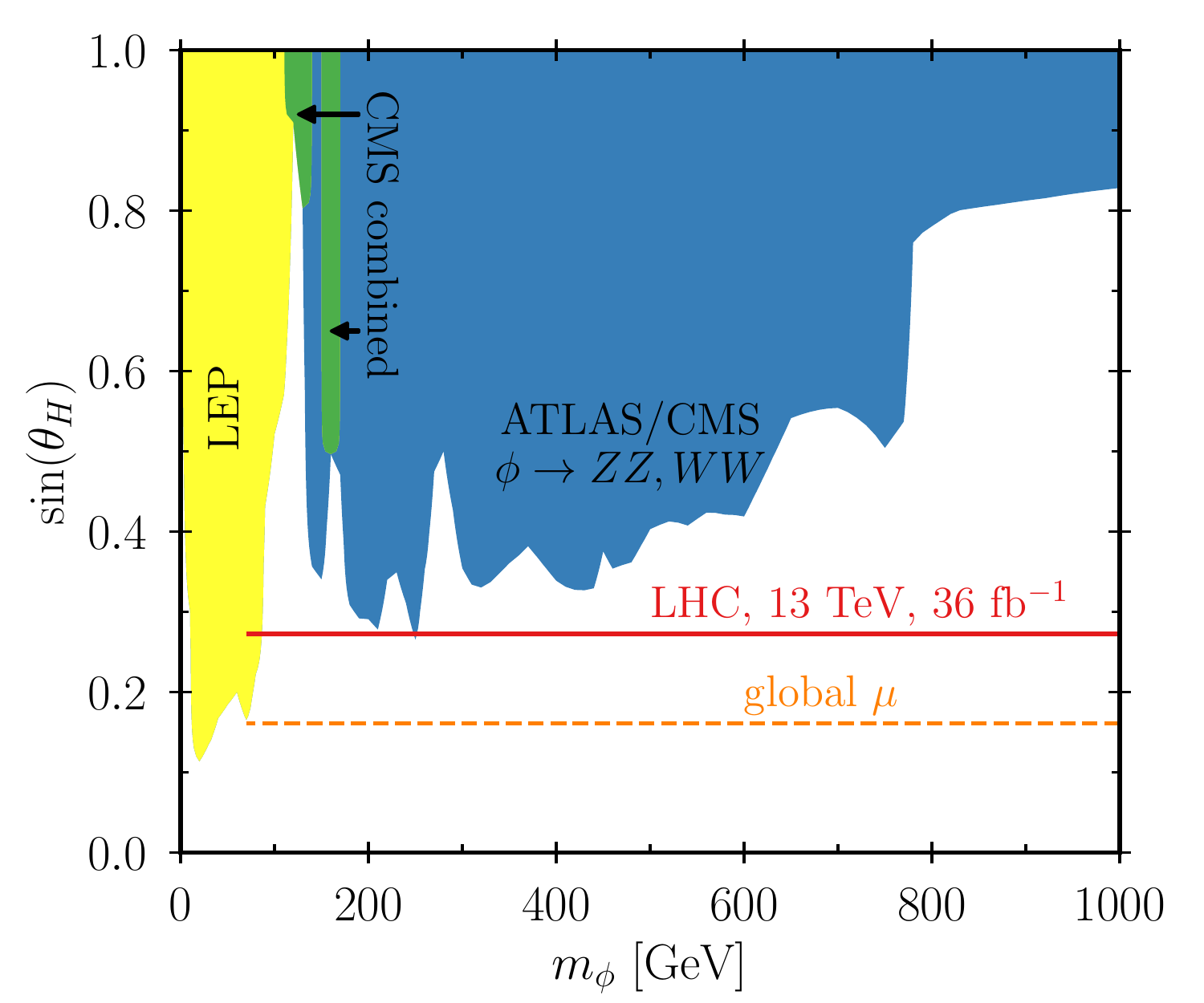}
		\caption{Exclusion bounds on the mass of the lepton number breaking scalar $\phi$ and the Higgs mixing angle $\theta_H$ from direct searches (colored regions) and signal strength measurements (colored lines).}
		\label{fig:HiggsConstraints}
	\end{minipage}
	\hfill
	\begin{minipage}[t]{.48\textwidth}
		\centering
		\includegraphics[width=\textwidth]{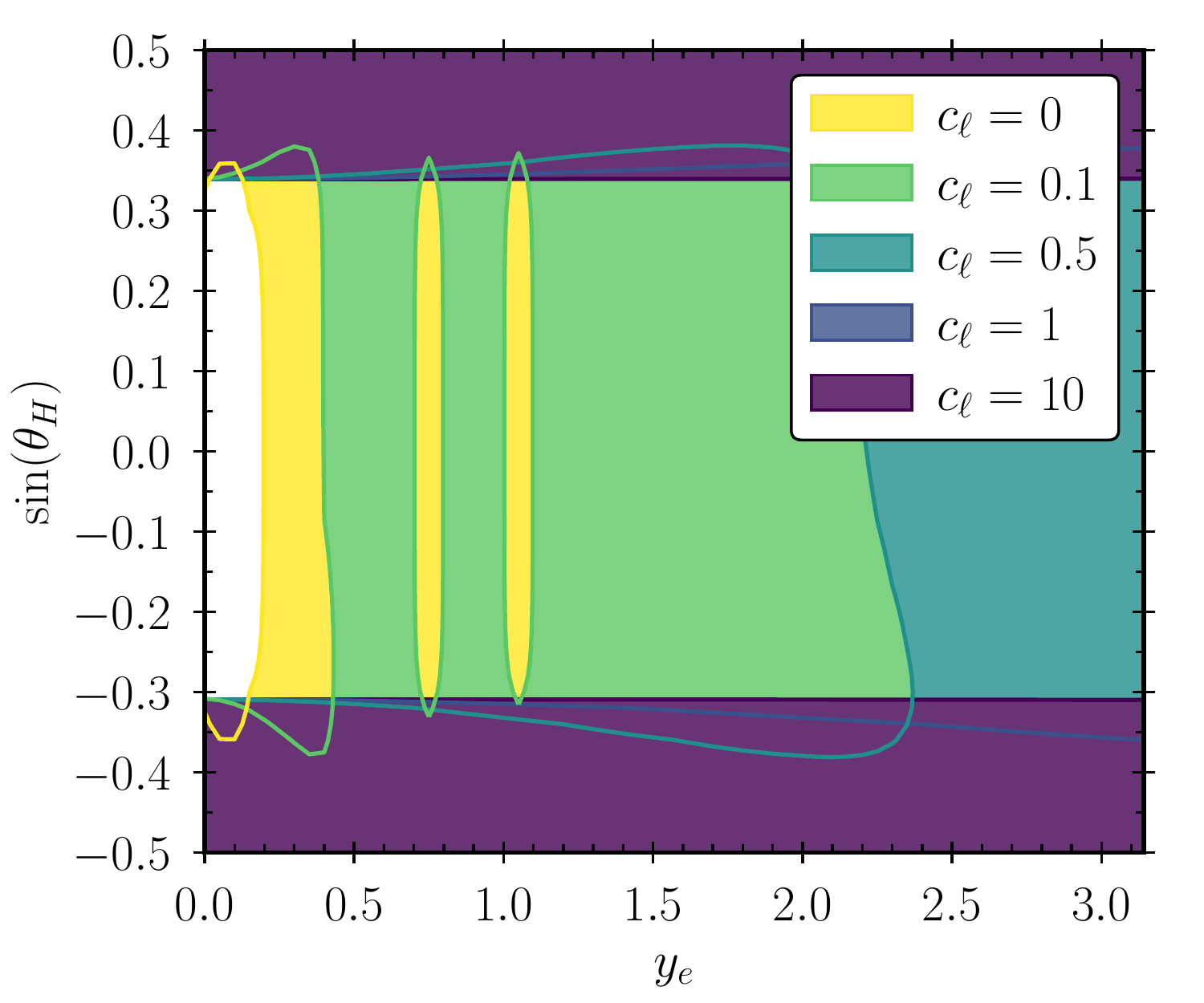}
		\caption{95\,\% exclusion limits from signal strength measurements by \experiment{ATLAS} and \experiment{CMS} on the Higgs mixing angle and the heavy electron Yukawa couplings.}
		\label{fig:HiggsLimitsYukawas}	
	\end{minipage}
\end{figure}

We evaluated constraints from direct Higgs searches using \software{HiggsBounds 4.3.1}~\cite{Bechtle:2013wla}. 
The corresponding limits from \experiment{LEP}~\cite{Schael:2006cr,Abbiendi:2002qp} (yellow), $\sqrt{s}=7$~TeV and $\sqrt{s}=8$~TeV searches with \experiment{ATLAS} and \experiment{CMS} for a Higgs boson in the $h\longrightarrow ZZ/WW$ channel \cite{CMS:bxa,CMS:xwa,Khachatryan:2015cwa,Aad:2015kna} (blue), and a combination of \experiment{CMS} 7 and 8~TeV searches in various final states~\cite{CMS:aya} (green) are shown as colored regions in figure~\ref{fig:HiggsConstraints}.
The red line indicates limits from signal strength measurements. These include measurements of Higgs boson properties in the $h\rightarrow 4\ell$ and $h\rightarrow\gamma\gamma$ channels by \experiment{ATLAS}~\cite{Aaboud:2017vzb,Aaboud:2018xdt}, and a \experiment{CMS} analysis combining different channels~\cite{CMS:2018lkl}, both at a center-of-mass energy of 13~TeV, as well as the combination of 7+8~TeV results from \experiment{ATLAS} and \experiment{CMS}~\cite{Khachatryan:2016vau}.
The dashed, orange line corresponds to the naive estimate~\eqref{eq:GlobalSignalStrengthLimit}.

The constraints on the Higgs mixing angle $\theta_H$ are shown in 
figure~\ref{fig:HiggsConstraints} for the parameter values given in table~\ref{tab:defaultParameters}. Signal strength measurements exclude $\sin\theta_H \gtrsim 0.27$, i.e.\ the limit~\eqref{eq:GlobalSignalStrengthLimit} from the global signal strength overestimates the exclusion reach. Direct searches for additional scalars provide somewhat weaker constraints of around $\sin\theta_H \gtrsim 0.4$ for a large range of $m_\phi$, but are stronger for scalar masses below the Higgs mass. The Higgs signal strength fits are more involved for $m_\phi$ near 125~GeV and for $m_\phi < 62.5$~GeV where the Higgs may decay into $\phi$-pairs. The signal strength constraint shown in Fig.~\ref{fig:HiggsConstraints} should be taken with a grain of salt in those regions. 

If the new leptons have sizeable couplings to the Higgs boson, the Higgs signal strengths in different channels can vary due to the loop contributions to $h\to \gamma\gamma$ and $h \to Z\gamma$ decays. Figure~\ref{fig:HiggsLimitsYukawas} shows the current \experiment{LHC} limits from~\cite{Khachatryan:2016vau,Aaboud:2017vzb,Aaboud:2018xdt,CMS:2018lkl} as a function of the mixing angle and the heavy electron Yukawa couplings $c_\ell$ and $y_e$.
If the $\Phi$ Yukawa coupling $c_\ell$ is small, the dark electrons gain their mass predominantly from electroweak symmetry breaking and hence strongly contribute to the $h\longrightarrow\gamma\gamma$ rate. Thus, the Yukawa coupling to the Higgs doublet $y_e$ is also restricted to be small.
For large $c_\ell$, the heavy electron contributions in~\eqref{eq:h2aa} are mass-suppressed, so that $y_e$ can take large values without modifying $\Gamma\left(h\longrightarrow\gamma\gamma\right)$ beyond the experimentally allowed limits.

Note that for $c_\ell = 0$ the charged, heavy lepton masses are given by $m_{e_{4/5}} = \frac{y_e v_H}{\sqrt{2}}$. 
The \experiment{LEP} limit on the mass (see section~\ref{sec:Lepton-constraints}) then constrains the Yukawa coupling to $y_e > 0.57$, hence the entire region allowed by $h\to \gamma\gamma$ and $h \to Z\gamma$ is excluded in this case.
Similarly,~\eqref{eq:e45mass} implies $|y_e| < 0.24$ for $c_\ell=0.1$ and $v_\Phi = 2$~TeV.   
Further note that the exclusion for $c_\ell = 10$ is shown despite severly challenging the bounds of perturbativity to illustrate the constraints in the limit of large $c_\ell$.

\subsection{Constraints on heavy leptons}
\label{sec:Lepton-constraints}

Whereas the DM candidate is mostly an SM singlet, the remaining heavy leptons carry electroweak charge and can hence be produced at colliders. 
Direct searches for charged heavy leptons at \experiment{LEP} set a lower limit of roughly 100~GeV on the ${e_4}$ and ${e_5}$ mass~\cite{Achard:2001qw}.
\experiment{LHC} limits on the dark leptons can be obtained by recasting supersymmetry (SUSY) searches for electroweakly produced charginos and neutralinos.

\begin{figure}
	\begin{minipage}[t]{.48\textwidth}
		\centering
		\includegraphics[width=\textwidth]{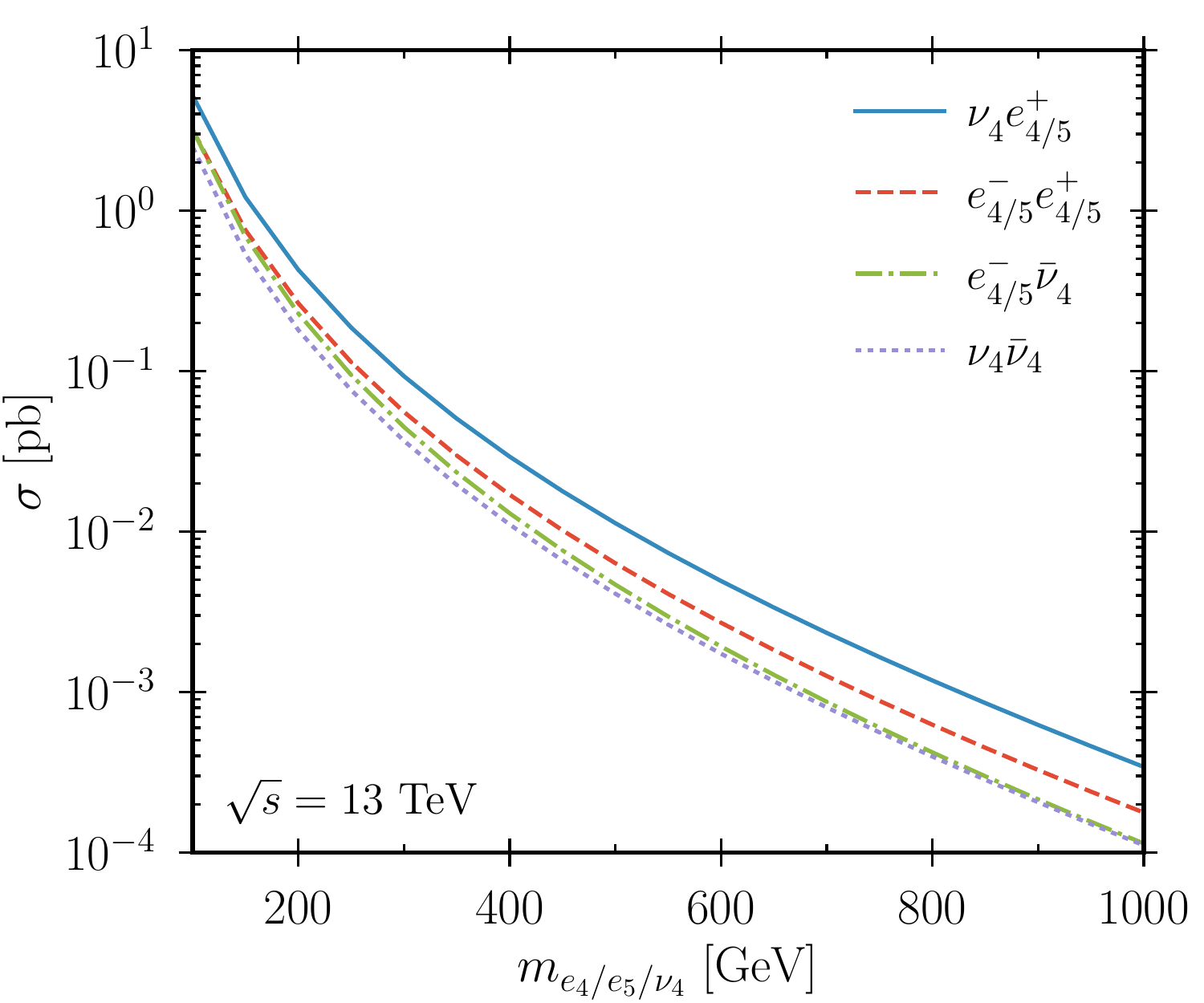}
		\caption{Production cross sections for heavy lepton pairs at the \experiment{LHC} with a center-of-mass energy of 13~TeV, assuming equal masses for all heavy leptons.}
		\label{fig:LHC-HL-xsec}
	\end{minipage}
	\hfill
	\begin{minipage}[t]{.48\textwidth}
		\centering
		\includegraphics[width=\textwidth]{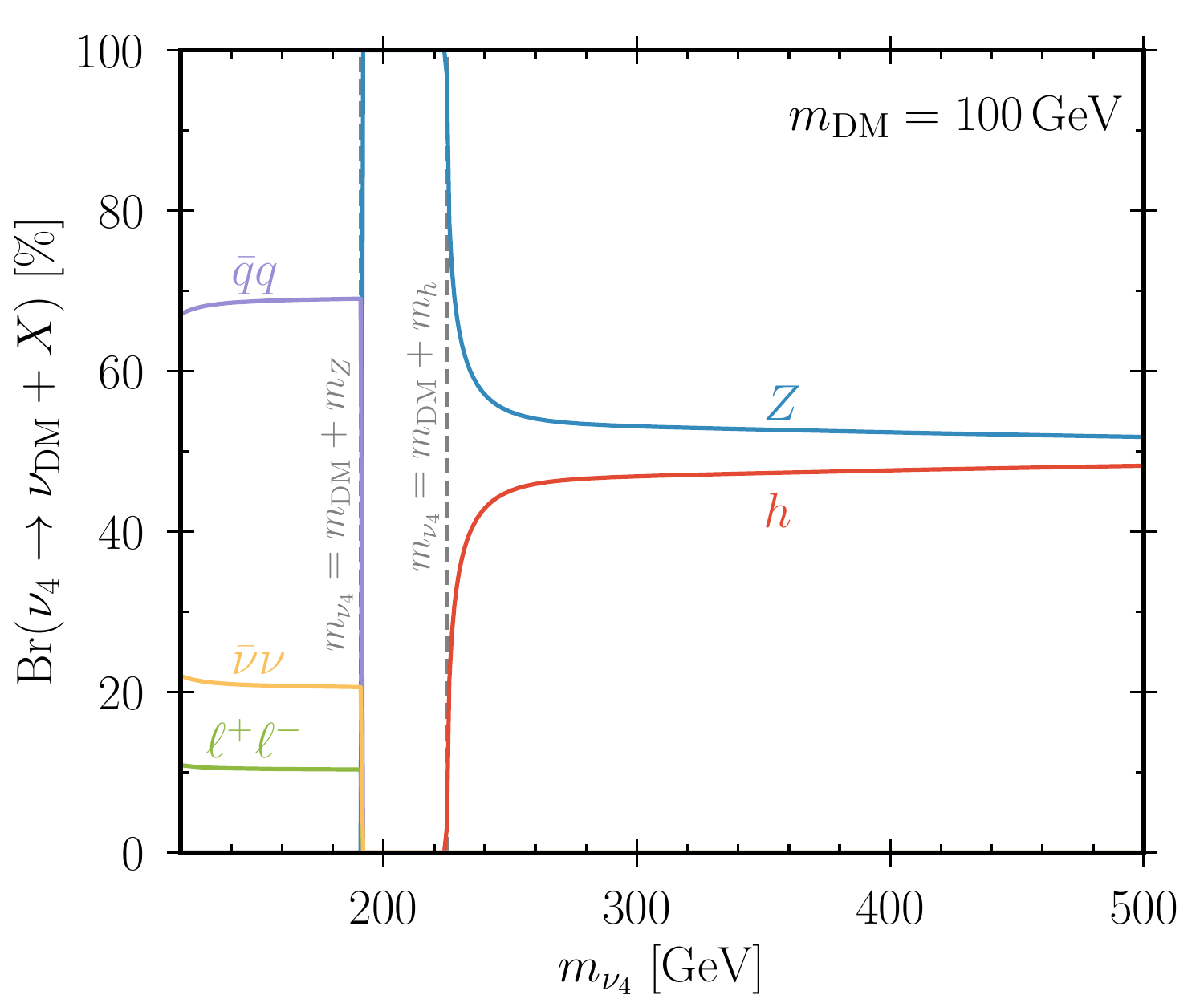}
		\caption{Branching ratios of the neutral heavy lepton, assuming $m_{e_4} = m_{e_5} \approx m_{\nu_4}$, $\sin\theta_\DM = 0.01$, and a DM mass of 100~GeV.}
		\label{fig:LHC-HL-Br}
	\end{minipage}
\end{figure}

Figure~\ref{fig:LHC-HL-xsec} shows the cross section for the production of two charged heavy leptons (dashed, red), heavy positrons with a heavy neutrino (solid blue), heavy electrons with a heavy anti-neutrino (dash-dotted, green), and pairs of heavy neutrinos (dotted, purple) in proton-proton collisions at $\sqrt{s} = 13$~TeV calculated using \software{CalcHep}~\cite{Belyaev:2012qa}.  
We here take the most optimistic scenario for heavy lepton production in which all heavy leptons have the same mass.
As the \DM\ mixing angle $\theta_\DM$ is restricted to be small by the direct detection constraints in section~\ref{sec:detection}, and therefore only has a negligible effect on the production cross section, it can be set to zero.

Due to the $U(1)$ symmetry that stabilizes the \DM, the heavy leptons can only decay amongst themselves or to dark matter.   
Consequently, to allow the lighter dark electron to decay, $\theta_\DM \neq 0$ is required.
Otherwise the model would have a charged \DM\ population and therefore be excluded.
However, even a (negligibly) small amount of \DM\ mixing is sufficient to let the exotic particles decay fast enough to avoid this problem.

The charged heavy leptons typically decay into \DM\ and a (potentially off-shell) $W$ boson. Depending on the masses, decays to other exotic leptons can also be possible, these are however suppressed by phase space.
The heavy neutrino can decay to \DM\ and a(n off-shell) $Z$ boson or, if $m_{\nu_4} > m_\DM + m_h$, to \DM\ and an SM Higgs boson. In the latter case, the branching rations to $\DM+Z$ and $\DM+h$ are roughly 50~\%.
Figure~\ref{fig:LHC-HL-Br} shows the branching ratios of $\nu_4$ as a function of the mass for a \DM\ mass of 100~GeV and a mixing angle of $\sin\theta_\DM = 0.01$.

At the \experiment{LHC}, the heavy leptons can be searched for by looking for missing transverse energy in association with $W$, $Z$, or $h$ bosons (or SM lepton pairs in the off-shell case). 
These searches have been performed by \experiment{ATLAS}~\cite{Aaboud:2018jiw} and \experiment{CMS}~\cite{Sirunyan:2018ubx} in the context of simplified SUSY models, using 36~fb$^{-1}$ of data recorded at the \experiment{LHC} with $\sqrt{s} = 13$~TeV.
They assume that the lightest neutralino $\tilde\chi_1^0$ is the lightest supersymmetric particle (LSP) and consider the process ${ p p \longrightarrow \tilde\chi_1^\pm \tilde\chi_2^0}$, where the lightest chargino $\tilde\chi_1^\pm$ decays to the LSP plus a $W$ boson, and the next-to-lightest neutralino $\tilde\chi_2^0$ to the LSP plus $Z$ or $h$.
The respective 95~\% exclusion bounds from \experiment{CMS} can be found in figure 8 of~\cite{Sirunyan:2018ubx}.

\begin{figure}
	\begin{minipage}[t]{.48\textwidth}
		\centering
		\includegraphics[width=\textwidth]{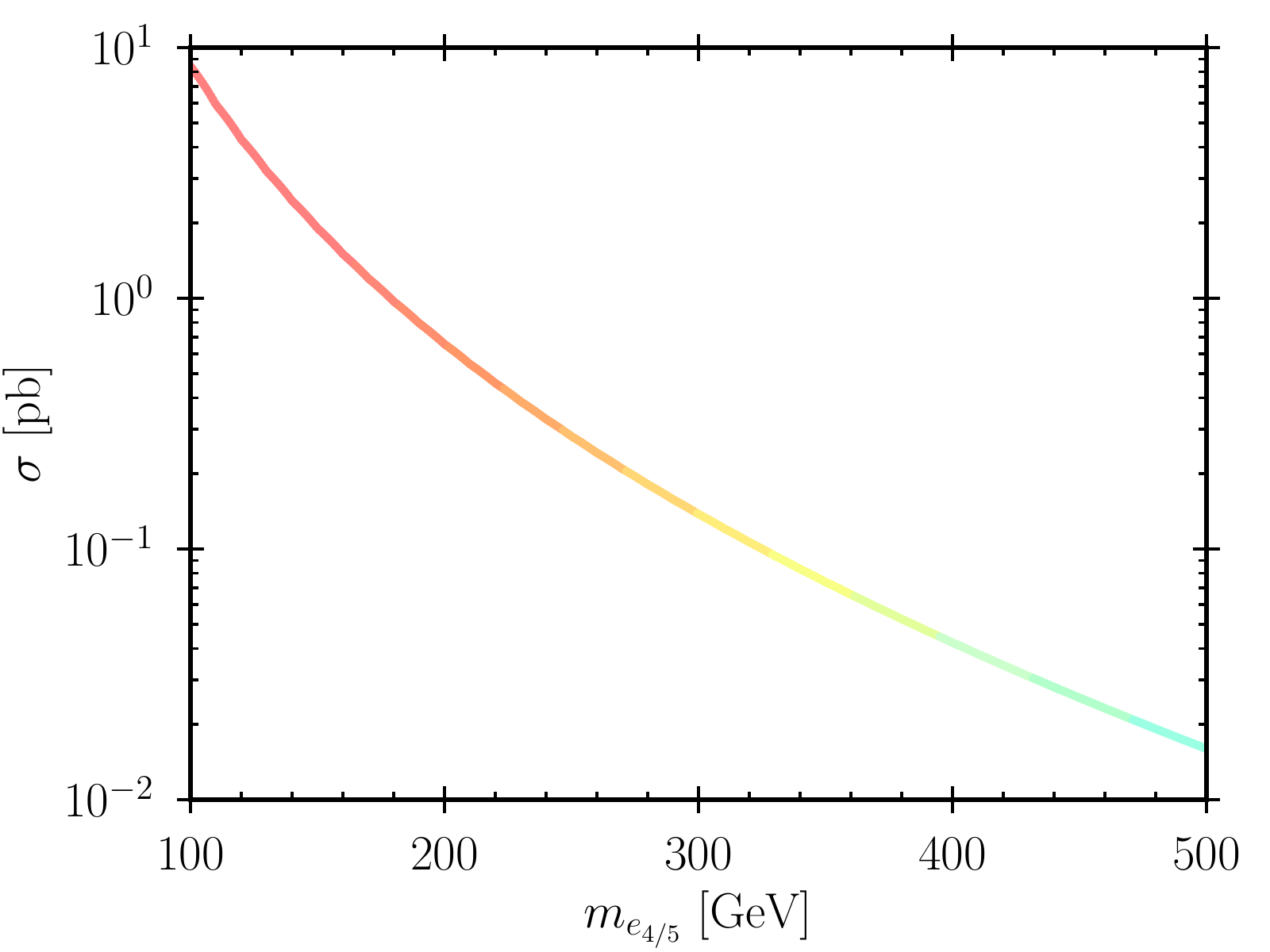}
		\caption{Cross section for $p p \longrightarrow e_{4/5}^\pm \nu_4$ production at $\sqrt{s}=13$~TeV as a function of the heavy lepton mass. The colors resemble the color-coding of the cross section in reference \cite{Sirunyan:2018ubx}.}
		\label{fig:LHC-xsec-SUSYsearch}
	\end{minipage}
	\hfill
	\begin{minipage}[t]{.48\textwidth}
		\centering
		\includegraphics[width=\textwidth]{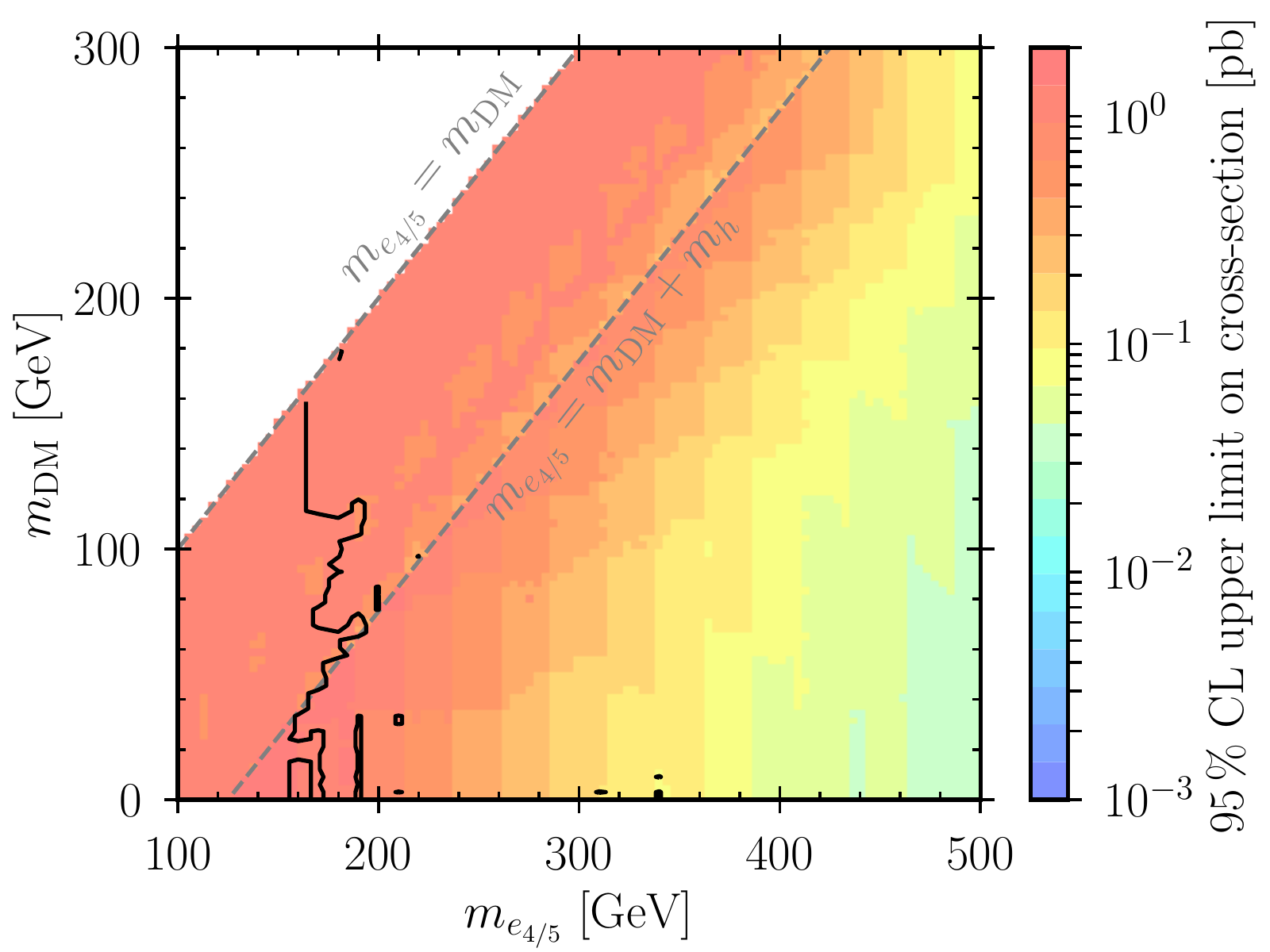}
		\caption{CMS 95\ \% exclusion limits from \cite{Sirunyan:2018ubx}, assuming equal masses for all heavy leptons and that they decay within the detector.}
		\label{fig:LHC-CMS-exclusionHL}
	\end{minipage}
\end{figure}

The production cross section for the corresponding process in our model is depicted in figure~\ref{fig:LHC-xsec-SUSYsearch}, again assuming $m_{e_4} = m_{e_5} = m_{\nu_4}$ ($\sin\theta_\DM = 0$).
The respective exclusion bounds from \experiment{CMS}~\cite{Sirunyan:2018ubx} are shown in figure~\ref{fig:LHC-CMS-exclusionHL}, taking the limits assuming $\mathrm{Br}\left(\tilde\chi_2^0\rightarrow\tilde\chi_1^0 Z\right) = 100\ \%$ for  $m_\DM < m_{e_4/e_5/\nu_4} < m_\DM + m_h$, and  $\mathrm{Br}\left(\tilde\chi_2^0\rightarrow\tilde\chi_1^0 Z\right) = \mathrm{Br}\left(\tilde\chi_2^0\rightarrow\tilde\chi_1^0 h\right) = 50\ \%$ for  $m_{e_4/e_5/\nu_4} > m_\DM + m_h$.
The \experiment{LHC} can currently exclude heavy lepton masses below $m_{e_4/e_5/\nu_4} \simeq 200$~GeV and DM masses below $m_\DM \simeq 150$~GeV. For the co-annihilation region discussed in section~\ref{sec:abundance}, the mass splitting between the charged states and the DM candidate becomes very small, so that the searches used here become inefficient. Instead it has been shown that these regions can be probed by mono-jet, mono-Z and disappearing track searches, with masses of up to 200~GeV reachable at the \experiment{LHC} and up to 1~TeV at a future hadron collider~\cite{Han:2013usa,Schwaller:2013baa,Baer:2014cua,Han:2014kaa,Low:2014cba,Mahbubani:2017gjh}.

\section{The lepton number breaking phase transition}
\label{sec:pt}
%

In the early Universe, spontaneously broken symmetries are typically restored due to thermal effects.
This can be seen by computing finite-temperature corrections to the effective potential of the scalar fields whose vacuum expectation values break the symmetries.
In the model at hand, these are the scalar $\hat{\phi}$ breaking the $U(1)_\ell$ lepton number gauge symmetry, and the Higgs field $\hat{h}$ which breaks $SU(2)_W \otimes U(1)_Y$.
As a consequence, a symmetry breaking phase transition (PT) occurred in the history of the Universe.

At high temperatures, the global minimum of the finite-temperature effective potential is at the origin, i.e.\ $SU(2)_W \otimes U(1)_Y \otimes U(1)_\ell$ is unbroken. 
When the temperature drops, the potential changes and at some point develops a minimum at non-vanishing field values.
Whether both fields develop non-zero VEVs at the same time or independently at different temperatures depends on the parameters of the model. 
Since the portal interaction between the two scalars is restricted to be small (see section~\ref{sec:detection} and \ref{sec:Higgs-constraints}), and due to the hierarchy of the VEVs ($v_H = 246$~GeV, $v_\Phi \gtrsim 1.9$~TeV), the lepton number breaking PT typically occurs first at temperatures in the TeV range, leaving the EW symmetry unbroken.
EW symmetry breaking subsequently occurs as in the SM at a temperature of $T\simeq 160$~GeV~\cite{DOnofrio:2015gop}.

The phase transition between the unbroken and the broken phase can proceed in basically two different ways.
In a first-order phase transition the scalar fields tunnel or fluctuate through a potential barrier between the two phases, whereas in a cross-over transition they smoothly evolve from one vacuum to the other.
The EWPT in the SM falls into the latter category~\cite{DOnofrio:2015gop}.

Here we are mainly interested in transitions of the first kind.
These have a much richer phenomenology, potentially giving rise to a stochastic background of gravitational waves as discussed in section~\ref{sec:gws}, or providing out-of-equilibrium dynamics which are, along with baryon number, $C$ and $CP$ violation, required by the Sakharov conditions for baryogenesis~\cite{Sakharov:1967dj}.

A cosmological first-order phase transition can occur when at some critical temperature $T_c$ the scalar potential develops two degenerate minima separated by a barrier.
The transition then proceeds through the nucleation of bubbles of the true vacuum (i.e.\ the global minimum at low temperatures) in the sea of the false vacuum (i.e.\ the global minimum at high temperatures).
The energy released from the potential difference of the two vacua drives the expansion of the bubbles, which collide and merge, and eventually fill the whole universe.

The bubble nucleation rate per unit volume at finite temperature is given by~\cite{Linde:1981zj,Quiros:1999jp,Caprini:2015zlo}
\be
	\Gamma(T) = A(T) \exp\left[-S_E(T)\right],
\ee
where $A(T)\sim T^4$ is a prefactor and $S_E$ is the Euclidean action
\be
	S_E (T) = \frac{S_3(T)}{T}  = \frac{1}{T} \int \dInt^3 x \left[ \frac{1}{2} \left(\nabla\phi\right)^2 + V(\phi,T)\right] ,
\ee
evaluated for the $O(3)$ symmetric bounce solution.
In the early Universe the process of bubble nucleation has to compete against the expansion of the Universe, given by the Hubble rate $H$. Furthermore, the surface energy would make nucleated bubbles collapse unless the energy gain from the potential difference is larger. Consequently, the transition does not occur immediately at the critical temperature, but at a lower temperature -- the so-called nucleation temperature $T_n$. It is defined by the temperature at which the probability to nucleate one sufficiently large bubble per Hubble volume is of order one. The bubbles nucleated within one Hubble patch proceed to expand and collide, until the entire volume is filled with the true vacuum.

\subsection{Finite temperature effective potential}
\label{sec:potential}

The daisy-resummed one-loop effective potential of the scalar fields at finite temperature is given by~\cite{Dolan:1973qd,Quiros:1999jp,Curtin:2016urg}
\be
	\label{eq:Veff}
	V_\eff(h,\phi,T) = V_0(h,\phi) + V_\textnormal{CW}(h,\phi) + V_\textnormal{c.t.}(h,\phi) + V_T(h,\phi,T) + V_\textnormal{ring}(h,\phi,T)\, .
\ee
Here, $h$ and $\phi$ are classical background fields,\footnote{Note that $h$ and $\phi$ do not correspond to the mass eigenstates but to the (gauge) interaction eigenstates $\hat h$ and $\hat\phi$. In the following, we drop the hats for convenience.}  and $V_0$ is the tree-level potential of $h$ and $\phi$ given by
\be
	\label{eq:V0}
	V_0(h,\phi) = - \frac{1}{2} \mu_H^2 h^2 + \frac{1}{4} \lambda_H h^4
	              - \frac{1}{2} \mu_\Phi^2 \phi^2 + \frac{1}{4} \lambda_\Phi \phi^4
	              + \frac{1}{4} \lambda_{p} h^2 \phi^2
	.
\ee

$V_\textnormal{CW}$ is the Coleman-Weinberg potential~\cite{Coleman:1973jx} which includes all temperature-independent one-loop corrections at vanishing external momentum. In dimensional regularization with $\overline{\textnormal{MS}}$ renormalization it reads
\be
	\label{eq:VCW}
	V_\textnormal{CW}(h,\phi) = \sum_i n_i \frac{m_i^4(h,\phi)}{64\pi^2} \left( \log\frac{m_i^2(h,\phi)}{\mu_R^2} - C_i\right) .
\ee
The sum runs over all particles that couple to the scalars. Since we compute the potential in Landau gauge this also includes Goldstone bosons. Here $n_i$ are the numbers of degrees of freedom of each particle, which include an additional factor $-1$ for fermions, and $m_i^2(h,\phi)$ are the field dependent masses. The constants $C_i$ are $C_i = \frac{3}{2}$ for scalars or fermions and $C_i = \frac{5}{6}$ for gauge bosons.

$V_\textnormal{c.t.}$ includes the non-$\overline{\textnormal{MS}}$ part of the counter-terms, i.e. the difference of the counter-terms to the usual $\overline{\textnormal{MS}}$ counter-terms.
These can be derived from the Coleman-Weinberg (and tree-level) potential and take the form
\be
	\label{eq:Vct}
	V_\textnormal{c.t.}(h,\phi) = - \frac{1}{2} \delta\mu_H^2 h^2 + \frac{1}{4} \delta\lambda_H h^4
	- \frac{1}{2} \delta\mu_\Phi^2 \phi^2 + \frac{1}{4}\delta\lambda_\Phi \phi^4
	+ \frac{1}{4} \delta\lambda_{p} h^2 \phi^2
	.
\ee
The exact form of the counter-term couplings depends on the renormalization conditions imposed.

The temperature-dependent one-loop corrections to the potential are contained in
\be
  \label{eq:VT}
	V_T(h,\phi,T) = \sum_i n_i \frac{T^4}{2\pi^2} J_{B/F}\left(\frac{m_i^2(h,\phi)}{T^2}\right) ,
\ee
where the thermal functions for bosons ($J_B$, ``$-$'' case) and fermions ($J_F$, ``$+$'' case) are
\be
	J_{B/F}(y^2) = \int\limits_0^\infty dx\ x^2 \log\left[ 1 \mp \exp\left(-\sqrt{x^2+y^2}\right) \right] .
\ee
For $T^2 \gg m^2$, these can be expanded as~\cite{Dolan:1973qd,Quiros:1999jp,Curtin:2016urg}
\begin{align}
	\label{eq:highTbosons}
	J_B(y^2) =&\ - \frac{\pi^4}{45} + \frac{\pi^2}{12} y^2 - \frac{\pi}{6} \left(y^2\right)^{3/2} - \frac{1}{32} y^4 \log\frac{y^2}{a_b} + \mathcal{O}\left(y^6\right)\ , \\
	\label{eq:highTfermions}
	J_F(y^2) =&\ \frac{7 \pi^4}{360} - \frac{\pi^2}{24} y^2 - \frac{1}{32} y^4 \log\frac{y^2}{a_f} + \mathcal{O}\left(y^6\right)\ ,
\end{align}
where $a_b = 16 \pi^2 \exp(3/2-2 \gamma_E)$ and $a_f = \pi^2 \exp(3/2-2 \gamma_E)$.

Finally, the resummed daisy corrections are
\be
	V_\textnormal{ring}(h,\phi,T) = - \sum_{i} \bar{n}_i \frac{T}{12\pi} \left(\left[m^2(h,\phi) + \Pi(T)\right]_i^{3/2} - \left[m_i^2(h,\phi)\right]^{3/2}\right) ,
\ee
where $\bar{n}_i$ indicates that the sum only runs over scalars and the longitudinal gauge boson degrees of freedom, and $\Pi_i(T)$ are the leading thermal mass corrections.

\begin{figure}
	\centering
	
	\includegraphics[width=.48\textwidth]{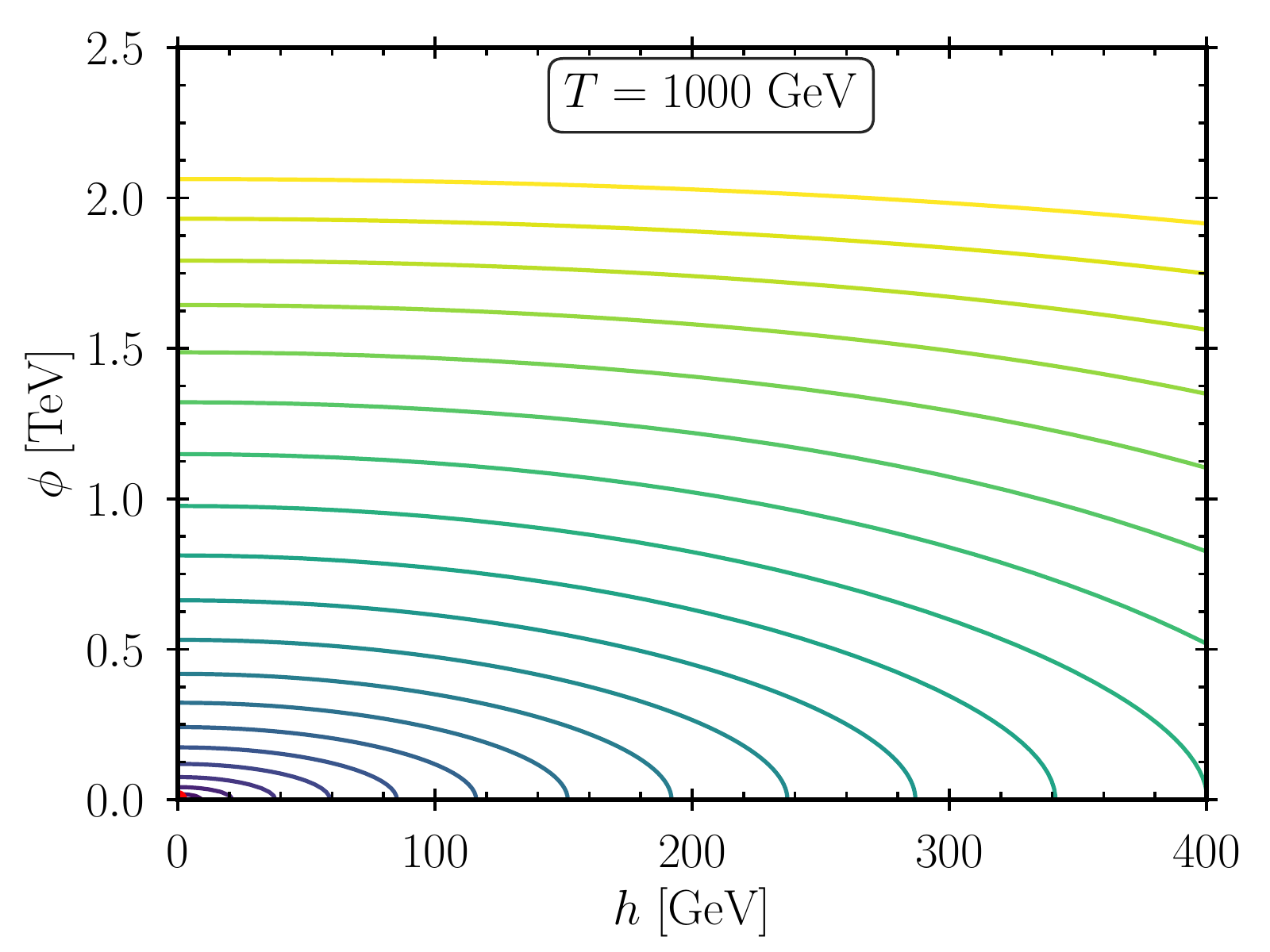}
	\hfill
	\includegraphics[width=.48\textwidth]{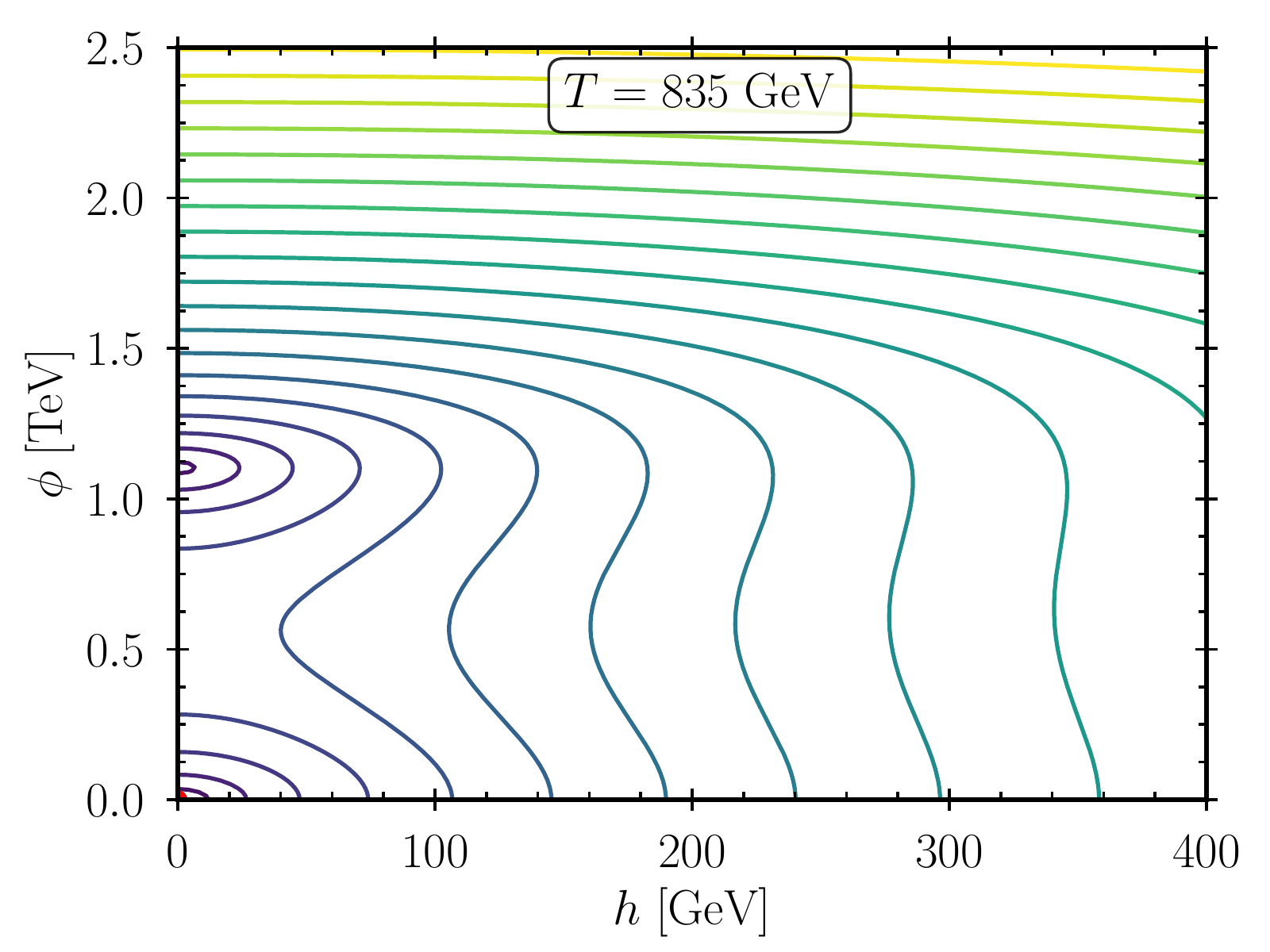}
	
	\includegraphics[width=.48\textwidth]{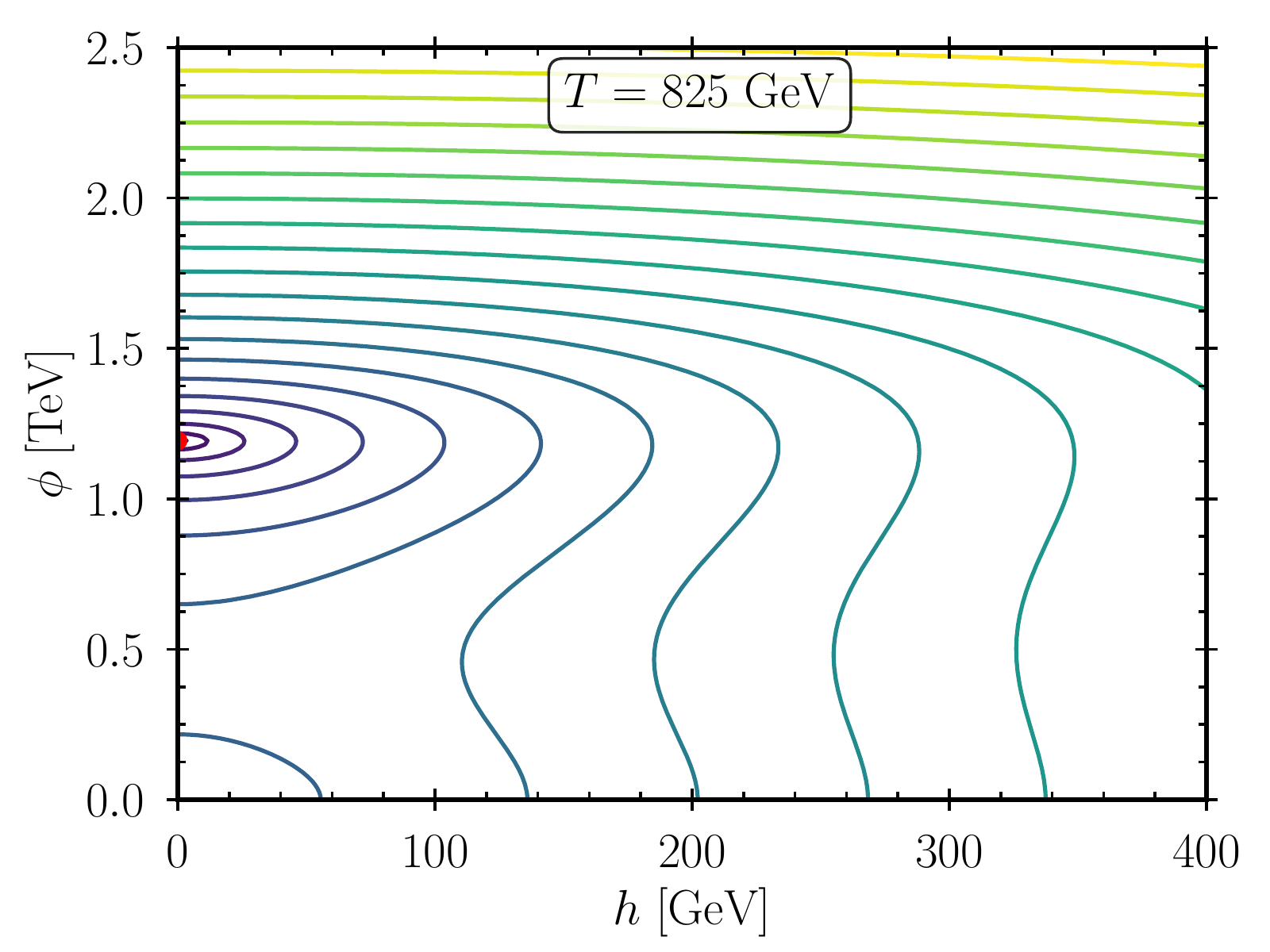}
	\hfill
	\includegraphics[width=.48\textwidth]{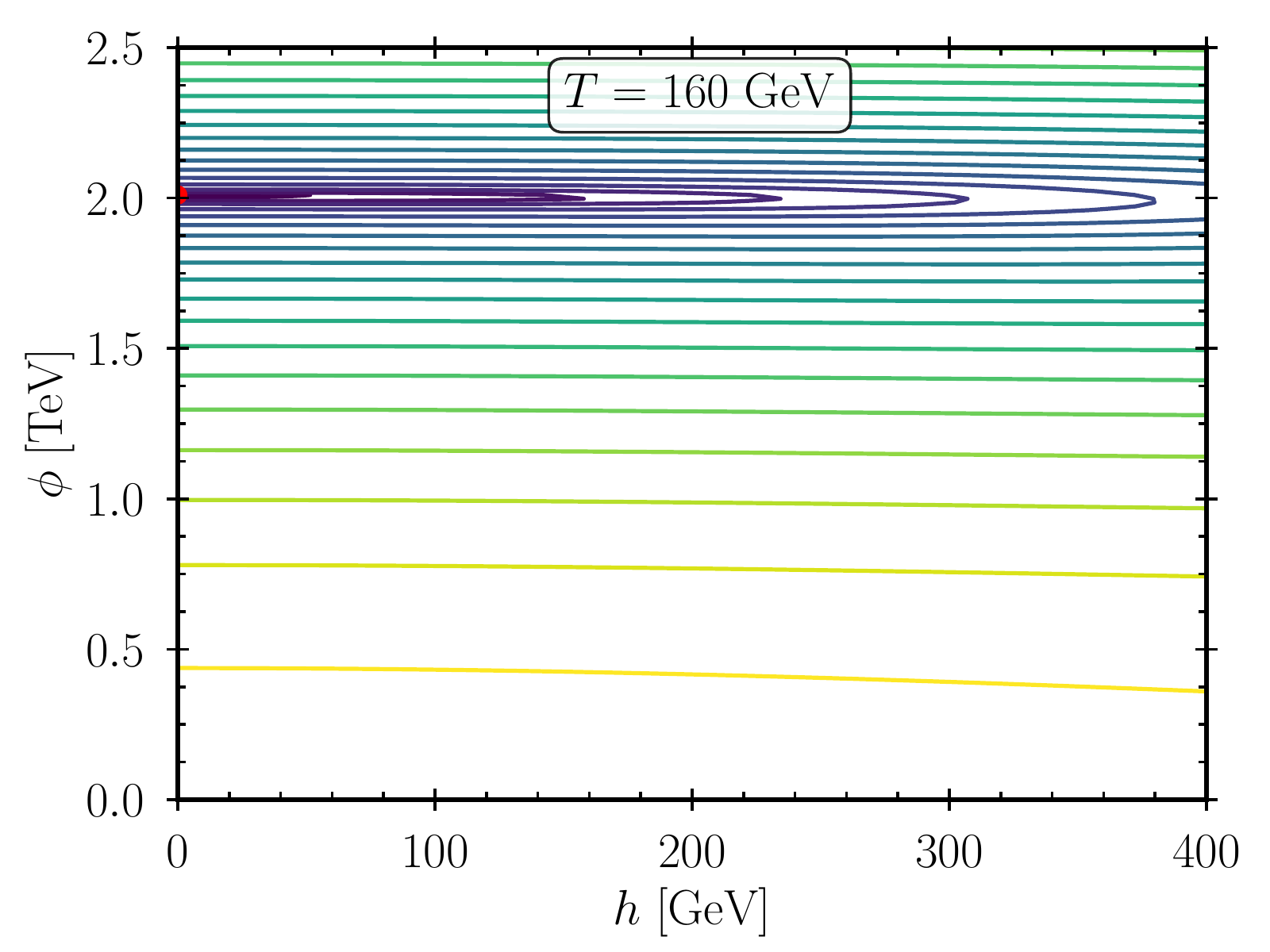}
	
	\includegraphics[width=.48\textwidth]{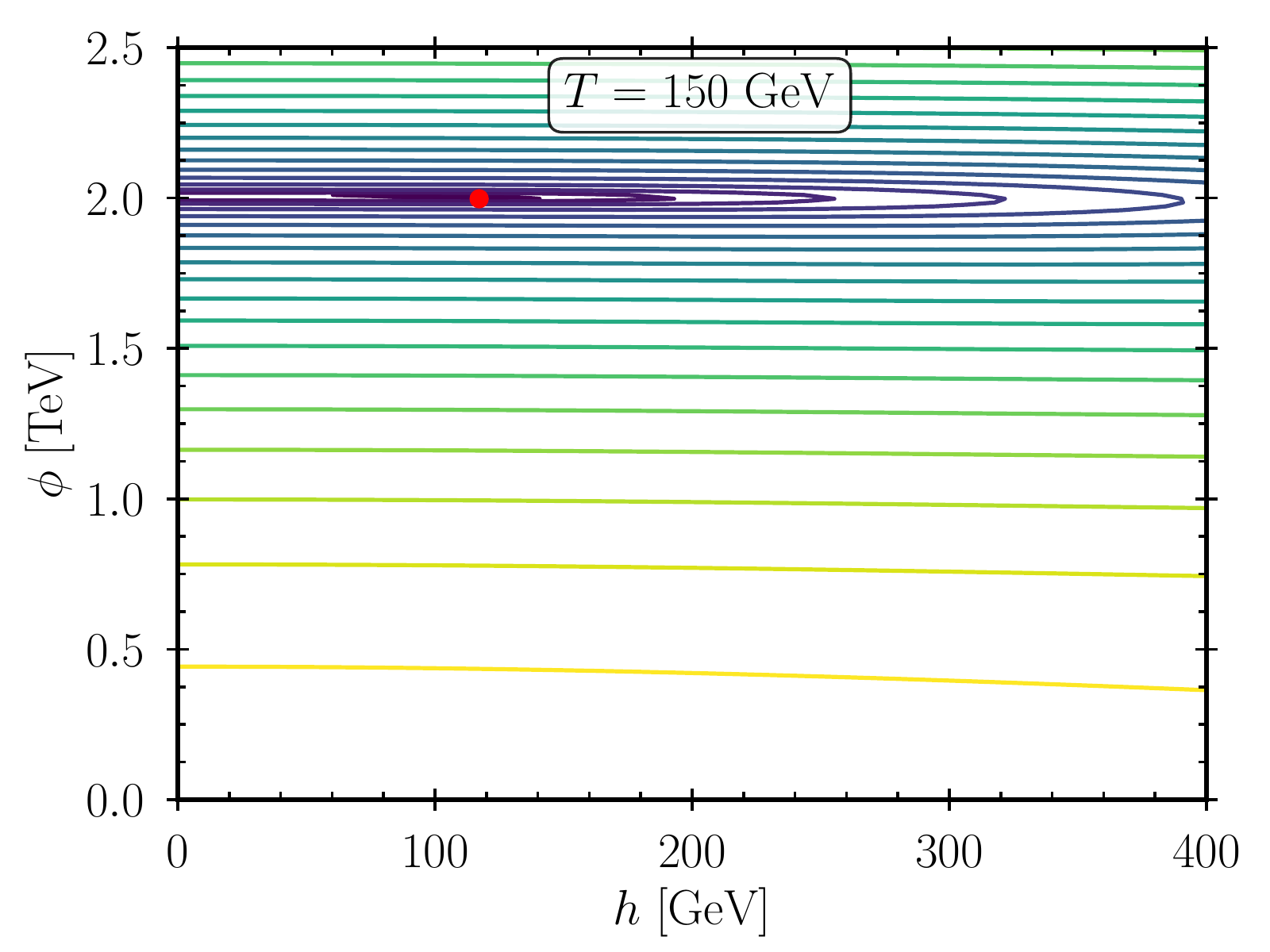}
	\hfill
	\includegraphics[width=.48\textwidth]{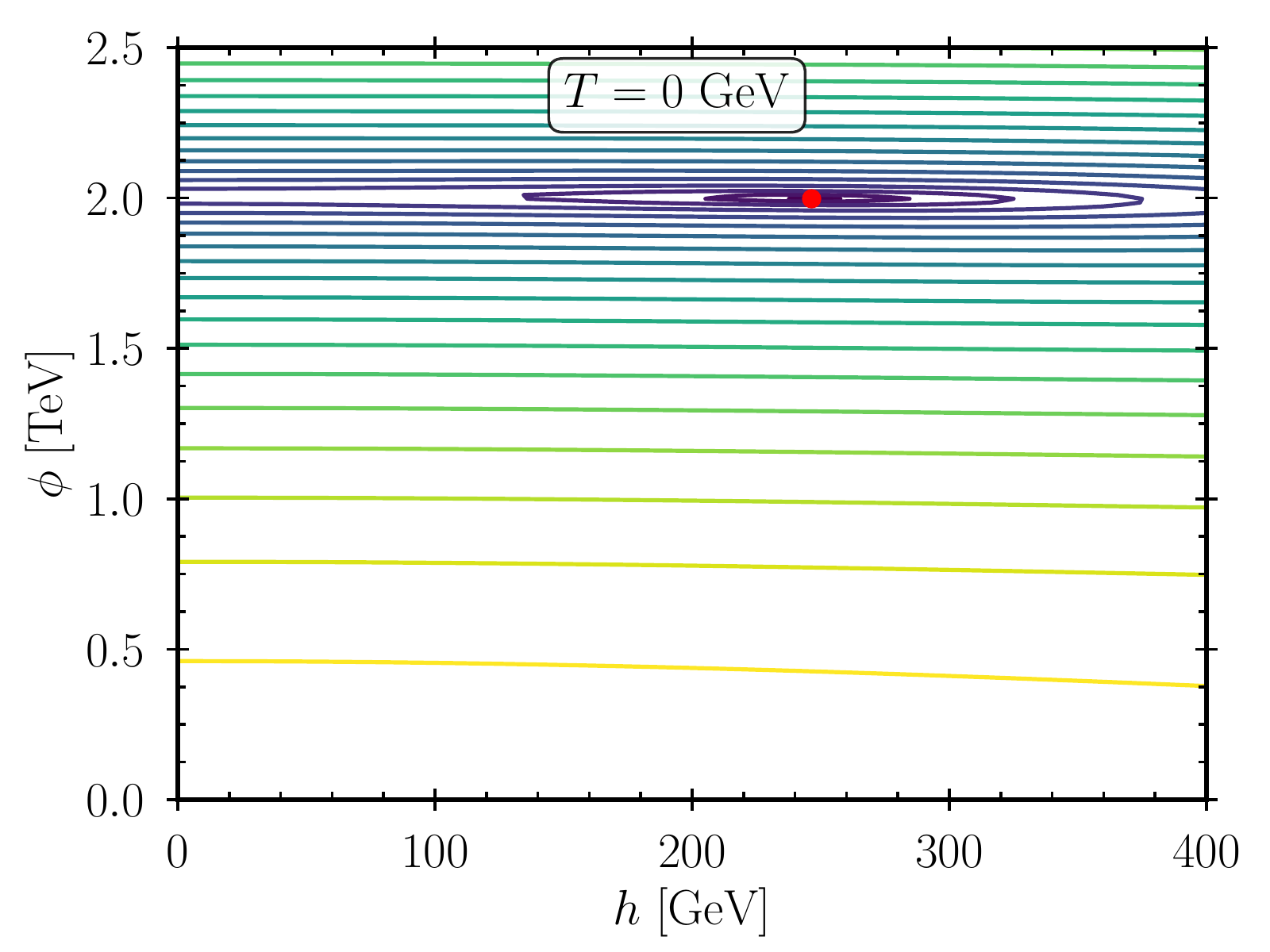}
	
	\caption{
		Effective potential as a function of temperature. 
		The colored equipotential lines correspond to $V_\eff = (30\ \textrm{GeV})^4, (60\ \textrm{GeV})^4, (90\ \textrm{GeV})^4, \ldots, (600\ \textrm{GeV})^4$ (from dark purple to yellow). 
		{\relpenalty=10000 \binoppenalty=10000 
		The red dot denotes the global minimum of the potential at  $V_\eff=0~\textrm{GeV}^4$.
		The model parameters are $v_\Phi = 2$~TeV, $m_\phi = 500$~GeV, $\sin\theta_H = 0.05$, $m_\Zp = 1.5$~TeV, $\epsilon=0$, $m_\DM = 590$~GeV, $\sin\theta_\DM=0$, $m_{e_4} = m_{e_5} = 650$~GeV, and $L^\prime = -1/2$.}
	}
	\label{fig:Veff}
\end{figure}

In figure~\ref{fig:Veff} we show the effective potential at different temperatures, to illustrate the individual steps of the symmetry breaking process. The model parameters are given in the figure caption. 
Counter-terms are set by the condition that the first and second derivatives of the zero-temperature effective potential are the same as for the tree-level potential. 

At high temperatures, the global minimum of the effective potential is in the symmetric (unbroken) vacuum $(h,\phi) = (0,0)$. 
As the Universe cools down, a second minimum starts to form at non-vanishing values of $\phi$. 
At $T_c \simeq 835$~GeV, the two minima are degenerate.
At lower temperatures, the second minimum $(h,\phi) \sim (0,1.1~\mathrm{TeV})$ is the global minimum and breaks lepton number, whereas the electroweak symmetry remains unbroken.
This minimum is separated from the symmetric minimum by a potential barrier.
Thus, to transition to the global minimum, the field has to tunnel (or remain in the symmetric vacuum until the barrier disappears).

As the Universe cools further, the minimum at the origin disappears  and the global minimum moves towards the zero-temperature lepton number breaking VEV $(h,\phi) = (0,2~\mathrm{TeV})$.
Subsequently, at $T\lesssim 160$~GeV, the minimum starts to shift to non-vanishing Higgs field values, breaking the electroweak symmetry in a cross-over transition.
Eventually, the Universe ends up in today's vacuum $(h,\phi) \simeq (246~\mathrm{GeV},2~\mathrm{TeV})$.

\subsection{A first-order lepton number breaking PT}
\label{sec:lepPT}

In this section, we examine the lepton number breaking phase transition in the limit of negligible portal coupling $\lambda_{p}$ between the SM Higgs and the scalar $\Phi$.
Further assuming that the kinetic mixing of the gauge bosons as well as the exotic Yukawa couplings $c_i$ and $y_i$ of the heavy leptons are small, we can study a simplified version of the effective potential in which only the lepton number breaking scalar and the lepton number gauge boson are considered.

In this case, the tree-level potential simplifies to
\be
	V_0(\phi) = - \frac{1}{2} \mu_\Phi^2 \phi^2 + \frac{1}{4} \lambda_\Phi \phi^4 \,.
\ee
Setting the lepton number breaking VEV to $v_\Phi = 2$~TeV, in agreement with the \experiment{LEP} constraint, the model is therefore fully specified by $m_{Z^\prime}$ and $m_\phi$. 

The field dependent masses of the scalar, the gauge boson, and the Goldstone boson are given by
\be
	m_\phi^2 = -\mu_\Phi^2 + 3 \lambda_\Phi \phi^2, \hspace{1eM}
	m_{\omega^0}^2 = -\mu_\Phi^2 + \lambda_\Phi \phi^2, \hspace{1eM}
	\textnormal{and} \hspace{1eM}
	m_{\Zp}^2 = 9 g_\ell^2 \phi^2 .
\ee
The thermal mass corrections are ($\Pi_\phi = \Pi_{\omega^0} = \Pi_\Phi$)
\be
	\Pi_\Phi = \left(\frac{1}{3} \lambda_\Phi + \frac{9}{4} g_\ell^2\right) T^2
	\hspace{1eM}\textnormal{and}\hspace{1eM}
	\Pi_{Z^\prime_L} = 3 g_\ell^2 T^2 + \frac{2}{3} g_\ell^2 T^2 \left(3 + L^{\prime\,2} + L^{\doubleprime\,2}\right) 
	,
	\label{eq:ThermalMasses}
\ee
where the first part of $\Pi_{Z^\prime_L}$ comes from the scalar and the second part from the SM and exotic leptons.
The subscript $L$ of $\Pi_{Z^\prime_L}$ indicates that only the longitudinal part of the \Zp\ boson receives a thermal correction.

We further use an on-shell scheme, imposing the conditions
\be
	\label{eq:VeffSimpleOnShell}
	\frac{\partial(V_\textnormal{CW}+V_\textnormal{c.t.})}{\partial\phi}\bigg|_{\phi=v_\Phi} \hspace{-1eM}= 0 
	\hspace{2eM}\textnormal{and}\hspace{2eM}
	\frac{\partial^2(V_\textnormal{CW}+V_\textnormal{c.t.})}{\partial\phi^2}\bigg|_{\phi=v_\Phi} \hspace{-1eM}= -\Delta\Sigma 
	.
\ee
This ensures that the VEV and the scalar mass at zero temperature remain at their tree-level values.
Here, $\Delta\Sigma\equiv\Sigma(m_\phi^2)-\Sigma(0)$ is the difference of the scalar self-energy evaluated at the tree-level mass and at zero-momentum, see appendix~\ref{sec:self-energy}. 
The second derivative of the Coleman-Weinberg potential in the vacuum suffers from logarithmic divergences originating from the vanishing Goldstone masses. These are IR divergencies and are due to the fact that the effective potential is evaluated at vanishing external momentum. However, the scalar self-energy at zero-momentum suffers from the same divergences, hence its presence in the second condition above~\cite{Delaunay:2007wb}.
The divergences in $\Delta\Sigma$ and $\partial^2 V_\textnormal{CW}/\partial \phi^2$ then cancel, ensuring that we obtain finite counter-terms. 

We use the numerical package \software{CosmoTransitions}~\cite{Wainwright:2011kj} to evaluate the effective potential and to analyze the phase transition.
Fixing the VEV to $v_\Phi = 2$~TeV (and setting the renormalization scale to $\mu_R=v_\Phi$), we identify the region in the $m_\phi-m_{\Zp}$ parameter space at which a first-order phase transition occurs.

In this model, the potential barrier between the vacua is generated by thermal corrections from gauge boson loops (note the cubic term in the high-$T$ expansion of the bosonic thermal integral~\eqref{eq:highTbosons}), i.e.\ the larger the gauge coupling (and hence also the \Zp\ mass)  the higher the barrier.
Increasing the scalar mass on the other hand increases the quartic coupling, which in turn reduces (the relative height of) the barrier.
Thus, first-order phase transitions can be obtained for $m_\Zp \gtrsim m_\phi$; strong transitions occur for $m_\Zp \gtrsim 2 m_\phi$.
The term ``strong'' here refers to transitions in which the VEV (or more precisely the distance between the two degenerate minima in field space) at the critical temperature is larger than the critical temperature itself, i.e.\ $v_\Phi(T_{c})/T_{c} \gtrsim 1$.
This measure is often employed in the context of baryogenesis~\cite{Quiros:1999jp}.

\begin{figure}
	\subfloat[Critical temperature $T_c$]{\includegraphics[width=.49\textwidth]{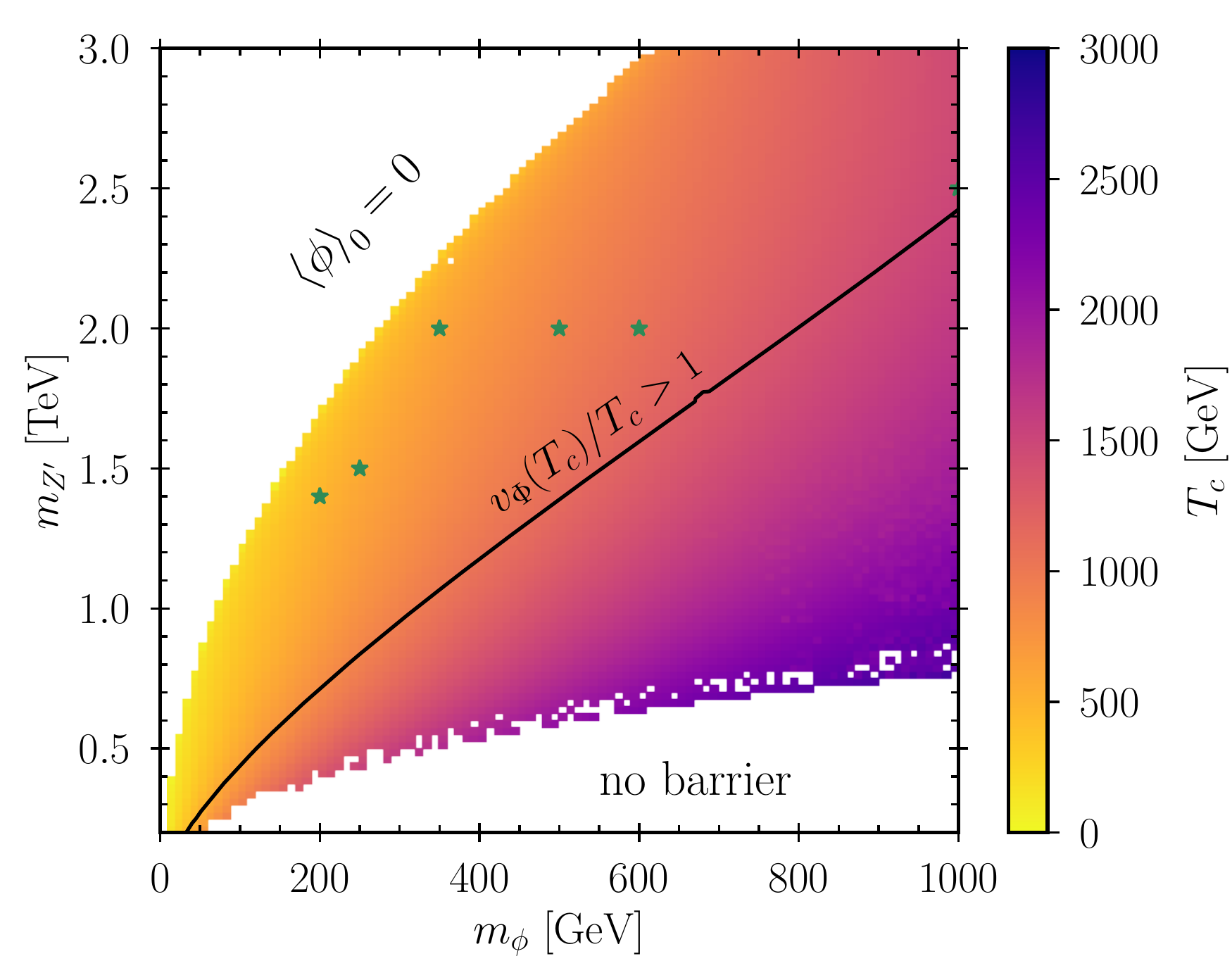}\label{fig:VeffTc}}
	\hfill
	\subfloat[Nucleation temperature $T_n$]{\includegraphics[width=.49\textwidth]{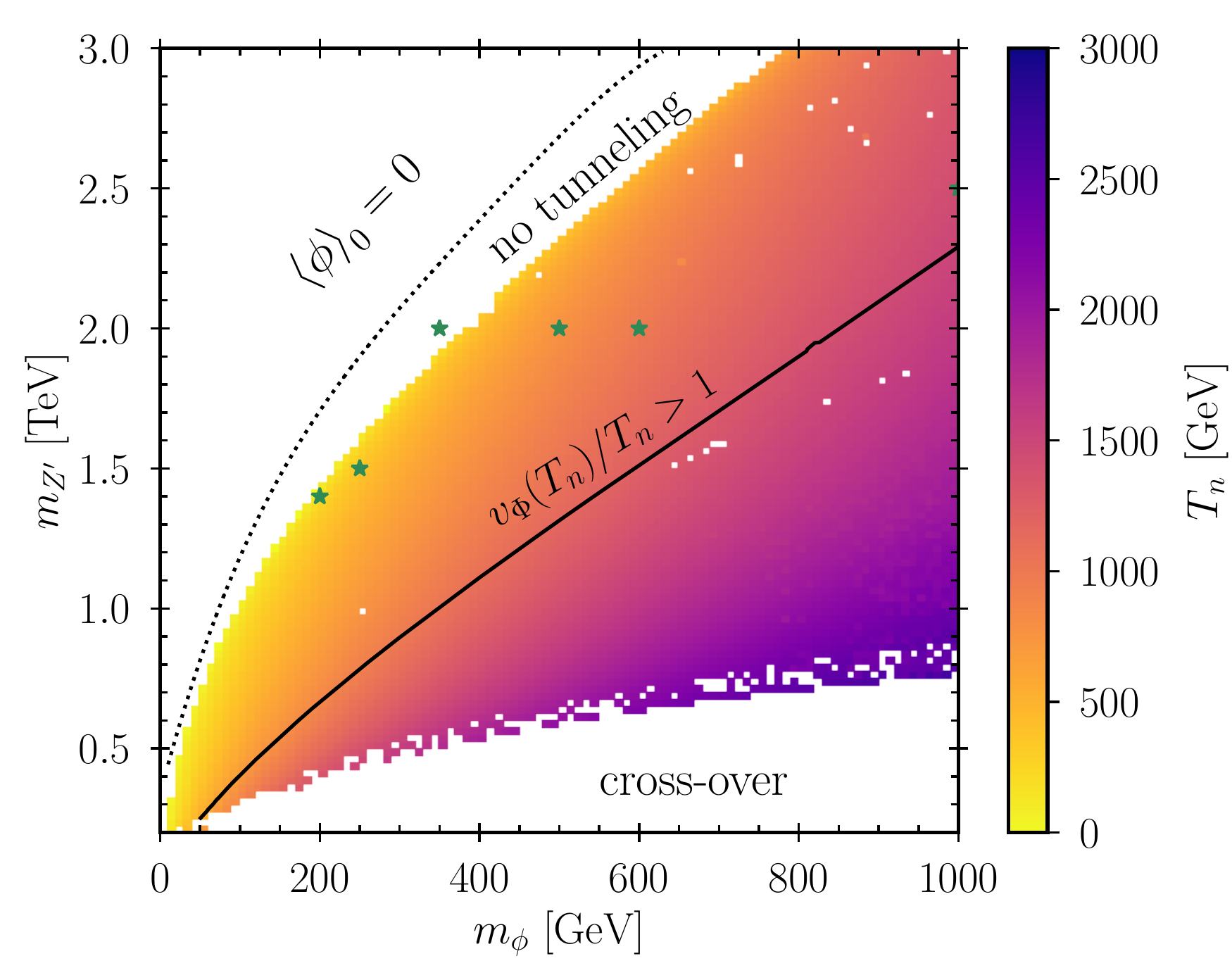}\label{fig:VeffTn}}
	\caption{Parameter points with two phases separated by a potential barrier at a critical temperature $T_c$ (left), and points that give rise to a cosmological first-order phase transition with a nucleation temperature $T_n$ (right). The colored regions above the solid, black line feature strong transitions with $v_\Phi(T)/T> 1$ at $T_c$ or $T_n$, respectively. The dotted line in the right plot denotes the border at which $\phi=0$ becomes a global minimum. The values of $T_c$ and $T_n$ for the benchmark points marked by a star can be found in appendix~\ref{sec:GWbenchmark}.}
	\label{fig:VeffT}
\end{figure}

Figure~\ref{fig:VeffTc} shows the regions in the $m_\phi-m_{\Zp}$ plane in which the effective potential develops degenerate minima at a critical temperature $T_c$ for $L^\prime = -1/2$.
The colors indicate the corresponding $T_c$.
In the colored region above the black line, the measure $v_\Phi(T_c)/T_c$ implies strong transitions.
The parameter points which actually lead to a first-order phase transition through bubble nucleation are shown in figure~\ref{fig:VeffTn} along with the corresponding nucleation temperature\footnote{Note that isolated, white dots in the colored regions do not indicate that particular parameter points do not feature a first-order PT, but are due to numerical artifacts.}, again for $L^\prime = -1/2$.
Here, the black line indicates  $v_\Phi(T_n)/T_n = 1$ evaluated at the nucleation temperature.

Although the renormalization conditions~\eqref{eq:VeffSimpleOnShell} ensure that the zero-temperature potential has a minimum at $\phi = v_\Phi$, this minimum is not necessarily the global minimum.
In particular, if the gauge boson mass $m_{\Zp}$ is much bigger than the scalar mass $m_\phi$, the potential develops a global zero-temperature minimum at $\phi = 0$, i.e.\ the Coleman-Weinberg corrections restore the symmetry already at $T=0$.\footnote{%
	Of course, the physical scalar and \Zp\ masses become $m_\phi = 0$ and $m_\Zp = 0$ in this region. Hence, the $x$ and $y$ axes should be interpreted as $\sqrt{2 \lambda_\Phi v_\Phi^2}$ and $3 g_\ell v_\Phi$ respectively, where $v_\Phi = 2$~TeV has no physical meaning.
}
This is the case in the white area labeled by ``$\left<\phi\right>_0=0$'' above the colored region in figure~\ref{fig:VeffTc} (and above the dotted line in figure~\ref{fig:VeffTn}), which is of course excluded since it would imply the existence of a second massless gauge boson with significant couplings to leptons. 
Furthermore, even a global minimum at $\phi=v_\Phi$ does not automatically ensure that today's Universe has transitioned to the true vacuum. 
If the barrier is very large with a small potential difference between the two vacua, which is the case close to the region in which the potential has a global zero-temperature minimum at $\phi=0$, the tunneling probability is too low. 
Therefore the field is stuck in the false vacuum and does not tunnel.
This corresponds to the parameter region labeled ``no tunneling'' in figure~\ref{fig:VeffTn}.\footnote{Note that \software{CosmoTransitions} only evaluates the thermal tunneling probability. Quantum tunneling is not taken into account.}

On the other hand, for $m_\Zp \lesssim m_\phi$ no significant barrier is induced and there is no temperature at which the potential has degenerate minima.
Also, if the potential barrier separating the phases is very shallow, it might disappear before bubbles are nucleated. 
In both cases the transition occurs without tunneling as a cross-over\footnote{We here rely on the ability of \software{CosmoTransitions} to identify cross-over transitions. A proper determination of whether a transition is cross-over may involve non-perturbative calculations and is beyond the scope of this paper. Here, we are mainly interested in the region where strong first-order transitions occur.} and no gravitational waves are generated. 
This happens in the areas labeled ``no barrier'' or ``cross-over'' in figure~\ref{fig:VeffT}.

So far, to simplify the parameter space to two dimensions we neglected the contributions from the dark Yukawa couplings of the fourth and fifth generation leptons to the effective potential.
However, if the leptons are heavy, the Yukawa couplings are large and the potential can be modified significantly. This is in particular the case for large \Zp\ masses, where the exotic leptons are required to be heavy in order to obtain the correct DM relic abundance. 

For simplicity, we here again assume that the exotic electrons and the exotic neutrino have equal masses, $m_\textnormal{HL} \equiv m_{e_4} = m_{e_5} = m_{\nu_4}$, i.e.\ that the SM Higgs Yukawa couplings $y_\nu$ and $y_e$ in Equation~\eqref{eq:yukawa} vanish.
The field dependent masses are then given by
\be
	m_\DM = \frac{c_\nu}{\sqrt{2}} \phi\ , \hspace{2eM} m_\HL = \frac{c_\ell}{\sqrt{2}} \phi\ .
\ee
The Yukawa couplings further contribute to the scalar thermal mass correction~\eqref{eq:ThermalMasses}, which becomes
\be 
	\Pi_\Phi = \left(\frac{1}{3} \lambda_\Phi + \frac{9}{4} g_\ell^2 + \frac{1}{12} c_\nu^2 + \frac{1}{4} c_\ell^2\right) T^2\ .
\ee

\begin{figure}
	\centering
	\subfloat[][$m_\DM = 200$~GeV, $m_\HL=210$~GeV]{\includegraphics[width=.49\textwidth]{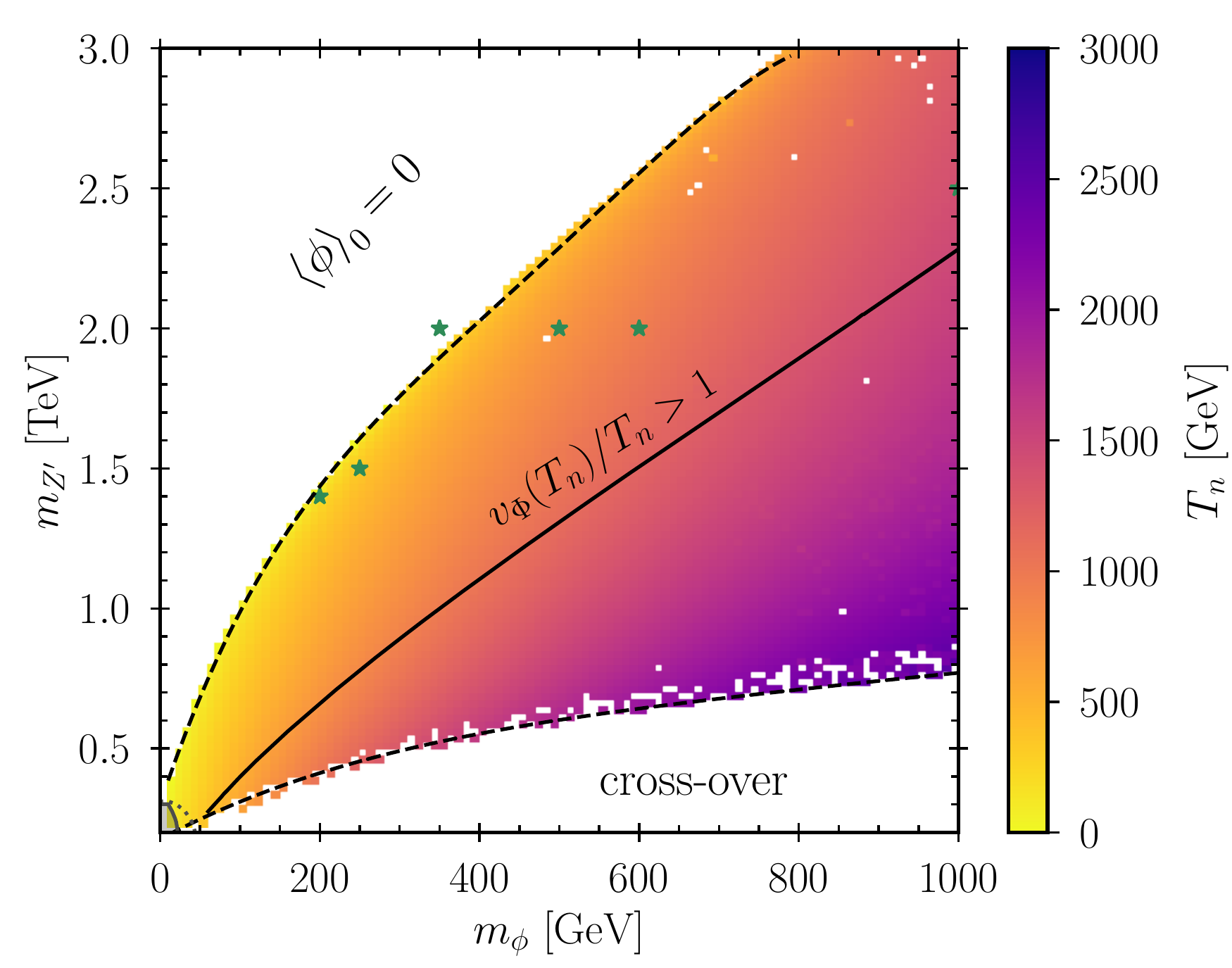}\label{fig:VeffTnFermionsLight}}
	\hfill
	\subfloat[][$m_\DM = 500$~GeV, $m_\HL=1$~TeV]{\includegraphics[width=.49\textwidth]{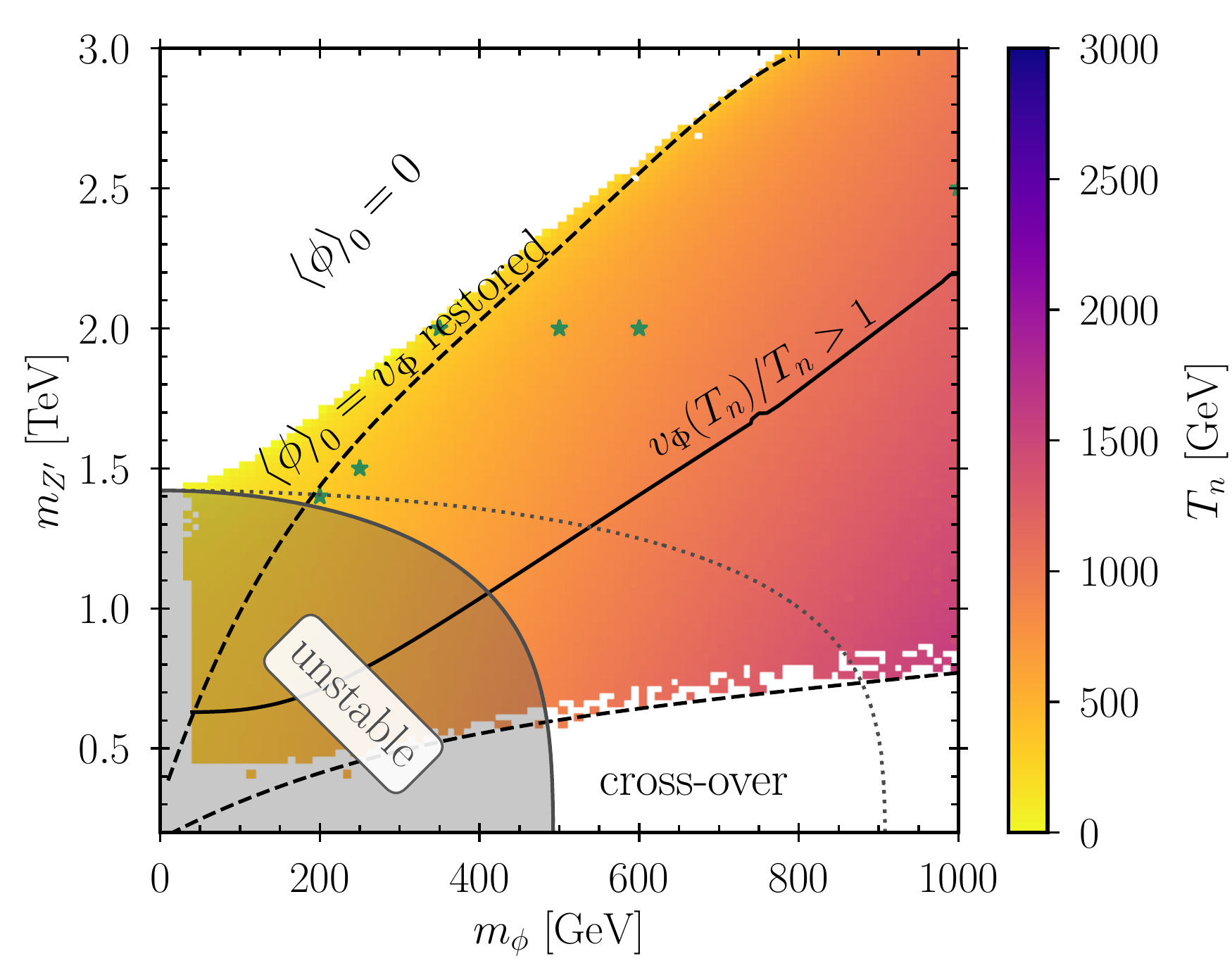}\label{fig:VeffTnFermionsHeavy}}
	\caption{Parameter points giving rise to a cosmological first-order phase transition with a nucleation temperature $T_n$, including the contribution from dark matter and the exotic leptons. The dashed lines indicate the corresponding region neglecting the fermion contributions (cf.\ figure~\ref{fig:VeffTn}). In the gray shaded region, the potential becomes unstable below $\phi=100$~GeV; above the gray, dotted line it is stable up to $\phi = 10^6$~TeV. The stars indicate benchmark points listed in appendix~\ref{sec:GWbenchmark}.}\label{fig:VeffTnFermions}
\end{figure}

Figure~\ref{fig:VeffTnFermions} shows the values of the scalar and $\Zp$ masses that lead to a first-order phase transition with the corresponding nucleation temperature, for two different choices of the DM and heavy lepton masses.
The region that gives rise to a first-order transition for vanishing fermion couplings (cf.\ figure~\ref{fig:VeffTn}) is indicated by the dashed lines. 

As expected, for light dark leptons (low Yukawa couplings) the situation changes only marginally with respect to the case assuming vanishing Yukawas.
However, for higher fermion masses, the region that yields a first-order PT changes, and the nucleation temperature decreases.

In the parameter region labeled ``$\left<\phi\right>_0 = v_\Phi$ restored'' in figure~\ref{fig:VeffTnFermionsHeavy}, the bosonic loop corrections to the zero-temperature potential induce a global minimum at $\phi=0$ in the absence of fermions.
If the dark sector leptons are included, their contributions have the opposite sign and partially cancel the bosonic ones, and the global minimum at $\phi=v_\Phi$ is restored.
Hence, the region allowing for a first-order PT is extended.
On the other hand, if the fermionic corrections overcome the bosonic ones at high field values, the potential is destabilized as it is not bounded from below.
This occurs for low \Zp\ and $\phi$ masses.
The gray shaded regions are excluded since the potential becomes unstable at field values below $\phi = 100$~TeV, i.e. $V_\eff(100~\textnormal{TeV}) < V_\eff(\left<\phi\right>_0)$ at $T=0$.
Above the gray, dotted curve the potential is stable even up to $\phi = 10^9$~GeV. 
Note however that a reliable evaluation of the potential at such high field values requires the inclusion of renormalization group effects.

At high temperatures, the loop corrections from the fermions give a positive contribution $\sim \phi^2 T^2$, whereas they do not contribute to the cubic terms (note that there is no $y^3$ term in~\eqref{eq:highTfermions}).
As a consequence, the finite-temperature corrections restore the symmetric minimum at lower temperatures, reducing the nucleation temperature.

\begin{figure}
	\centering
	\includegraphics[width=.5\textwidth]{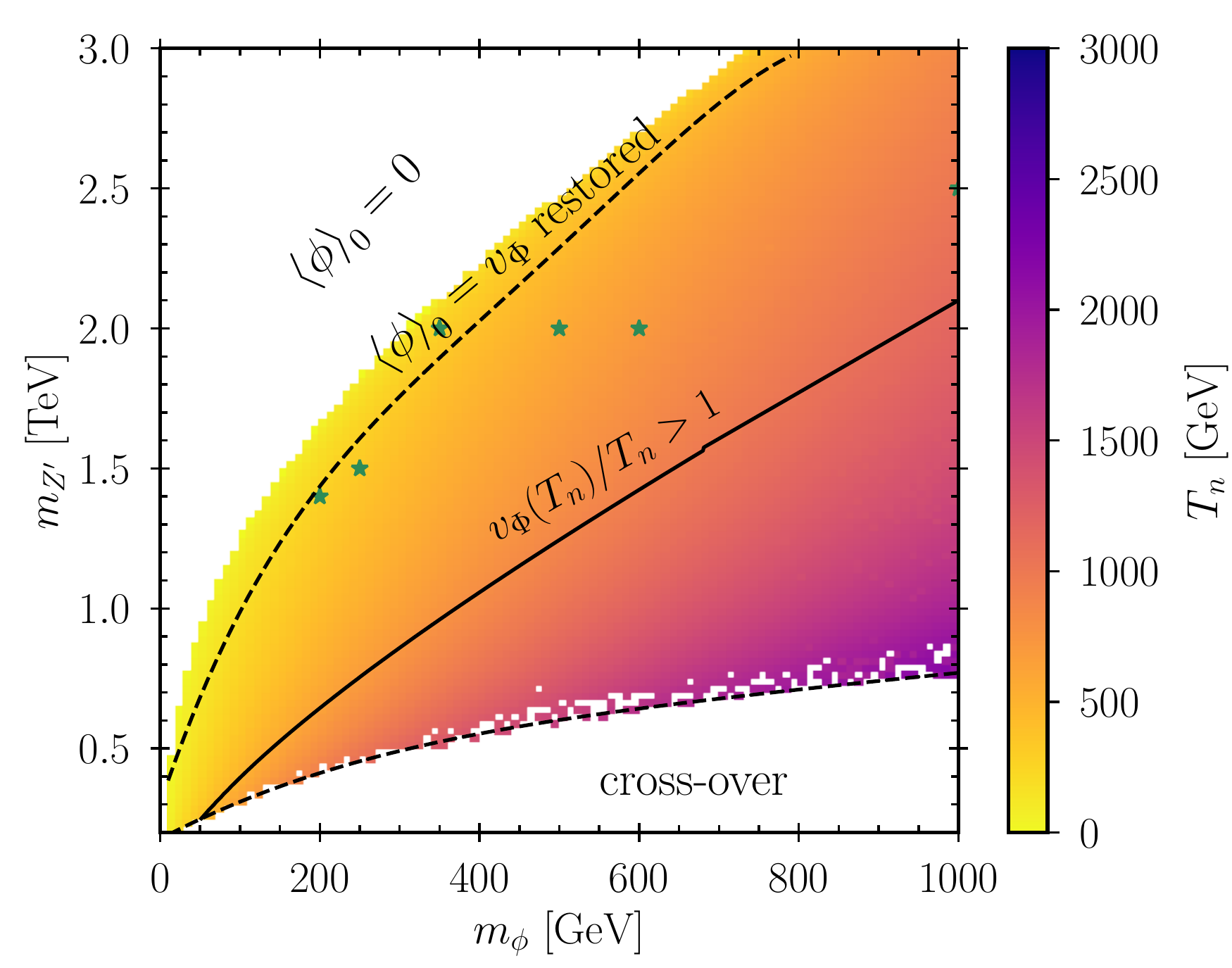}
	\caption{Same as figure \ref{fig:VeffTnFermions}, but at each parameter point the DM mass is set to a value that yields the correct relic abundance. The masses of $e_4$, $e_5$, and $\nu_4$ are set to $m_\HL = 1.5 \times m_\DM$.}
	\label{fig:VeffTnDM}
\end{figure}

Finally, to properly connect to the DM picture let us require that the DM candidate has the correct thermal abundance.\footnote{Note that the DM in figure~\ref{fig:VeffTnFermionsLight} already has the measured abundance by co-annihilation for most values of $m_\Zp$, cf.\ the green line in figure~\ref{fig:RDCA}.}
Figure~\ref{fig:VeffTnDM} shows the nucleation temperature for the corresponding phase transitions, assuming that $m_\HL = 1.5 \times m_\DM$.
At each parameter point in the $m_\phi-m_\Zp$ plane we use \software{micrOMEGAs} to find the value of the DM mass that yields the measured abundance, picking the value below the \Zp\ resonance (i.e.\ we are sitting on the upper branch of the blue line in figure~\ref{fig:DMRelicDensity}).
Again, the dashed lines indicate the parameter region that provides a first-order PT if the fermions are neglected.

As the DM mass required to obtain the correct abundance increases with the \Zp\ mass and is mostly independent of the scalar mass, the effects of including the vector-like leptons are stronger for larger \Zp\ masses.
Hence, the fermionic corrections restore the $T=0$ minimum at $\phi=v_\Phi$ for high $m_\Zp$, whereas this effect is absent in the low $m_\Zp$ range. 
Furthermore, since $m_\DM < m_\Zp/2$ (and $m_\HL = 1.5 \times m_\DM$), the bosonic contributions are sufficiently large to circumvent the destabilizing effects of the fermionic corrections in the full parameter space shown in figure~\ref{fig:VeffTnDM}.

\section{Gravitational waves}
\label{sec:gws}
%

A first-order phase transition proceeds through thermal fluctuations or quantum tunneling by the nucleation of bubbles of the broken phase in the sea of the unbroken phase.
The bubbles then expand and collide, producing a stochastic background of gravitational waves.
If the phase transition is sufficiently strong first-order, these gravitational waves may be detectable by future observatories such as \experiment{LISA}~\cite{Caprini:2015zlo,Audley:2017drz}.

In the following we will briefly review the relevant parameters and the calculation of the GW spectrum, and explain how we evaluate the detectability of the resulting spectra.
See~\cite{Caprini:2015zlo,Leitao:2015fmj,Weir:2017wfa,Caprini:2018mtu} for some recent, more comprehensive reviews.
The corresponding results are presented in section~\ref{sec:GWresults}.

\subsection{Transition parameters} 
\label{sec:PTparams}

For a phase transitions occurring at a temperature $T_\ast$, the GW spectrum depends on four parameters. The most important ones are the transition temperature $T_\ast$ which we approximate by the nucleation temperature $T_n$, the inverse time duration $\beta$ of the transition divided by the Hubble rate
\be
  \frac{\beta}{H_\ast} = T_\ast \frac{\dInt S_E}{\dInt T}\bigg|_{T_\ast},
\ee
and the ratio of the latent heat $\mathcal{L}(T_\ast)$ released by the transition (i.e.\ the difference of the energy densities in the two vacua) to the radiation density
\be 
  \alpha = \frac{\mathcal{L}(T_\ast)}{g_\ast \pi^2 T_\ast^4 / 30},
\ee
where $g_\ast$ is the effective number of relativistic degrees of freedom.

The spectrum further depends on the bubble wall velocity $v_w$.
For phase transitions occurring in a plasma (in contrast to those in a vacuum dominated epoch) the expanding bubbles are exposed to a friction force exerted by the particles of the plasma.
If the friction is sufficient to compensate the driving force from the energy liberated by tunneling, the bubble walls reach a terminal velocity, otherwise they accelerate perpetually with $v_w$ approaching the speed of light.
This latter case is referred to as run-away~\cite{Bodeker:2009qy}.
In this case, the energy transferred to the plasma saturates, and the surplus energy accelerates the scalar bubbles, giving rise to a non-negligible GW spectrum from bubble collisions (cf.~section~\ref{sec:GWspec}).
It has however been argued that the presence of vector bosons that gain a mass in the PT produces friction from transition radiation that prevents the bubble from such a behavior~\cite{Bodeker:2017cim}.
Still, bubble velocities of $v_w \sim1 $ can be reached.

As the detectability of the generated stochastic background eventually only mildly depends on the wall velocity, we simply take $v_w = 1$ in the following.
See appendix~\ref{sec:vW} for a brief study of the impact of the wall velocity on the detectability. 

\subsection{GW spectrum} 
\label{sec:GWspec}

Cosmological first-order phase transitions can produce gravitational waves mainly via three mechanisms:
the collision of the bubble walls themselves, acoustic production from sound shells surrounding the bubble walls, and turbulence effects in the plasma.
The corresponding GW spectra can be simulated numerically or estimated analytically~\cite{Huber:2008hg,Hindmarsh:2015qta,Caprini:2009yp}, and are then expressed in terms of quantities that can be derived from the effective potential by fitting to the corresponding results.
The full spectrum is given by
\be
\OMEGA{GW}(f) \simeq \OMEGA{$\phi$}(f) + \OMEGA{sw}(f) + \OMEGA{turb}(f) ,
\ee
where $\Omega_\textnormal{GW}(f) = \frac{1}{\rho_c} \frac{\dInt \rho_\mathrm{GW}}{\dInt \log f}$ is the spectral GW energy density normalized to the critical energy density $\rho_c = 3 H^2/(8\pi G)$, and $H = h \times 100\,\mathrm{km}\,\mathrm{Mpc}^{-1}\,\mathrm{s}^{-1}$ is the Hubble rate.

The first contribution $\OMEGA{$\phi$}(f)$ to the GW spectrum comes from the scalar bubbles nucleated during the transition.
Due to the spherical symmetry of the bubbles, the potential energy that is exempted when the field tunnels cannot be released directly into gravitational waves. 
It can thus only drive the expansion of the bubbles.
However, when bubbles collide, their spherical symmetry is broken and gravitational waves are generated.
The corresponding spectrum can be computed using the envelope approximation~\cite{Kosowsky:1992vn}. 
In this approximation, the GWs are sourced from the fraction $\kappa_\phi$ of the latent heat stored in thin shells around the bubble walls, taking into account only the uncollided part of the shells, i.e.\ the envelope of the collided bubbles.

Based on numerical simulation, the corresponding GW spectrum is given by~\cite{Huber:2008hg}
\be
  \OMEGA{$\phi$}(f) = 1.67 \times 10^{-5} \left(\frac{H_\ast}{\beta}\right)^2 \left(\frac{\kappa_\phi \alpha}{1+\alpha}\right)^2 \left(\frac{100}{g_\ast}\right)^{\frac{1}{3}} \left(\frac{0.11 v_w^3}{0.42 + v_w^2}\right) S_{\phi}(f) 
\ee
with
\be
  S_\phi(f) = \frac{3.8 \left(f/f_\phi\right)^{2.8}}{1+2.8 \left(f/f_\phi\right)^{3.8}} ,
\ee
where $f_\phi$ is the peak frequency red-shifted to today
\be
  f_\phi = 16.5\,\mu\textnormal{Hz} \left(\frac{0.62}{1.8 - 0.1 v_w + v_w^2}\right) \left(\frac{\beta}{H_\ast}\right) \left(\frac{T_\ast}{100\,\textnormal{GeV}}\right) \left(\frac{g_\ast}{100}\right)^\frac{1}{6} ,
\ee
and $\kappa_\phi = \rho_\phi/\rho_{\textnormal{vac}}$ is the efficiency factor for the conversion of the vacuum energy density into scalar gradient energy.
The scalar contribution can typically be neglected, unless the transition proceeds in the run-away regime.
Here, due to the friction exerted by the lepton number gauge boson the bubbles do not run away~\cite{Bodeker:2017cim}.
We thus set $\kappa_\phi=0$ and neglect the scalar contribution in the following. 

The second source comes from acoustic production.
Since the scalar bubbles expand in the primordial plasma of the early Universe, their expansion excites the particles in the plasma and induces sound waves.
If the wall velocity is less than the speed of sound in the fluid, $c_s = 1/\sqrt{3}$, the sound shell forms a deflagration in front of the wall, whereas for supersonic bubble walls, the shell becomes a detonation behind the wall.
Similar to the scalar bubbles, the sound shells radiate gravitational waves upon collision.
However, in contrast to the former case, it is not sufficient to consider only the overlap of collided shells.
The collided parts of the shells do not disappear, but pass through one another, giving rise to a long-lasting (compared to the scalar bubble collisions) source of GW with a lifetime on the order of a Hubble time~\cite{Hindmarsh:2015qta}.
As a consequence, the sound-wave contribution typically dominates the GW spectrum.

Again, the spectrum is obtained by fits to numerical simulations, giving~\cite{Hindmarsh:2015qta}
\be
  \label{eq:GWSW}
  \OMEGA{sw}(f) = 2.65 \times 10^{-6} \left(\frac{H_\ast}{\beta}\right) \left(\frac{\kappa_v \alpha}{1+\alpha}\right)^2 \left(\frac{100}{g_\ast}\right)^{\frac{1}{3}} v_w S_{\textnormal{sw}}(f) 
\ee
with the spectral shape
\be
  S_{\textnormal{sw}}(f) = \left(f/f_{\textnormal{sw}}\right)^3 \left(\frac{7}{4+3\left(f/f_{\textnormal{sw}}\right)^2}\right)^{3.5} 
\ee
and
\be
  f_{\textnormal{sw}} = 19\,\mu\textnormal{Hz} \frac{1}{v_w} \left(\frac{\beta}{H_\ast}\right) \left(\frac{T_\ast}{100\,\textnormal{GeV}}\right) \left(\frac{g_\ast}{100}\right)^\frac{1}{6} .
\ee
The efficiency factors for the conversion of the vacuum energy density into bulk motion for relativistic bubbles in the non-runaway case is given by~\cite{Espinosa:2010hh}
\be
\kappa_v \simeq \alpha \left( 0.73 + 0.083 \sqrt{\alpha} + \alpha \right)^{-1}  .
\ee

Last but not least, GWs are also generated by turbulences.
The expanding bubbles can induce eddies in the fluid. 
Since the plasma is ionized, these give rise to magnetohydrodynamic (MHD) turbulences.
However, simulations indicate that turbulence effects are negligible compared to the sound waves~\cite{Hindmarsh:2015qta}.

Assuming Kolmogorov-type turbulences, the turbulently generated spectrum has the form~\cite{Caprini:2009yp,Binetruy:2012ze}
\be
  \label{eq:GWturb}
  \OMEGA{turb}(f) = 3.35 \times 10^{-4} \left(\frac{H_\ast}{\beta}\right) \left(\frac{\kappa_{\textnormal{turb}} \alpha}{1+\alpha}\right)^{\frac{3}{2}} \left(\frac{100}{g_\ast}\right)^{\frac{1}{3}} v_w S_{\textnormal{turb}}(f) ,
\ee
where
\be
  S_{\textnormal{turb}}(f) = \frac{\left(f/f_{\textnormal{turb}}\right)^3}{\left[1+\left(f/f_{\textnormal{turb}}\right)\right]^{11/3} ( 1+ 8\pi f/h_\ast )}
\ee 
with the red-shifted Hubble rate at GW generation
\be
  h^\ast = 16.5\,\mu\textnormal{Hz} \left(\frac{T_\ast}{100\,\textnormal{GeV}}\right) \left(\frac{g_\ast}{100}\right)^{\frac{1}{6}}
\ee
and 
\be
  f_\textnormal{turb} = 27\,\mu\textnormal{Hz} \frac{1}{v_w} \left(\frac{\beta}{H_\ast}\right) \left(\frac{T_\ast}{100\,\textnormal{GeV}}\right) \left(\frac{g_\ast}{100}\right)^\frac{1}{6} .
\ee
For the efficiency factor for the conversion into MHD turbulence, simulations suggest $\kappa_{\textnormal{turb}} = \epsilon \kappa_v$ with $\epsilon \sim 5 - 10 \%$. We here follow~\cite{Caprini:2015zlo} and take $\epsilon = 0.05$.

\begin{figure}
	\begin{minipage}[c]{.7\textwidth}
		\centering
		\includegraphics[width=\textwidth]{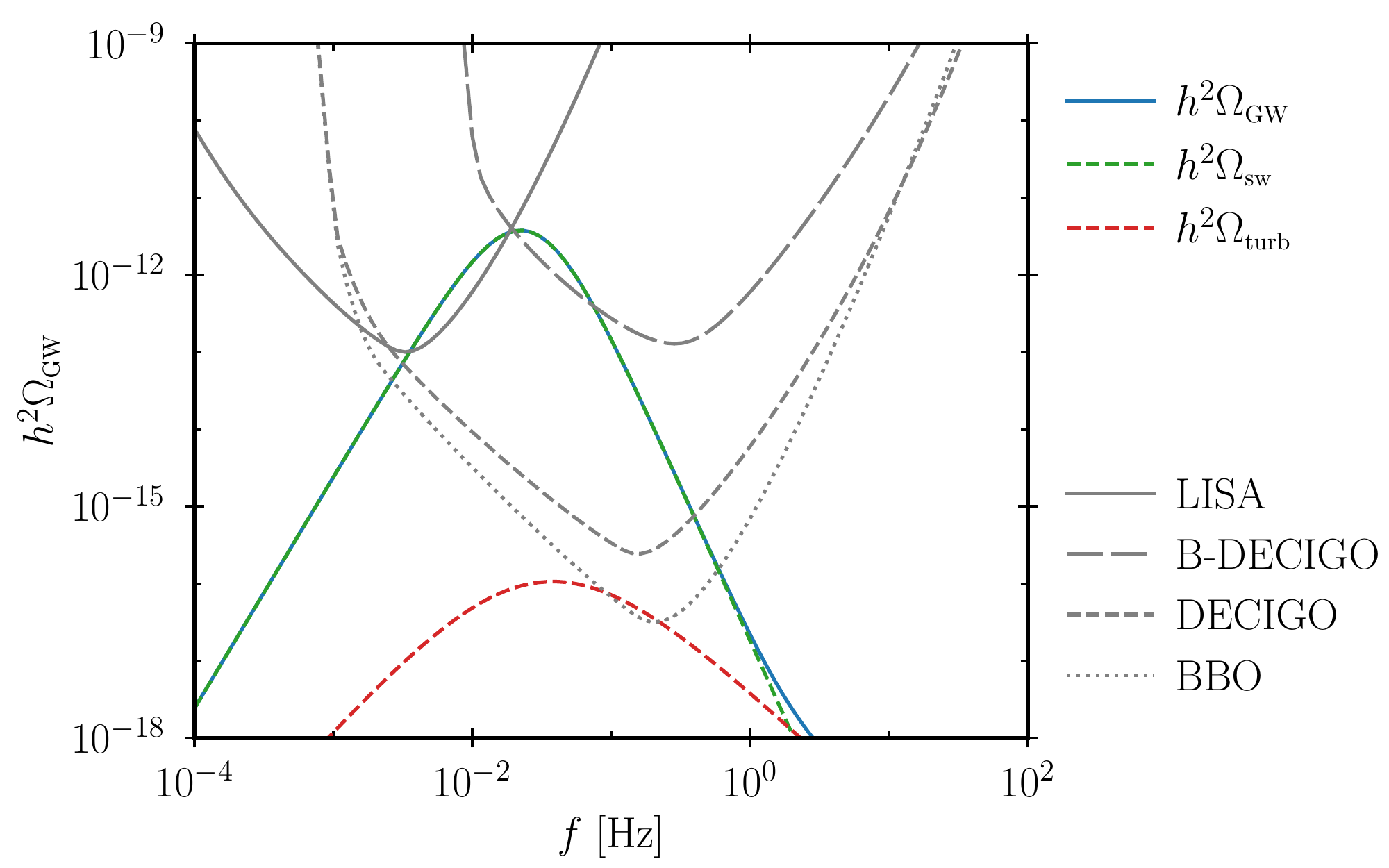}
	\end{minipage}
	\hfill
	\begin{minipage}[c]{.28\textwidth}
		\centering
		\renewcommand{\arraystretch}{1.3}
		\begin{tabular}{c c}
			$m_\phi$ & 200~GeV \\
			$m_\Zp$ & 1.4~TeV \\
			$v_\Phi$ & 2~TeV \\
			$L^\prime$ & $-\frac{1}{2}$ \\
			\hline\hline
			$T_c$ & 487~GeV \\
			$T_n$ & 198~GeV \\
			$\alpha$ & 0.18 \\
			$\beta/H$ & 570
		\end{tabular}
	\end{minipage}
	\caption{GW spectrum (blue line) from the $U(1)_\ell$ breaking phase transition for $m_\phi = 200$~GeV and $m_{Z^\prime} = 1.4$~TeV. The contributions from different production mechanisms are indicated by the dashed lines. The gray lines indicate the (power-law integrated) sensitvity of \experiment{LISA} (solid), \experiment{B-DECIGO} (long dashed), \experiment{DECIGO} (dashed), and \experiment{BBO} (dotted).}\label{fig:GWspectrum}
\end{figure}

An example of the GW spectrum generated by the lepton number breaking phase transition for a scalar mass of $m_\phi = 200$~GeV and a gauge boson mass of $m_{Z^\prime} = 1.4$~TeV (with $v_\Phi=2$~TeV and $L^\prime = -1/2$, neglecting the heavy lepton contributions) is shown in figure~\ref{fig:GWspectrum} (blue curve).
The contributions from acoustic production (green) and MHD turbulence (red) are indicated by dashed lines.
For this choice of parameters, the transition occurs at a nucleation temperature of $T_n \sim 200$~GeV with a peak frequency of the spectrum of $f \sim22$~mHz.
The spectra for further benchmark points and the corresponding values of $T_c$, $T_n$, $\alpha$, and $\beta/H$ can be found in appendix~\ref{sec:GWbenchmark}.

\subsection{Detectability}
\label{sec:GWdet}

The stochastic GW background generated by a first-order PT can in principle be detected by GW observatories.
Whereas current earth-based experiments such as LIGO and Virgo are not sufficiently sensitive to constrain first-order PTs in the frequency range we are interested in (due to seismic noise),
space-based interferometers provide promising prospects for the detection of a stochastic background.
We therefore evaluate detection prospects for \experiment{LISA}~\cite{Audley:2017drz}, \experiment{B-DECIGO}~\cite{Isoyama:2018rjb}, \experiment{DECIGO}~\cite{Seto:2001qf} and \experiment{BBO}~\cite{Crowder:2005nr}.

The \experiment{Laser Interferometer Space Antenna (LISA)}~\cite{Audley:2017drz} is an upcoming space-based GW interferometer consisting of six laser links on three satellites arranged in a regular triangle with an arm length of $2.5\times 10^9$~m and a mission life-time of 4 years.
Its launch is scheduled for 2034~\cite{Caprini:2018mtu}.
The \experiment{Big Bang Observer (BBO)}~\cite{Crowder:2005nr} is a proposed follow-up experiment consisting of four \experiment{LISA}-like detectors, again arranged in a regular triangle. 
The fourth detector is co-located with one of the other detectors, forming a star-of-David configuration.
A similar configuration has been proposed for the Japanese \experiment{Deci-Hertz Interferometer Gravitational Wave Observatory (DECIGO)}~\cite{Seto:2001qf}, which is planned to be launched in the 2030's~\cite{Sato:2017dkf}.
Its scaled-down version \experiment{B-DECIGO (Basic/Base-DECIGO)}~\cite{Isoyama:2018rjb} supposedly starts in the late 2020's~\cite{Sato:2017dkf}.

The detectability of a stochastic background by a given experiment can be evaluated by calculating the signal-to-noise ratio (SNR) $\rho$.
If the SNR is greater than a threshold value $\rho_\textnormal{thr}$, we consider the background as detectable.

The SNR can be calculated from the GW spectrum and the experiment's noise spectrum.
For a single detector such as \experiment{LISA} and \experiment{B-DECIGO}, it is given by~\cite{Thrane:2013oya,Caprini:2015zlo}
\be
	\rho = \sqrt{T \int\limits_{f_\textnormal{min}}^{f_\textnormal{max}} \dInt f\ \left[\frac{\OMEGA{GW}(f)}{\OMEGA{n}(f)}\right]^2}\ , 
\ee
whereas for a network of detectors like \experiment{BBO} or \experiment{DECIGO} it reads~\cite{Thrane:2013oya,Caprini:2018mtu}
\be
\rho = \sqrt{2 T \int\limits_{f_\textnormal{min}}^{f_\textnormal{max}} \dInt f\ \left[\frac{\OMEGA{GW}(f)}{\OMEGA{n}(f)}\right]^2}\ .
\ee
Here, $T$ is the mission life-time in seconds, $(f_\textnormal{min},f_\textnormal{max})$ is frequency range of the experiment, and $\OMEGA{n}$ is the experiment's (effective) noise.
For the noise curves we use the expressions given in \cite{Cornish:2018dyw}~(\experiment{LISA}), \cite{Yagi:2011yu}~(\experiment{BBO}), \cite{Yagi:2013du}~(\experiment{DECIGO}), and \cite{Isoyama:2018rjb}~(\experiment{B-DECIGO}).
We further assume $T=4$~yrs and $\rho_\textnormal{thr} = 10$~\cite{Caprini:2015zlo} ($\rho_\textnormal{thr} = 8$~\cite{Isoyama:2018rjb} for \experiment{B-DECIGO}).

The sensitivity curves of the experiments are shown as gray lines in figure~\ref{fig:GWspectrum}. 
Note that these curves are not the noise curves, but the power-law integrated curves~\cite{Thrane:2013oya}.
These indicate that a GW background should be detectable by the experiment if the spectrum touches or reaches into the region above the respective sensitivity curve.
Consequently, the spectrum shown in the figure can be probed by all four experiments.

\subsection{Results} 
\label{sec:GWresults}

\begin{figure}
	\centering
	\subfloat[][$\alpha$]{\includegraphics[width=.49\textwidth]{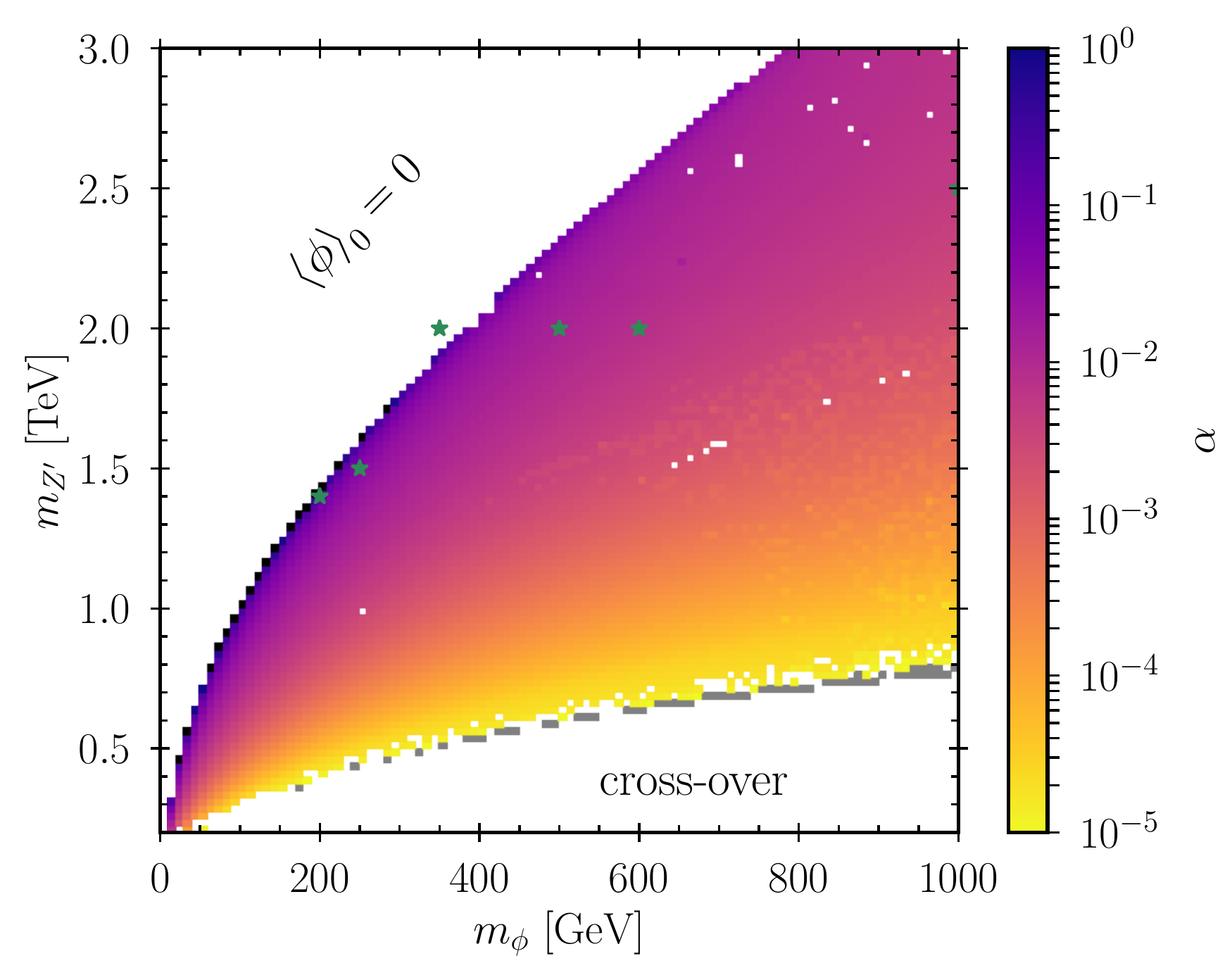}\label{fig:GWalpha}}
	\hfill
	\subfloat[][$\beta/H_\ast$]{\includegraphics[width=.49\textwidth]{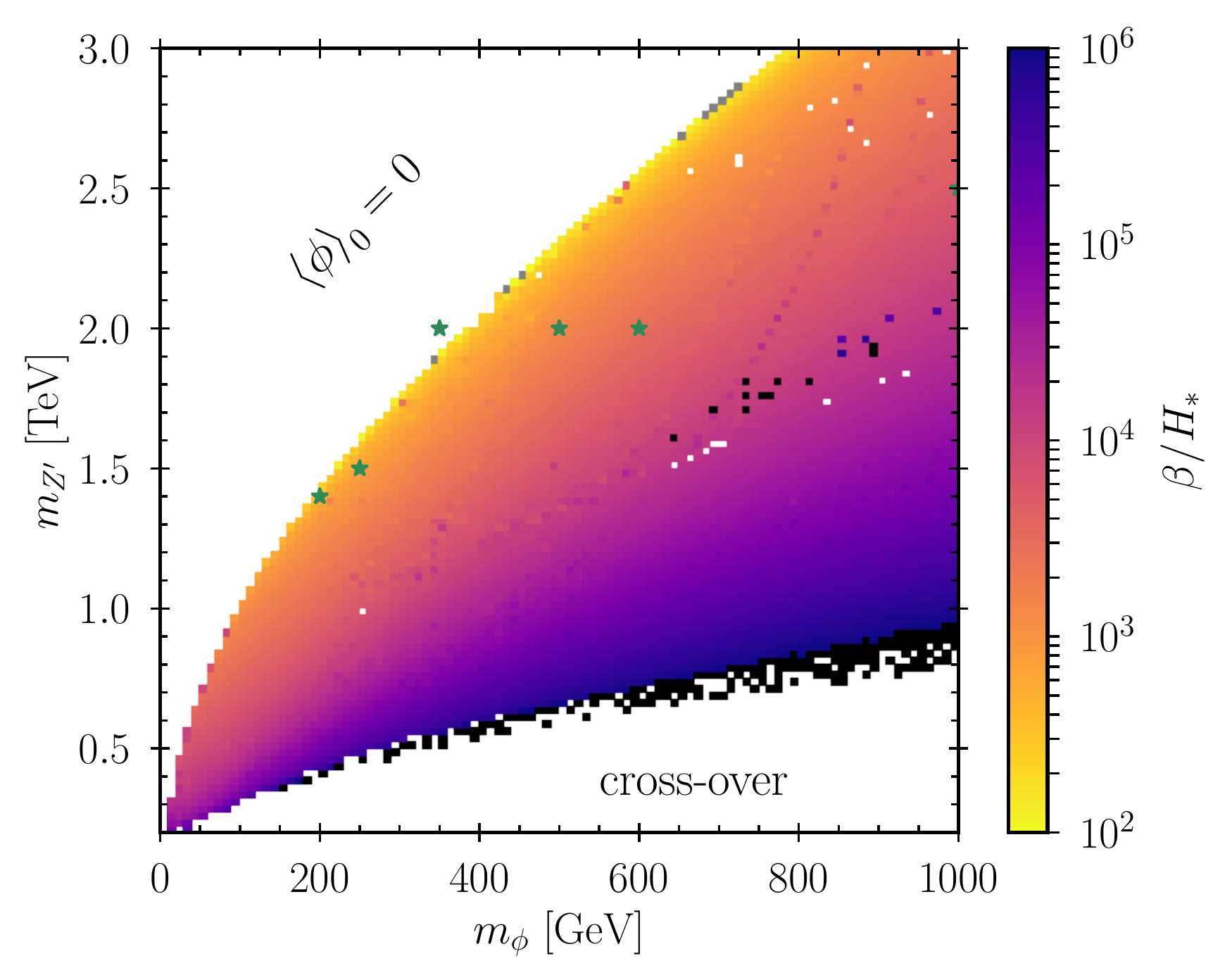}\label{fig:GWbeta}}
	\caption{$\alpha$ and $\beta/H_\ast$ as a function of the $U(1)_\ell$ breaking scalar mass $m_\phi$ and the $U(1)_\ell$ gauge boson mass $m_{Z^\prime}$ in the case of negligible portal coupling $\lambda_{p}$. Gray points correspond to underflows ($\alpha<10^{-5}$ or $\beta/H_\ast < 10^2$), black dots are overflows ($\alpha > 1$ or $\beta/H_\ast > 10^6$). See appendix~\ref{sec:GWbenchmark} for the values of the starred benchmark points.}\label{fig:GWAlphaBeta}
\end{figure}

Figure~\ref{fig:GWAlphaBeta} shows the parameters $\alpha$ and $\beta/H_\ast$ for the lepton number breaking phase transition as a function of the $\phi$ and $Z^\prime$ mass in the simplified case of negligible portal coupling $\lambda_{p}$ considered in section~\ref{sec:lepPT}, calculated using \software{CosmoTransitions}~\cite{Wainwright:2011kj}.
Most choices of masses give rise to a rather short first-order phase transition (high $\beta/H_\ast$) with few energy released (low $\alpha$). 
However, large values of $\alpha$ and small values of $\beta/H_\ast$ can be obtained in the $m_\Zp \gtrsim 2 m_\phi$ region, which is the region we identified to give rise to strong first-order PTs in section~\ref{sec:lepPT}.
As the amplitude of the sound-wave contribution to the GW spectrum~\eqref{eq:GWSW} is proportional to $\alpha^2/(1+\alpha)^2$ and $H_\ast/\beta$, this is indeed the region in which the stochastic background can be expected to be detectable.

\begin{figure}
	\centering
	\subfloat[neglecting Yukawa terms\label{fig:GWsensitivity1}]{\includegraphics[width=.49\textwidth]{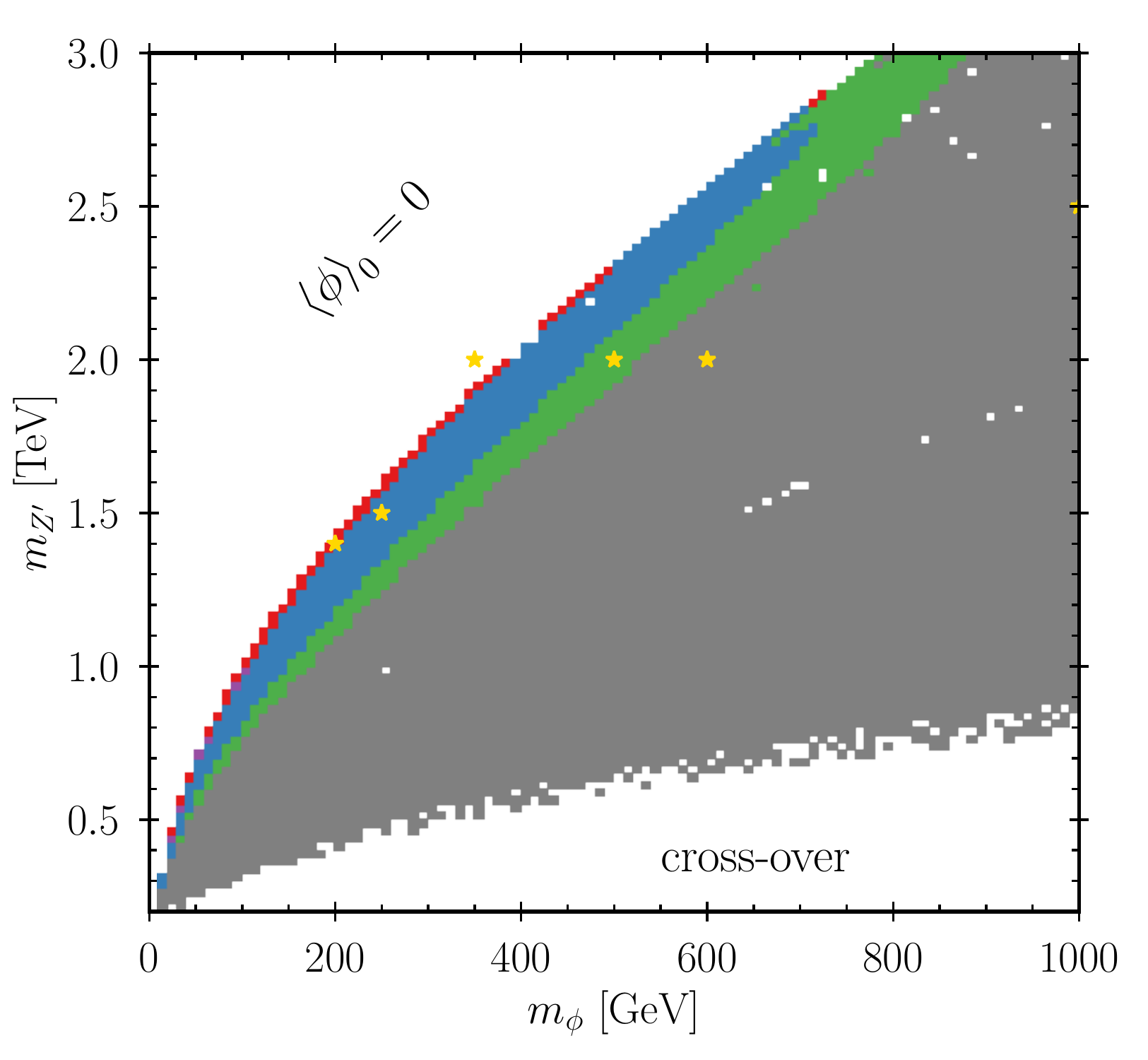}}
	\hfill
	\subfloat[$m_\DM = 150$~GeV, $m_\HL=200$~GeV \label{fig:GWsensitivity_fermionsLight}]{\includegraphics[width=.49\textwidth]{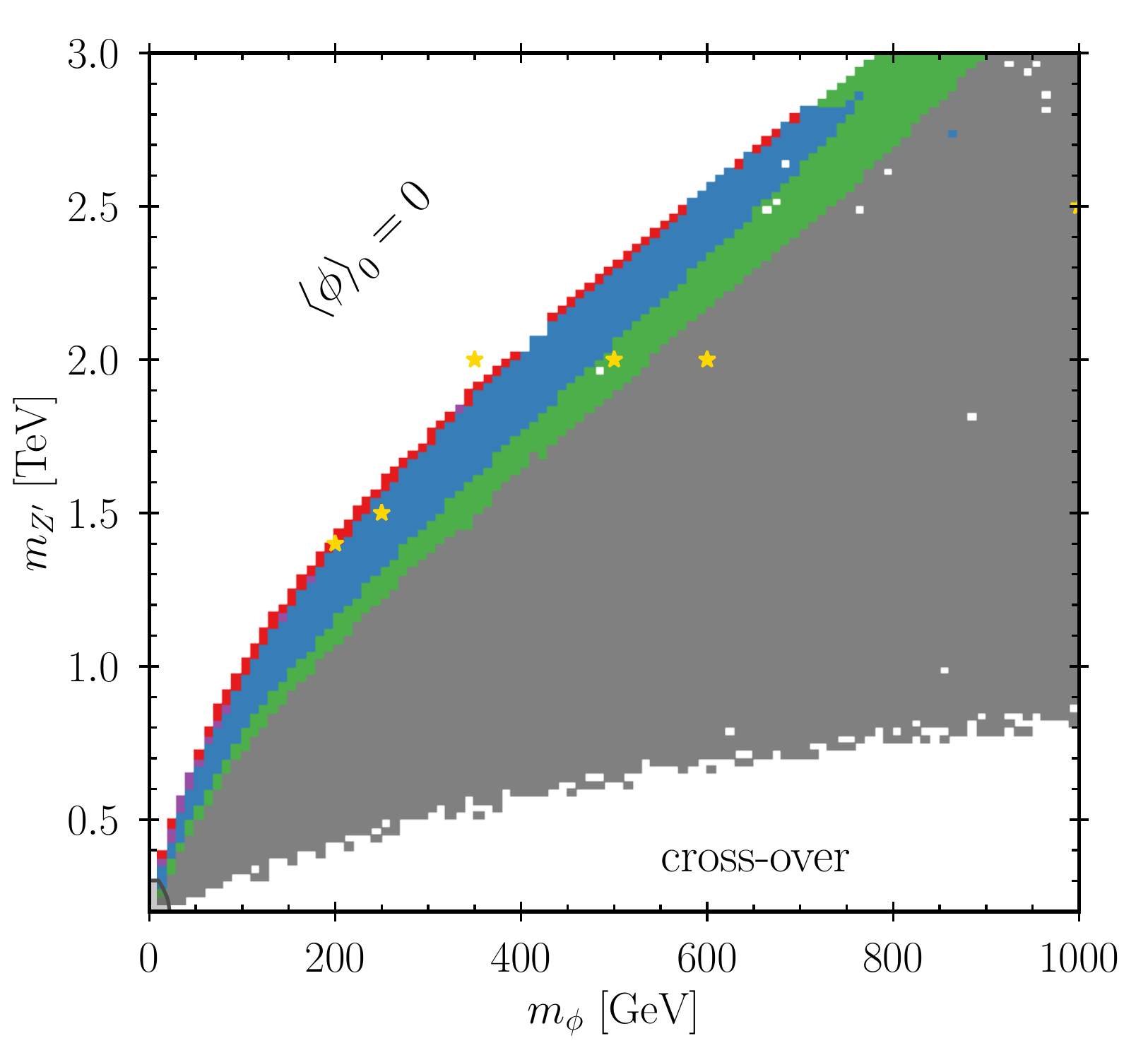}}
	
	\subfloat[$m_\DM = 500$~GeV, $m_\HL=1$~TeV \label{fig:GWsensitivity_fermionsHeavy}]{\includegraphics[width=.49\textwidth]{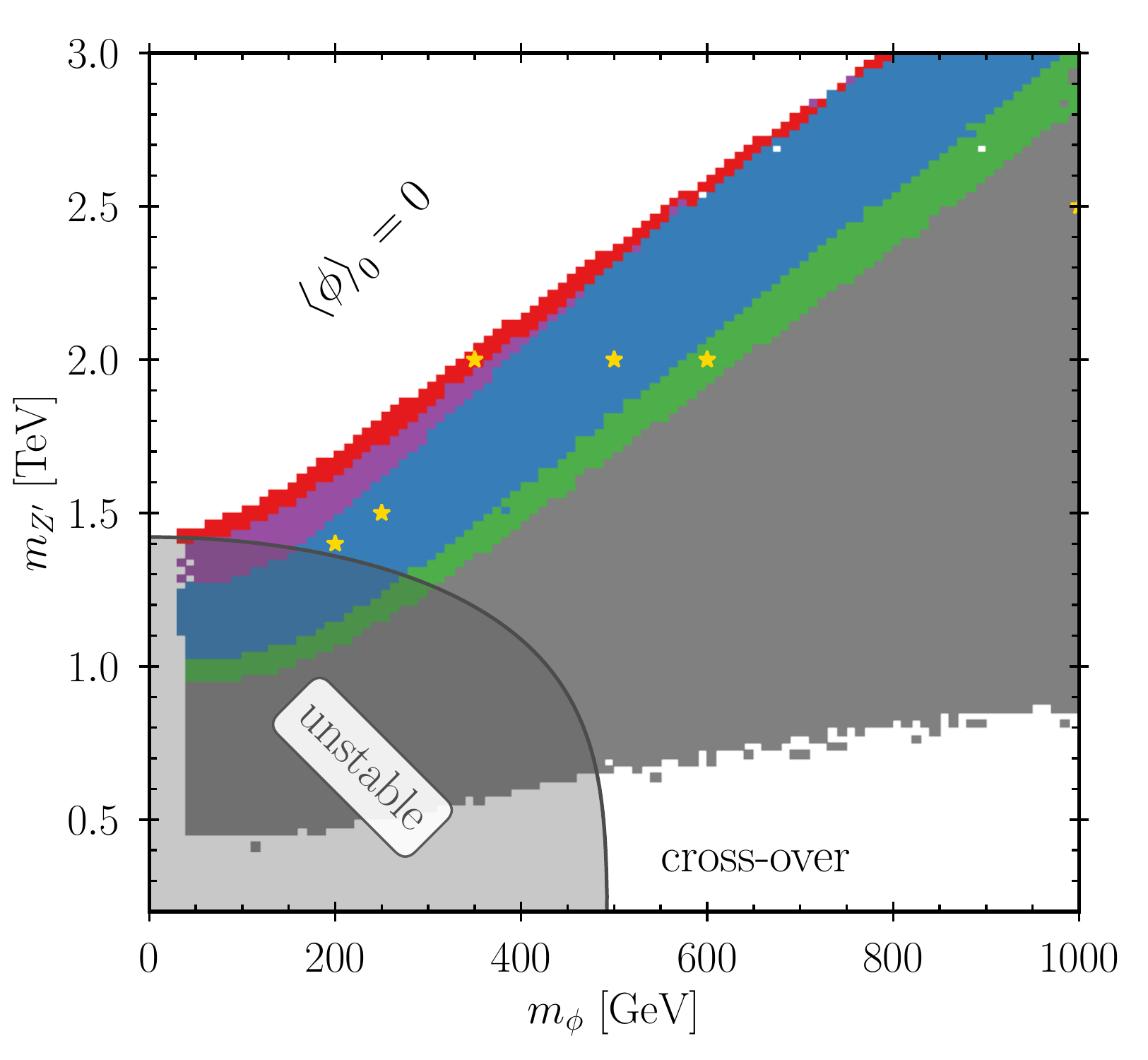}}
	\hfill
	\subfloat[$\OMEGA{\DM} = 0.1198,\ m_\HL = 1.5 \times m_\DM$ \label{fig:GWsensitivity_DM}]{\includegraphics[width=.49\textwidth]{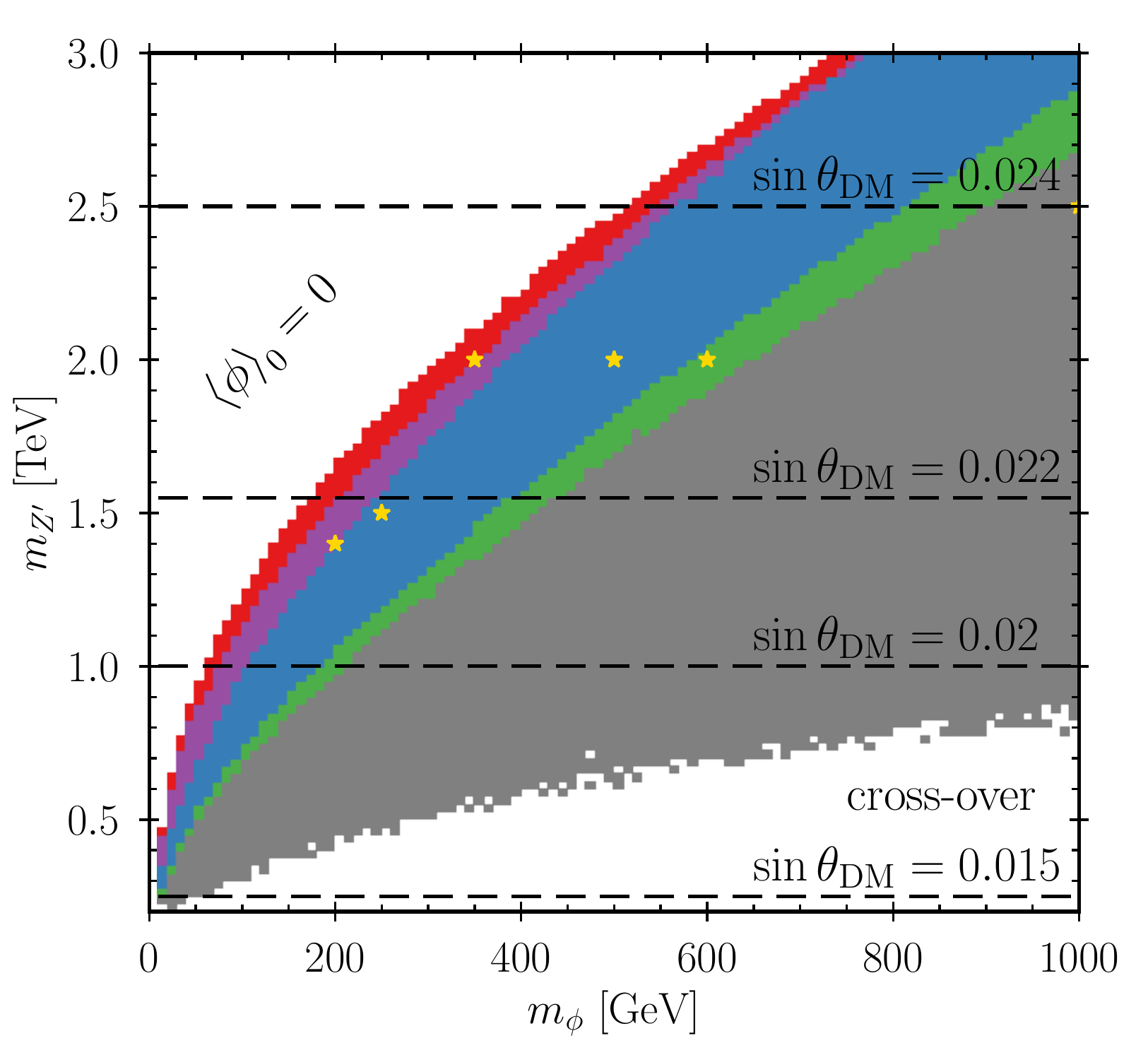}}
	
	\vspace{.25\baselineskip}
	\includegraphics[width=.8\textwidth]{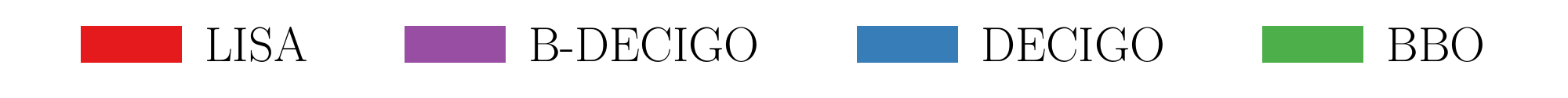}
	
	\caption{
		Sensitivity of space-based GW observatories to the stochastic background from the lepton number breaking PT (for negligible portal couplings) in the $m_\phi-m_\Zp$ plane.
		In the colored regions, the GW can be detected by \experiment{LISA} (red), \experiment{B-DECIGO} (purple), \experiment{DECIGO} (blue), and \experiment{BBO} (green), respectively. 
		In the gray region, the stochastic background is not detectable.	
		The SNRs of the points marked by a star are listed in appendix~\ref{sec:GWbenchmark}.
		The gray shaded regions (below the dark gray, solid line) in figures~\ref{fig:GWsensitivity_fermionsLight} and~\ref{fig:GWsensitivity_fermionsHeavy} are excluded as the potential is unstable.
		In figure~\ref{fig:GWsensitivity_DM}, the DM mass is set to a value reproducing the measured DM abundance at each parameter point. The dashed lines indicate the exclusion reach of \experiment{XENON1T} for different values of the DM mixing angle~$\theta_\DM$.
		\label{fig:GWsensitivity}
	}	
\end{figure}

The corresponding parameter points for which the stochastic GW background generated by the lepton number breaking PT is accessible to space-based GW interferometers are depicted in figure~\ref{fig:GWsensitivity1}.
The red, purple, blue, and green regions can be detected by \experiment{LISA}, \experiment{B-DECIGO}, \experiment{DECIGO}, and \experiment{BBO}, respectively, whereas in the gray region the generated GW background in not detectable.
If a parameter point is detectable by more than one experiment, the color corresponds to the experiment named first in the list above.

If the $\phi$ Yukawa couplings (and the Higgs portal coupling) are neglected, \experiment{LISA} and \experiment{B-DECIGO} are only sensitive in the $m_\Zp \gg m_\phi$  margin of the parameter space that has a first-order PT.
\experiment{DECIGO} can probe a small portion of the parameter space, also only close to the $m_\Zp \gg m_\phi$ edge, \experiment{BBO} is slightly more sensitive.
Still, the majority of the parameter space is inaccessible to GW experiments.

So far we neglected the contributions of the heavy leptons to the effective potential. 
Including them can significantly improve the detection prospects.
The detectability of the stochastic GW background for two different choices of heavy lepton masses are shown in figures~\ref{fig:GWsensitivity_fermionsLight} and~\ref{fig:GWsensitivity_fermionsHeavy}.
For light dark leptons (small Yukawas) with $m_\DM = 200$~GeV and $m_\HL = 210$~GeV ($m_\HL \equiv m_{e_4} = m_{e_5} = m_{\nu_4}$), the detectable part of the parameter space barely changes.
For heavier leptons with  $m_\DM = 0.5$~TeV and $m_\HL = 1$~TeV however, a significant fraction of the first-order transitions can be probed.
Still, the sensitivity of \experiment{LISA} is restricted to a band near the $m_\Zp \gg m_\phi$ edge of the phase-transition region, however the size of the detectable region increases notably compared to the case with vanishing Yukawa couplings.
\experiment{B-DECIGO} can probe additional parameter points, mostly for low scalar masses.
\experiment{DECIGO} and \experiment{BBO} can reach $m_\Zp \gtrsim 1$~TeV for $m_\phi \sim 50$~GeV and $m_\Zp \gtrsim 2.8$~TeV for $m_\phi \sim 1$~TeV.

Last but not least, figure~\ref{fig:GWsensitivity_DM} shows the detectability of the GW from the lepton number breaking PT requiring that the DM accounts for the full thermal abundance measured by \experiment{Planck}, assuming $m_\HL = 1.5 \times m_\DM$ as in figure~\ref{fig:VeffTnDM}.
Again, the effects of the dark leptons significantly enhance the parameter space to which future space-based GW observatories are sensitive.
The dashed lines indicate the exclusion reach of \experiment{XENON1T} (cf.\ figure~\ref{fig:DDsinDM}) for DM mixing angles of $\sin\theta_\DM = 0.015,\ 0.02,\ 0.22$ and $0.024$.
The white region in the upper left part of the plot is excluded as the lepton number gauge group remains unbroken.
Although not specifically mentioned, this applies to all sub-figures of figure~\ref{fig:GWsensitivity}.

\section{Summary}
\label{sec:summary}
%

In this work we have studied an extension of the SM in which lepton number is promoted to a $U(1)$ gauge group.
We updated the collider and dark matter limits in~\cite{Schwaller:2013hqa} and added further constraints.
In particular, we considered the lepton number breaking phase transition.
We investigated the parameter region in which the transition is of first order and evaluated the detectability of the generated stochastic background of gravitational waves.

Beside the lepton number gauge boson and the scalar field that spontaneously breaks $U(1)_\ell$, the model features additional leptonic states.
Their presence is forced upon us by the necessity to cancel the gauge anomalies associated with $U(1)_\ell$.
The lightest of these additional dark leptons is stable, unless specific values of the lepton number charges are assumed.
Hence, the model naturally provides a dark matter candidate.

Assuming that the DM is a thermal relic, we identified the regions of the parameter space in which the DM candidate can account for the full abundance measured by \experiment{Planck}.
We found that the correct relic density can be reproduced for a broad extent of DM masses in the $\mathcal{O}(100~\textnormal{GeV})$ to TeV range.
This typically requires choosing $m_\Zp \sim 2 m_\DM$.

Direct and indirect detection experiments put limits on the DM interactions with SM fields.
Direct detection constrains the various mixings that can give rise to DM-quark interactions.
These are the SM doublet admixture into the singlet DM characterized by $\theta_\DM$, the kinetic mixing parameter $\epsilon$ of the lepton number and hypercharge gauge groups, and the mixing angle $\theta_H$ between the SM and the lepton number Higgs. \experiment{XENON1T} can exclude $\epsilon$ and $\theta_\DM$ in the percent range, and $\sin\theta_H \sim \mathcal{O}(0.1)$; \experiment{LZ} and \experiment{DARWIN} can improve the limits by roughly an order of magnitude.
Indirect detection on the other hand mainly probes \Zp\ mediated DM-SM interactions. 
However, even the next-generation \experiment{CTA} is only sensitive for lepton number charges as large as $L^\prime = 3/2$.

We also investigated collider limits. 
The most important ones are \experiment{LEP} limits.
These put a lower bound on the lepton number breaking scalar VEV $v_\Phi \gtrsim 1.88$~TeV, and exclude \Zp\ masses below $m_\Zp \simeq 200$~GeV, as well as charged exotic leptons below 100~GeV.
The \experiment{LHC} can only put limits on the \Zp\ mass if the kinetic mixing is $\epsilon \sim \mathcal{O}(0.01-0.1)$, the high-luminosity \experiment{LHC} with $3~\textnormal{ab}^{-1}$ can reach $\epsilon \sim 10^{-3}$ for low \Zp\ masses and even exclude $m_\Zp \lesssim 500$~GeV if $\epsilon=0$.
Current \experiment{LHC} measurements further exclude Higgs mixing angles $\sin\theta_H \gtrsim 0.27$ and constrain the exotic leptons' Yukawa couplings.

Finally, we investigated the lepton number breaking phase transition in the early Universe.
If the portal coupling between the SM and dark Higgs is sufficiently small, the lepton number and electroweak phase transitions happen independently from one another.
Due to the VEV hierarchy imposed by the \experiment{LEP} constraints, the former typically occurs first. 
We found that in a large fraction of the parameter space the lepton number transition is first order.
It can thus generate a stochastic background of gravitational waves.
We calculated the corresponding GW spectrum and evaluated the detection prospects for future space-based GW observatories.

Whereas \experiment{LISA} can only probe a rather small fraction of the parameter space, its possible successors \experiment{BBO} and \experiment{DECIGO} are able to explore a significant fraction of the parameter points that give rise to a first-order PT.
Notably, the exotic leptons significantly enhance the detection prospects, particularly when requiring that the measured relic abundance is reproduced.
This is due to two effects.
First, the presence of additional particles lowers the nucleation temperture, and second, the fermionic contributions restore the broken minimum in a part of the parameter space in which the bosonic corrections alone would shift the vacuum back to the origin.

Further interesting effects may arise if one considers non-vanishing portal couplings between the dark and SM scalar sectors.
The transition can then proceed diagonally in field space, breaking the electroweak and lepton number gauge groups simultaneously.
We leave this subject for future work.
Another possible direction would be the investigation of the phase transition in the context of Baryogenesis.
Also, a more elaborate evaluation of exclusion or discovery prospects at future colliders such as the \experiment{ILC} or a 100~TeV $pp$ collider is needed.

\acknowledgments 

We thank M.~Baker, M.~Breitbach, T.~Konstandin, T.~Opferkuch and S.~Westhoff for helpful discussions, and N.~Christensen for help with FeynRules. 
Our work is supported by the DFG Cluster of Excellence PRISMA (EXC 1098). E.\ M.\ is a recipient of a fellowship through GRK Symmetry Breaking (DFG/GRK 1581). We also acknowledge the use of TikZ-Feynman~\cite{Ellis:2016jkw} and TikZ-FeynHand~\cite{Dohse:2018vqo} for drawing Feynman diagrams.

\begin{appendix}

\section{Goldstone divergences}
\label{sec:self-energy}

In this appendix we address the cancellation of the IR divergence in the second derivative of the effective potential, originating from the vanishing Goldstone mass in Landau gauge.
We follow the treatment in~\cite{Delaunay:2007wb}.
This leads to the renormalization condition~\eqref{eq:VeffSimpleOnShell}.
We calculate the self-energy $\Sigma (p^2)$ of the lepton number breaking scalar $\phi$ in Landau gauge, using dimensional regularization in $D=4-2\epsilon$ dimensions.
More precisely, we are interested in the difference of the self-energy evaluated at the scalar mass $p^2 = m_\phi^2$ and at $p^2=0$, $\Delta\Sigma \equiv \Sigma(m_\phi^2) - \Sigma(0)$, where $p^2$ is the external momentum squared. 

\subsection{Scalar self-energy}

We consider the Lagrangian
\be
	\Lag = \Dcov_\mu \Phi^\dagger \Dcov^\mu \Phi + \mu_\Phi^2 \Phi^\dagger \Phi - \lambda_\Phi \left(\Phi^\dagger\Phi\right)^2\,,
\ee
where $\Dcov_\mu \Phi = \partial_\mu \Phi - i g_\ell L_\Phi \Zp \Phi$. 
We rewrite the complex scalar $\Phi$ in terms of its real and imaginary parts and its vacuum expectation value $v_\Phi$ as $\Phi = \frac{1}{\sqrt{2}} \left(v_\Phi + \phi + i \omega^0\right)$.

\begin{figure}
	\centering
	\includegraphics[width=.25\textwidth]{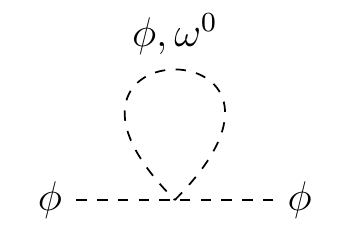}
	\hspace{1cm}
	\includegraphics[width=.25\textwidth]{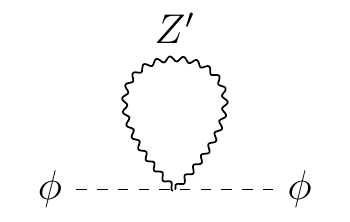}
	\caption{Tadpole diagrams contributing to the self-energy of $\phi$.}
	\label{fig:self-energy-tadpole}
\end{figure}

The (bare) self-energy of $\phi$ receives contributions from loops of $\phi$ itself, the Goldstone boson $\omega^0$, and the gauge boson \Zp. 
As the tadpole diagrams depicted in figure~\ref{fig:self-energy-tadpole} are independent of the external momentum, we only need to evaluate the remaining bubble diagrams.
These are
\begin{align}
	- i \Sigma_0^{S}(p^2) =&\ \mathfig{\includegraphics[width=4cm]{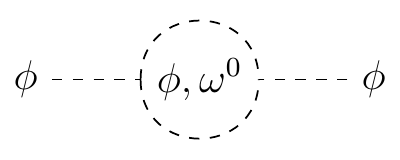}}\,,\\
	- i \Sigma_0^{\Zp}(p^2) =&\ \mathfig{\includegraphics[width=4cm]{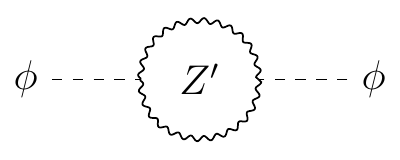}} + \mathfig{\includegraphics[width=4cm]{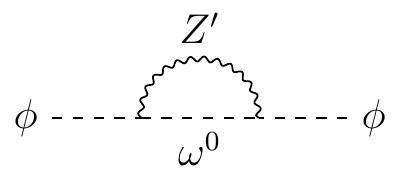}\vspace{5.5mm}}\,.
\end{align}

We perform the Passarino-Veltman reduction~\cite{Passarino:1978jh} of the corresponding integrals using \software{FeynCalc 9.3.0}~\cite{Mertig:1990an,Shtabovenko:2016sxi}, yielding
\begin{align}
	\label{eq:selfenergy-sigmaScalar}
	\Sigma_0^{S}(p^2) =&\ - \frac{\lambda_S^2 v_\Phi^2}{32 \pi^2} B_0\left(p^2,m_S^2,m_S^2\right)\,,\\
	\Sigma_0^{\Zp}(p^2) =&\ - \frac{L_\Phi^2 g_\ell^2}{32 \pi^2 m_\Zp^2} \Big\{
			\left[p^4 - 4 m_\Zp^2 p^2 + 12 m_\Zp^4 \right] B_0\left(p^2,m_\Zp^2,m_\Zp^2\right)
			\notag \\ &\hspace{10eM}
			- p^4 B_0\left(p^2,0,0\right)
			- 2 p^2 A_0\left(p^2\right)
			- 8 m_\Zp^4
		\Big\}\,,
\end{align}
where $\lambda_S = 6 \lambda_\Phi$ for $S=\phi$ and $\lambda_S = 2 \lambda_\Phi$ for $S=\omega^0$, respectively.
The scalar one- and two-point functions $A_0$ and $B_0$ are defined by
\begin{align}
	A_0\left(m^2\right) =&\ \frac{(2\pi\mu_R)^{4-D}}{i \pi^2} \int \dInt^D\!q\ \frac{1}{q^2 - m^2 + i \varepsilon} \,,\\
	B_0\left(p^2,m_1^2,m_2^2\right) =&\ \frac{(2\pi\mu_R)^{4-D}}{i \pi^2} \int \dInt^D\!q\ \frac{1}{q^2 - m_1^2 + i \varepsilon} \frac{1}{(q+p)^2 - m_2^2 + i \varepsilon} \,,
\end{align}
where $\mu_R$ is the renormalization scale.
The full (bare) self-energy is then given by
\be
	\label{eq:selfenergy-sigma0}
	\Sigma_0\left(p^2\right) = \Sigma_0^\phi\left(p^2\right) + \Sigma_0^{\omega^0}\left(p^2\right) + \Sigma_0^\Zp\left(p^2\right) + \textnormal{tadpole contributions}\,.
\ee

The renormalized self-energy is related to the bare one by
\be
	\Sigma_R\left(p^2\right) = \Sigma_0\left(p^2\right) + \delta m^2 - p^2 \delta Z\,,  
\ee
where $\delta m^2$ and $\delta Z$ are the mass and field renormalization counter-terms for $\phi$,
\be
	\Lag \supset \delta Z \partial_\mu \phi \partial^\mu \phi - \delta m^2 \phi^2\,.
\ee
When calculating $\Delta\Sigma$, $\delta m^2$ cancels in the difference but $\delta Z$ remains, i.e.
\be
	\label{eq:selfenergy-deltaSigma}
	\Delta\Sigma = \Sigma_0\left(m_\phi^2\right) - \Sigma_0\left(0\right) - m_\phi^2 \delta Z\,.
\ee
In particular, this means that $\Delta\Sigma$ is independent of the renormalization conditions we impose on the counter-terms $\delta m^2$ and $\delta\lambda$ when calculating the effective potential.
We can now fix $\delta Z$ by requiring canonical normalization of the field $\phi$, i.e.
\be
	\label{eq:selfenergy-deltaZ}
	\frac{\partial \Sigma_R}{\partial p^2}\left(m_\phi^2\right) = 0
	\hspace{2eM}\Longrightarrow\hspace{2eM}
	\delta Z = \frac{\partial \Sigma_0}{\partial p^2}\left(m_\phi^2\right)\,.
\ee
Again, the tadpole diagrams in figure~\ref{fig:self-energy-tadpole} do not contribute as they are independent of $p^2$.

Finally, the difference in the self-energy used in the renormalization condition~\eqref{eq:VeffSimpleOnShell} is obtained by plugging~\eqref{eq:selfenergy-sigma0} and~\eqref{eq:selfenergy-deltaZ} into~\eqref{eq:selfenergy-deltaSigma}.
We use \software{LoopTools 2.13}~\cite{Hahn:1998yk,vanOldenborgh:1989wn} to evaluate the finite part of the scalar integrals and their derivatives.

\subsection{On-shell renormalization of the effective potential}

The momentum-dependent mass of $\phi$ is given by 
\be
	m_\phi^2\left(p^2\right) = m_{\phi,R}^2 + \Sigma_R\left(p^2\right)\,,
\ee
where $m_{\phi,R}$ is the renormalized mass parameter in the Lagrangian, which is related to the physical (pole) mass $m_\phi^2 \equiv m_\phi^2(m_\phi^2)$ by
\be
	m_{\phi,R}^2 = m_\phi^2 - \Sigma_R\left(m_\phi^2\right)\,.
\ee

Since the effective potential is defined at vanishing external momentum, we now impose the conditions
\begin{align}
	\left.\frac{\partial V_\eff(\phi,T=0)}{\partial \phi}\right|_{\phi=v_\Phi} =&\ 0\,, \\
	\left.\frac{\partial^2 V_\eff(\phi,T=0)}{\partial^2 \phi}\right|_{\phi=v_\Phi} =&\ m_\phi^2(0) = m_\phi^2 - \Delta\Sigma\,.
\end{align}
We now further want the VEV $v_\Phi$ and the scalar mass $m_\phi$ to be identical to the values inferred from the tree-level potential, i.e.
\be
	\left.\frac{\partial V_0(\phi)}{\partial \phi}\right|_{\phi=v_\Phi} = 0\,, 
	\hspace{2eM}
	\left.\frac{\partial^2 V_0(\phi)}{\partial^2 \phi}\right|_{\phi=v_\Phi} = m_\phi^2\,,
\ee
hence, using $V_\eff(\phi,T=0) = V_0(\phi) + V_\textnormal{CW}(\phi) + V_\textnormal{c.t.}(\phi)$, we obtain the renormalization conditions~\eqref{eq:VeffSimpleOnShell}.

Note that $\Delta\Sigma$ has an IR divergence coming from the Goldstone contribution~\eqref{eq:selfenergy-sigmaScalar} to the self-energy at zero-momentum,
\be
	\Sigma_0^{\omega^0}(0) = -\frac{\lambda_\Phi^2 v_\Phi^2}{8 \pi^2} B_0\left(0,m_{\omega^0}^2,m_{\omega^0}^2\right)\,.
\ee
In Landau gauge $m_{\omega^0}=0$, but we keep it as a regulator.
Taking the analytic expression for the scalar two-point function from~\cite{tHooft:1978jhc,Denner:1991kt}, 
\begin{align}
	B_0\left(0,m_{\omega^0}^2,m_{\omega^0}^2\right) =&\ \Delta - \log\frac{m_{\omega^0}^2 - i \varepsilon}{\mu_R^2}\,,
\end{align}
where $\Delta = \frac{1}{\epsilon} - \gamma_E + \log 4\pi$, we obtain the IR divergent part
\be
	\label{eq:selfenergy-DeltaSigmaDivergence}
	- \Delta\Sigma = \frac{\lambda_\Phi^2 v_\Phi^2}{8 \pi^2} \log\frac{m_{\omega^0}^2}{\mu_R^2} + \textnormal{finite terms}\,.
\ee

On the other hand, the Goldstone contribution to the Coleman-Weinberg potential~\eqref{eq:VCW} is given by
\be
	V_\textnormal{CW}(\phi) \supset \frac{m_{\omega^0}^4(\phi)}{64\pi^2}\left[\log\frac{m_{\omega^0}^2(\phi)}{\mu_R^2} - \frac{3}{2}\right]\,,
\ee
where $m_{\omega^0}^2(\phi) = \lambda_\Phi \phi^2 - \mu_\Phi^2$ with $\mu_\Phi^2 = \lambda_\Phi v_\Phi^2$, and its derivatives are
\begin{align}
	\frac{\partial V_\textnormal{CW}}{\partial \phi} \supset&\ \frac{\lambda_\Phi \phi\ m_{\omega^0}^2(\phi)}{16\pi^2}\left[\log\frac{m_{\omega^0}^2(\phi)}{\mu_R^2} - 1\right]\,,\\
	\frac{\partial^2 V_\textnormal{CW}}{\partial \phi^2} \supset&\ \frac{\lambda_\Phi m_{\omega^0}^2(\phi)}{16\pi^2}\left[\log\frac{m_{\omega^0}^2(\phi)}{\mu_R^2} - 1\right] + \frac{\lambda_\Phi^2 \phi^2}{8\pi^2}\log\frac{m_{\omega^0}^2(\phi)}{\mu_R^2}\,.
\end{align}
Whereas the parts with the square brackets go to zero when taking the limit $\phi\longrightarrow v_\Phi$, the second part in the second derivative gives the same IR divergence we encountered in~\eqref{eq:selfenergy-DeltaSigmaDivergence}.
Hence, the IR divergences on both sides of the second condition in~\eqref{eq:VeffSimpleOnShell} cancel and we obtain IR-finite counter-terms.

\section{Bubble wall velocity}
\label{sec:vW}

\begin{figure}
	\centering
	\subfloat[][$v_w=0.9$]{\includegraphics[width=.49\textwidth]{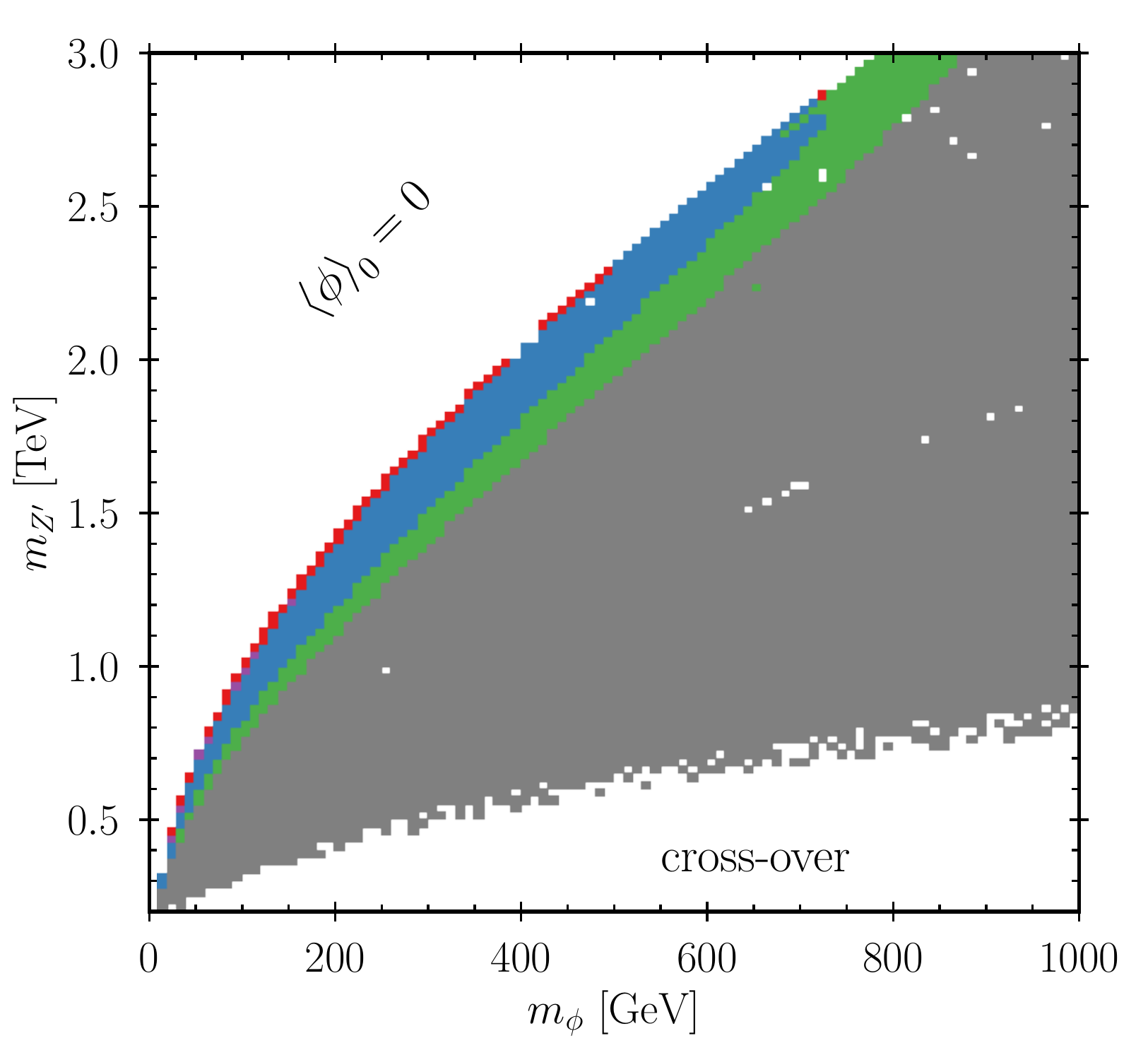}}
	\hfill
	\subfloat[][$v_w=0.6$]{\includegraphics[width=.49\textwidth]{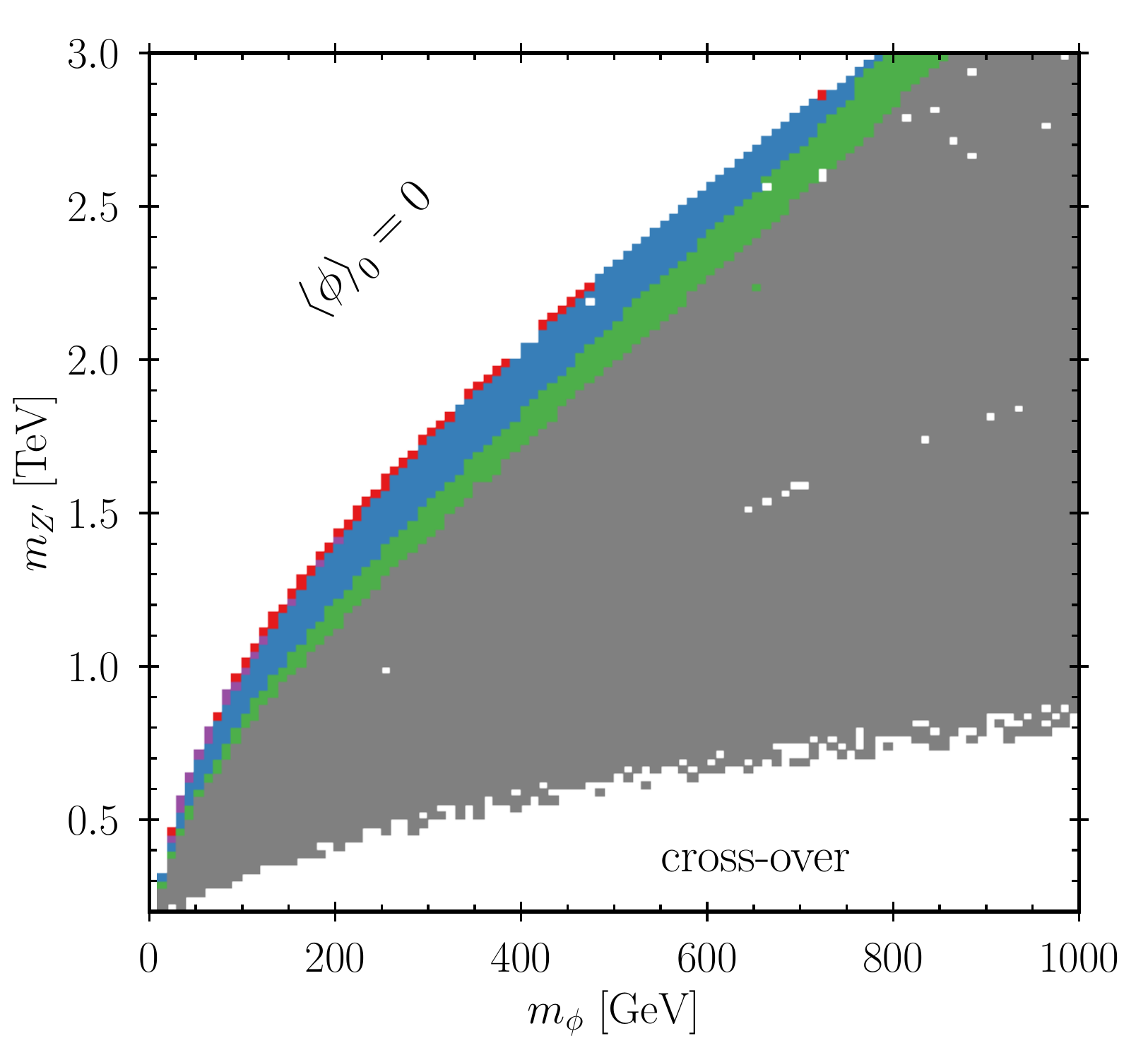}}
	\\
	\subfloat[][$v_w=0.3$]{\includegraphics[width=.49\textwidth]{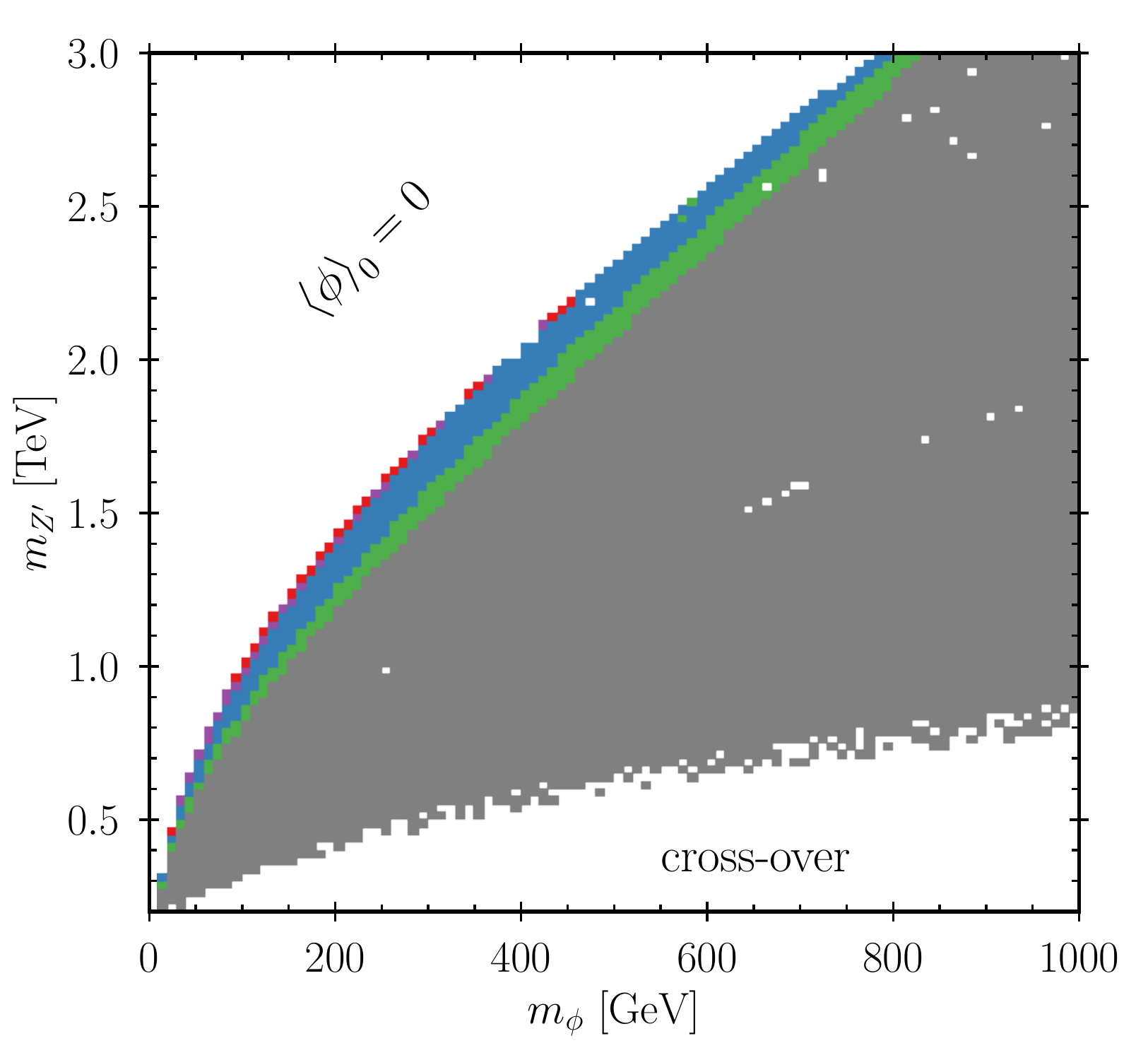}}
	\hfill
	\subfloat[][$v_w=0.1$]{\includegraphics[width=.49\textwidth]{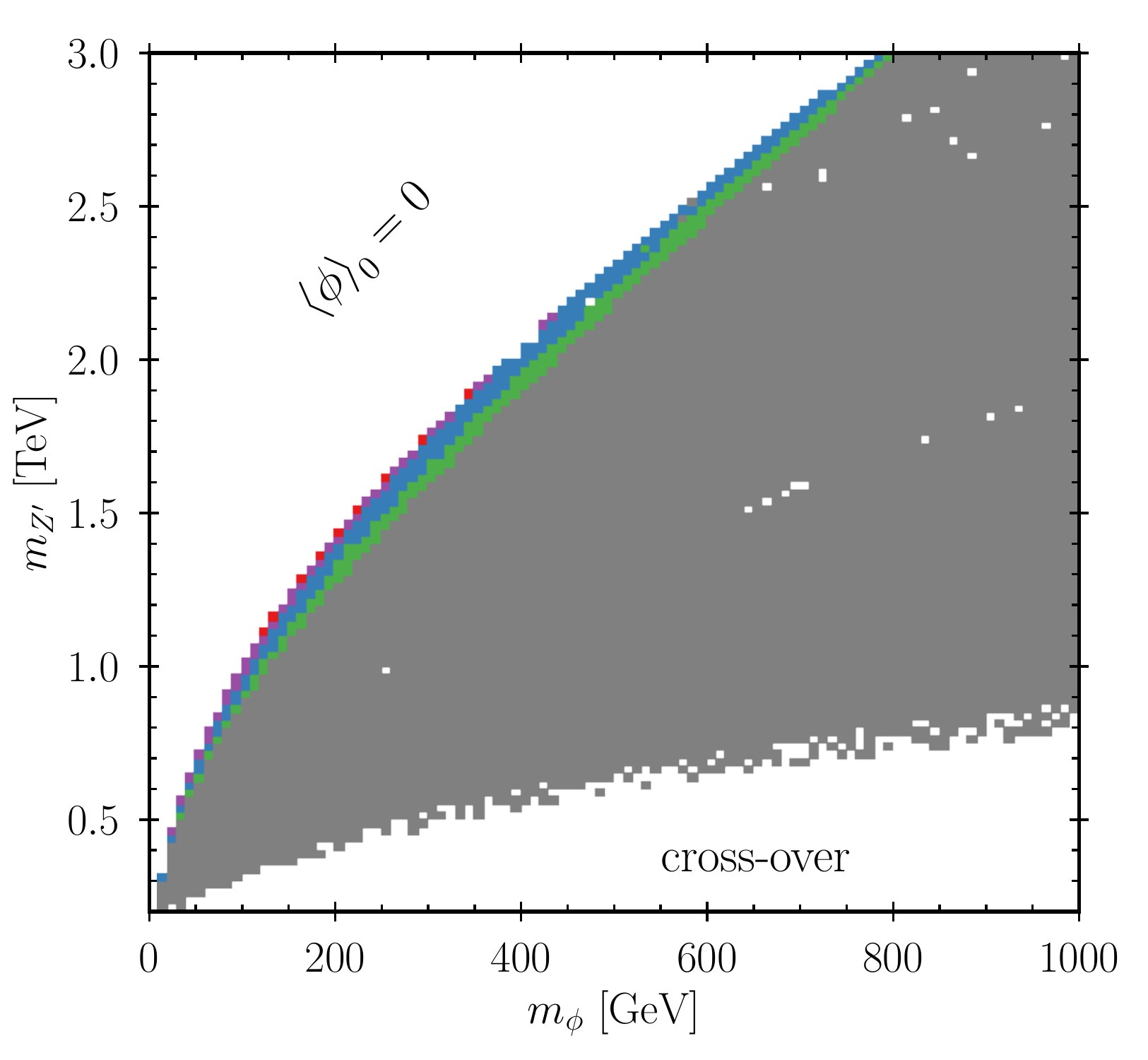}}
	\caption{
		Same as figure~\ref{fig:GWsensitivity1}, but with different wall velocities. Figure~\ref{fig:GWsensitivity1} assumes $v_w=1$.
	}\label{fig:GWsensitivity_vW}
\end{figure}

The wall velocity is rather difficult to compute and can in general not be determined from the finite-$T$ effective potential alone as it involves out-of-equilibrium dynamics.
It can be obtained by a microscopic treatment of the fluid solving Boltzmann equations or at the macroscopic level adding an effective friction term to the scalar equation of motion, see~\cite{Espinosa:2010hh,Konstandin:2014zta} and reference therein for more details, or \cite{Dorsch:2018pat} for recent results.
Assuming Chapman-Jouget detonations~\cite{Steinhardt:1981ct}, $v_w$ can be calculated as a function of $\alpha$, yielding values ranging from the speed of sound $c_s = \frac{1}{\sqrt{3}}$ in the plasma to the speed of light.
However, this assumption is not justified and typically incorrect~\cite{Laine:1993ey}. 

Since we expect that the bubbles do not run away in our model, the GW spectrum is given by $\OMEGA{GW}(f) = \OMEGA{sw}(f)+\OMEGA{turb}(f)$. 
For both sound wave and turbulence contribution (Equations~\eqref{eq:GWSW} and \eqref{eq:GWturb}) the amplitudes of the spectra are proportional to $v_w$ and the peak frequencies shift as $1/v_w$, i.e.\ order one changes in the wall velocity only have an order one effect on the spectrum and peak frequencies.
Hence, the detectability of the generated stochastic background eventually only has a mild dependence on $v_w$.
As a detectable signal further typically requires strong transitions with large wall velocities, we simply take the most optimistic estimate $v_w = 1$.

Figure~\ref{fig:GWsensitivity_vW} shows the dependence of the detectability on the bubble wall velocity in the case of negligible portal and Yukawa couplings.
Compared to figure~\ref{fig:GWsensitivity1} above where $v_w=1$ was assumed, here we show the sensitivity for a slightly lower wall velocity ($v_w=0.9$), a wall velocity close to the speed of sound ($v_w=0.6$), and subsonic bubbles ($v_w=0.3$ and $v_w=0.1$).
Again, the colored regions are detectable by LISA~(red), B-DECIGO~(purple), DECIGO~(blue) and BBO~(green). The GW spectra generated by the first-order phase transitions in the gray region are not detectable.
For supersonic bubbles, the detectable regions barely change when varying $v_w$.
Taking $v_w$ to subsonic values decreases the sensitivity visibly.

\section{GW benchmark points}
\label{sec:GWbenchmark}

We here present results for 24 benchmark points.
We consider the 6 combinations of scalar and $\Zp$ masses listed in table~\ref{tab:benchmarks}.
At each mass combination we take 4 different scenarios for the exotic leptons.
Scenario~A neglects the fermionic contributions, whereas B and C assume $m_\DM = 200$~GeV, $m_\HL = 210$~GeV and $m_\DM = 500$~GeV, $m_\HL = 1$~TeV, respectively. Again, $m_\HL \equiv m_{e_4} = m_{e_5}$.
Finally, scenario~C takes the values listed in table~\ref{tab:benchmarks}, for which the DM accounts for the full relic abundance measured by \experiment{Planck}.
All benchmarks assume $v_\Phi = 2$~TeV and $L^\prime = -1/2$. All mixing angles are set to zero.

\begin{table}
	\renewcommand{\arraystretch}{1.2}
	\centering
	\begin{tabular}{c|*{6}{C{3eM}}}
		\hline\hline
		benchmark & 1 & 2 & 3 & 4 & 5 & 6\\
		\hline
		$m_\phi$ [GeV] & 200   & 250   & 350   & 500   & 600    & 1000   \\
		$m_\Zp$ [TeV]  & 1.4   & 1.5   & 2.0   & 2.0   & 2.0    & 2.5    \\
		\hline
		$m_\DM$ [GeV]  & 550   & 580   & 720   & 720   & 720   & 835   \\
		$m_\HL$ [GeV]  & 825   & 870   & 1080  & 1080  & 1080  & 1250  \\
		\hline\hline
	\end{tabular}
	\caption{Benchmark values of the scalar and \Zp\ masses, along with the DM and exotic lepton masses for the benchmark points D1 -- D6. }
	\label{tab:benchmarks}
\end{table}

\begin{figure}
	\centering
	\parbox{.49\textwidth}{\includegraphics[width=.49\textwidth]{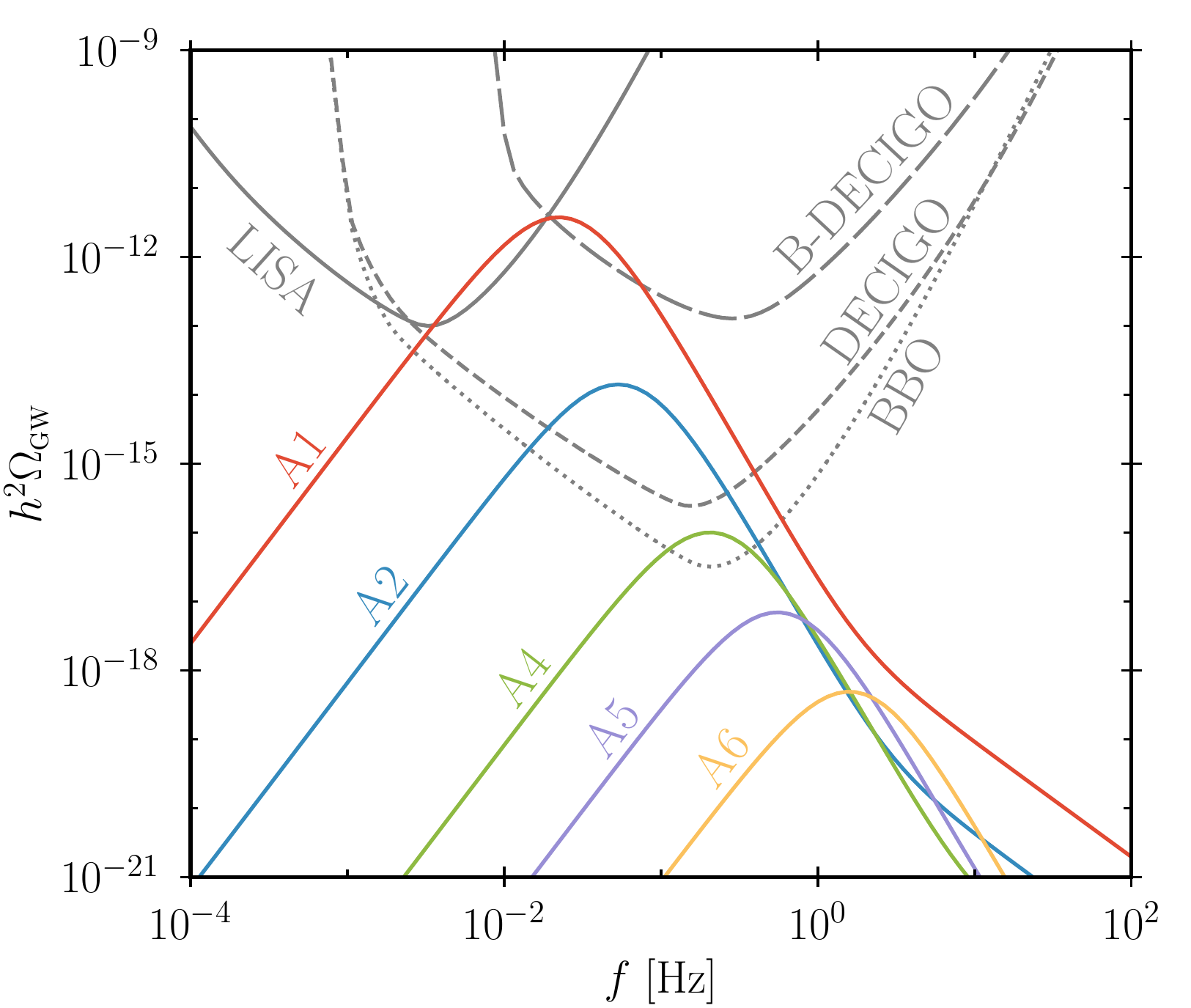}}
	\hfill
	\parbox{.49\textwidth}{\includegraphics[width=.49\textwidth]{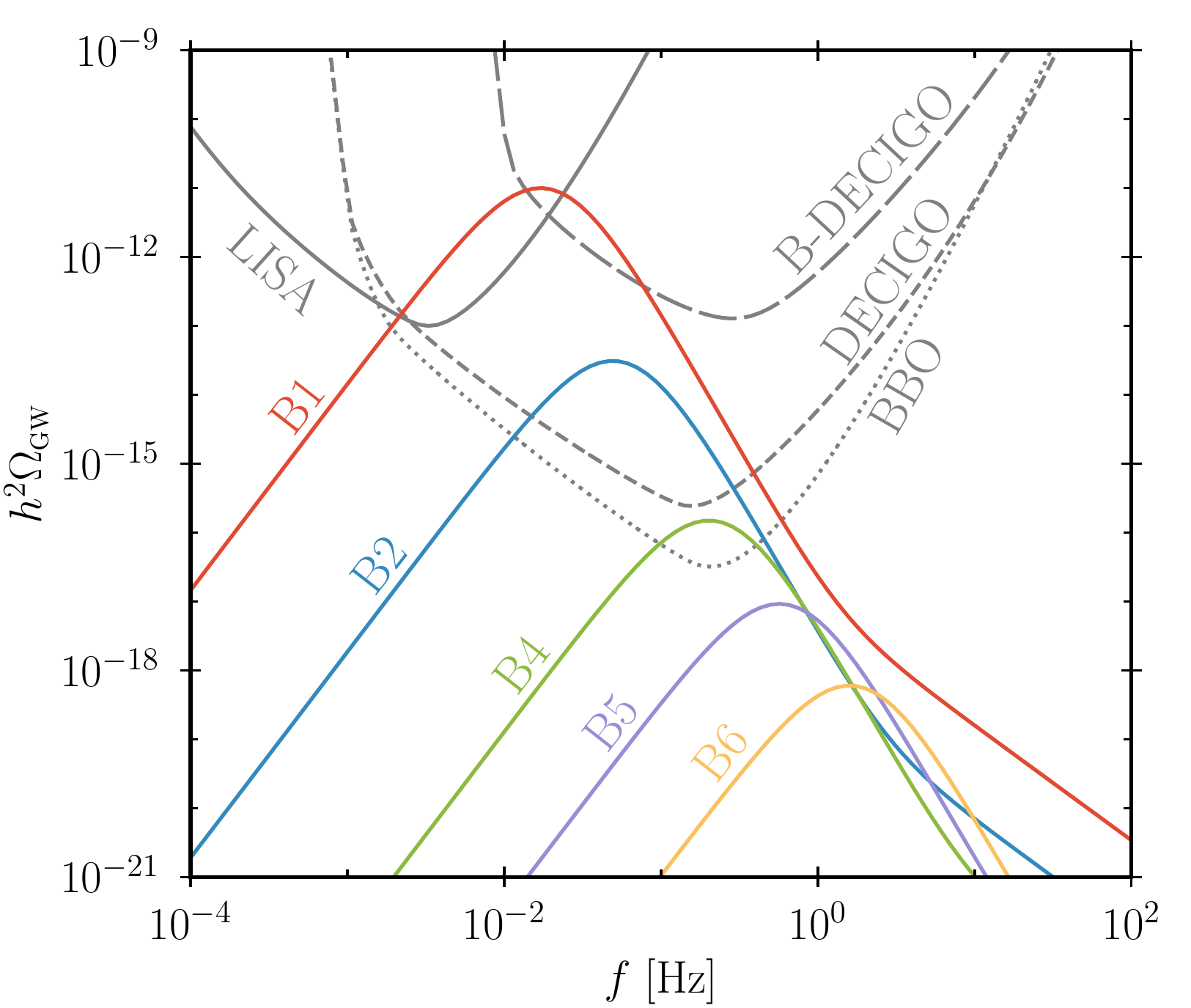}}
	\\
	\parbox{.49\textwidth}{\includegraphics[width=.49\textwidth]{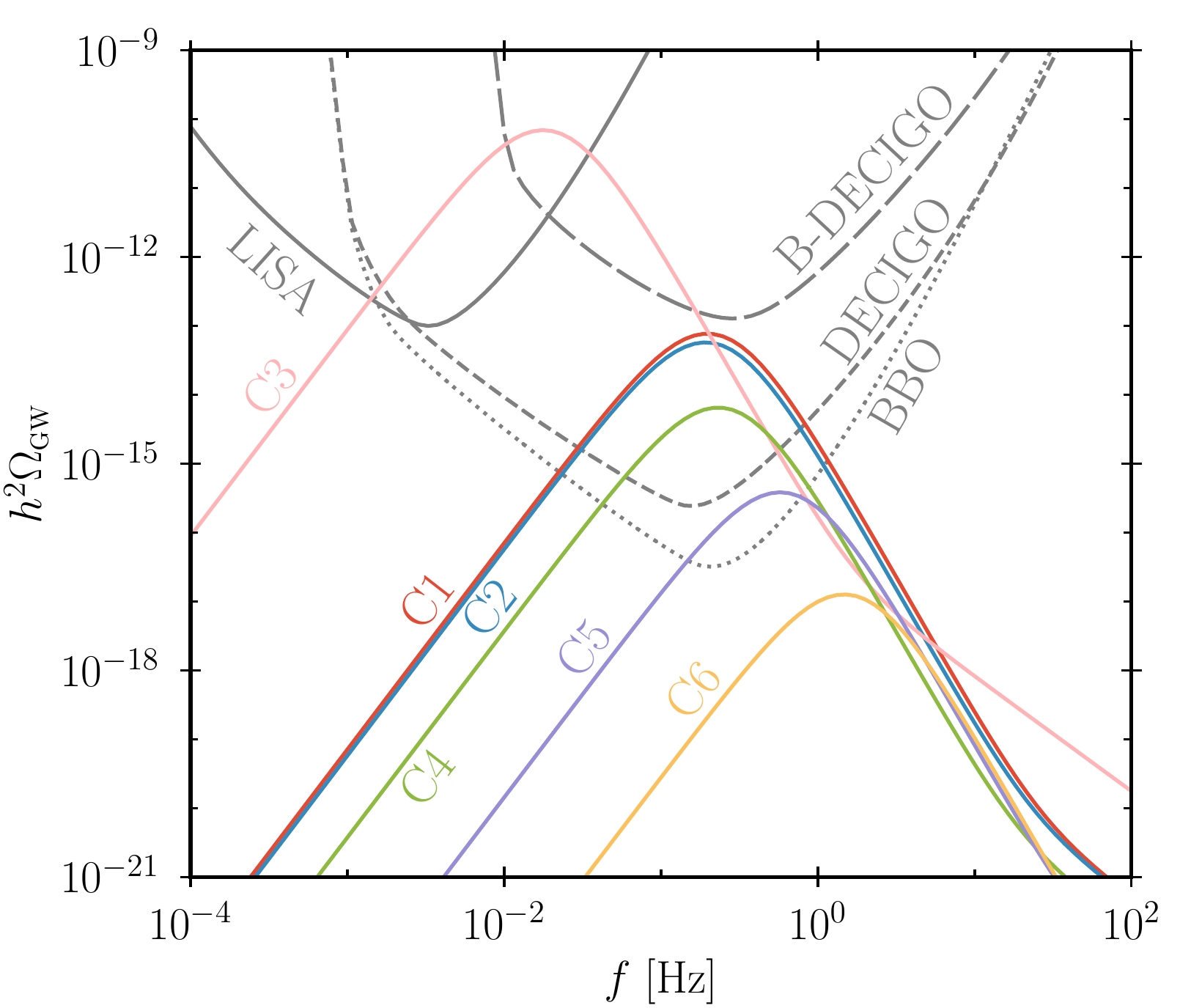}}
	\hfill
	\parbox{.49\textwidth}{\includegraphics[width=.49\textwidth]{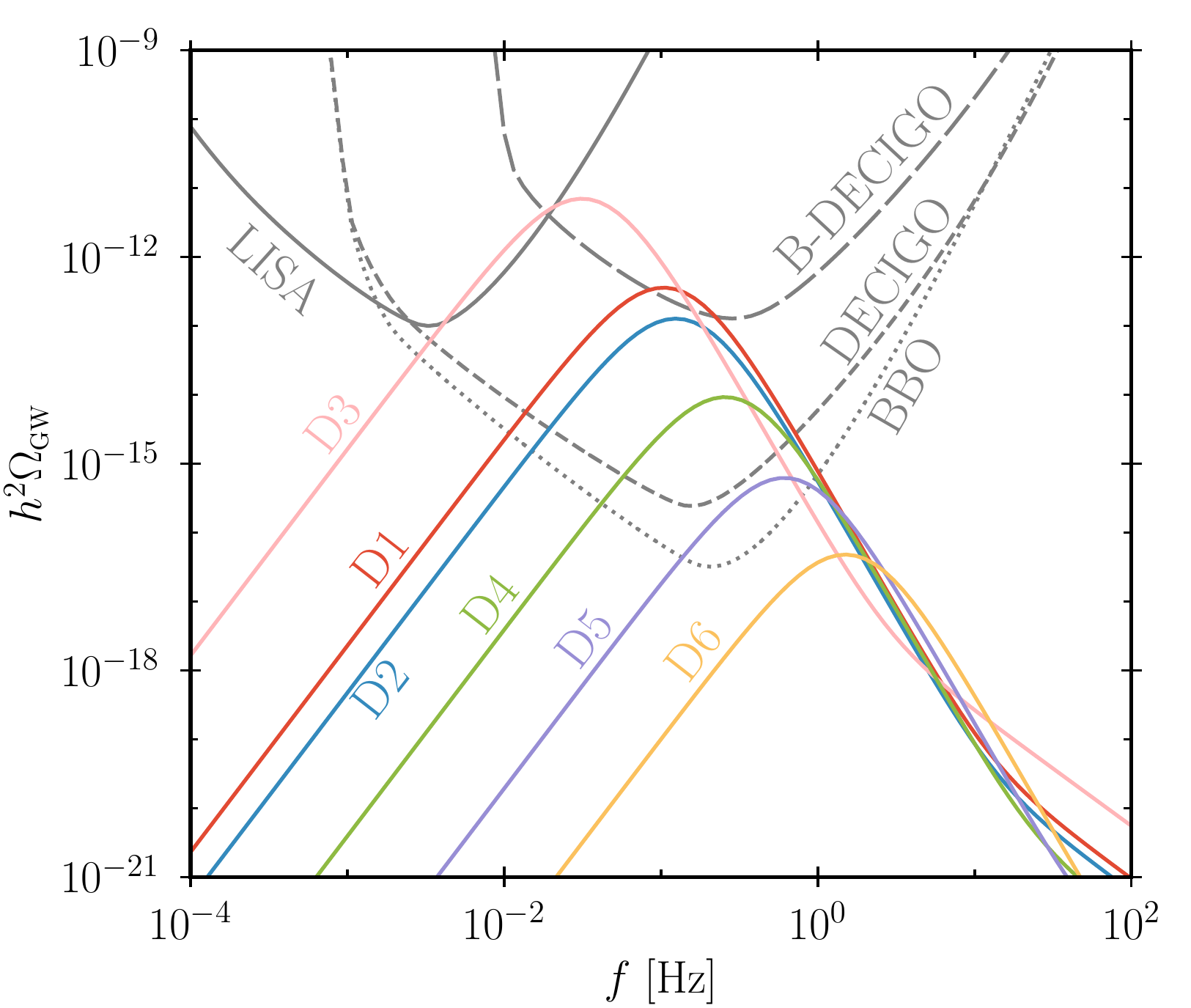}}
	\caption{GW spectra of the benchmark points (colored lines) and power-law integrated sensitivity curves for space-based GW observatories (gray lines).}
	\label{fig:benchmarkSpectra}
\end{figure}

We have computed the critical and nucleation temperatures $T_c$ and $T_n$ as well as the transition parameters $\alpha$ and $\beta$ for the considered benchmark points. Table~\ref{tab:benchmark-results} lists the corresponding results obtained using \software{CosmoTransitions}~\cite{Wainwright:2011kj}.
The respective GW spectra are depicted in figure~\ref{fig:benchmarkSpectra}, assuming a bubble wall velocity of $v_w = 1$. The sensitivities of \experiment{LISA}, \experiment{B-DECIGO}, \experiment{DECIGO}, and \experiment{BBO} are also shown, with the SNRs included in table~\ref{tab:benchmark-results}.

\begin{table}
	\renewcommand{\arraystretch}{1.2}
	\centering
	\begin{tabular}{c|*{4}{C{4eM}}|*{4}{C{3eM}}}
		\hline\hline
		   & $T_c$ [GeV] & $T_n$ [GeV] & $\alpha$ & $\beta$ & SNR1 & SNR2 & SNR3 & SNR4 \\
		\hline
		A1 & 487 & 198 & 0.18 & 570 & 22 & 9.9 & $>10^3$ & $>10^3$ \\
		A2 & 576 & 360 & 0.041 & 750 & 0.009 & 0.14 & 170 & 730 \\
		A4 & 907 & 770 & 0.013 & 1300 & 0 & 0.005 & 3.3 & 27 \\
		A5 & 1068 & 987 & 0.0082 & 2800 & 0 & 0 & 0.068 & 0.85 \\
		A6 & 1507 & 1447 & 0.0050 & 5500 & 0 & 0 & 0 & 0.007 \\
		\hline
		B1 & 451 & 188 & 0.23 & 460 & 110 & 18 & $>10^3$ & $>10^3$ \\
		B2 & 541 & 342 & 0.051 & 720 & 0.025 & 0.26 & 310 & 1300 \\
		B4 & 875 & 745 & 0.015 & 1400 & 0 & 0.007 & 5.0 & 40 \\
		B5 & 1036 & 959 & 0.0090 & 3000 & 0 & 0 & 0.87 & 1.1 \\
		B6 & 1476 & 1418 & 0.0053 & 5700 & 0 & 0 & 0 & 0.009 \\
		\hline
		C1 & 382 & 330 & 0.096 & 3000 & 0.001 & 3.4 & $>10^3$ & $>10^3$ \\
		C2 & 427 & 368 & 0.085 & 2600 & 0.001 & 2.5 & $>10^3$ & $>10^3$ \\
		C3 & 446 & 216 & 0.44 & 410 & 680 & 130 & $>10^3$ & $>10^3$ \\
		C4 & 652 & 580 & 0.044 & 2000 & 0 & 0.31 & 200 & $>10^3$ \\
		C5 & 777 & 733 & 0.025 & 4000 & 0 & 0.014 & 3.5 & 44 \\
		C6 & 1149 & 1112 & 0.012 & 6700 & 0 & 0 & 0.017 & 0.23 \\
		\hline
		D1 & 361 & 271 & 0.13 & 1900 & 0.037 & 8.5 & $>10^3$ & $>10^3$ \\
		D2 & 417 & 337 & 0.096 & 1900 & 0.008 & 3.7 & $>10^3$ & $>10^3$ \\
		D3 & 446 & 279 & 0.22 & 570 & 18 & 31 & $>10^3$ & $>10^3$ \\
		D4 & 634 & 571 & 0.050 & 2300 & 0 & 0.46 & 260 & $>10^3$ \\
		D5 & 751 & 711 & 0.029 & 4300 & 0 & 0.022 & 5.1 & 65 \\
		D6 & 1050 & 1019 & 0.017 & 7400 & 0 & 0 & 0.061 & 0.082 \\
		\hline\hline
	\end{tabular}
	\caption{Phase transition parameters at the benchmark points and corresponding signal-to-noise ratios for \experiment{LISA} (SNR1), \experiment{B-DECIGO} (SNR2), \experiment{DECIGO} (SNR3), and \experiment{BBO} (SNR4). Benchmarks A3 and B3 do not provide a phase transition and are thus omitted.}
	\label{tab:benchmark-results}
\end{table}

\end{appendix}
\clearpage

\bibliographystyle{JHEP}
\bibliography{GaugedL}
\end{document}